\documentclass[preprint,12pt,authoryear]{cas-sc}
\usepackage{natbib}
\setcitestyle{numbers,square}
\usepackage{booktabs}
\usepackage{array, caption, threeparttable}
\usepackage[font=small,labelfont=bf,labelsep=none]{caption}
\usepackage{geometry}
\usepackage{algorithmic}
\usepackage{algorithm}
\usepackage{bigstrut,multirow,rotating,booktabs}
\usepackage{longtable,booktabs}

\def\tsc#1{\csdef{#1}{\textsc{\lowercase{#1}}\xspace}}
\tsc{WGM}
\tsc{QE}

\let\printorcid\relax 

\shortauthors{T. LOU et al.}

\title [mode = title]{A competitive game optimization algorithm for Unmanned Aerial Vehicle path planning}

\begin{abstract}
To solve the Unmanned Aerial Vehicle (UAV) path planning problem, a meta-heuristic optimization algorithm called competitive game optimizer (CGO) is proposed. In the CGO model, three phases of exploration and exploitation, and candidate replacement, are established, corresponding to the player's search for supplies and combat, and the movement toward a safe zone. In the algorithm exploration phase, Levy flight is introduced to improve the global convergence of the algorithm. The encounter probability which adaptively changes with the number of iterations is also introduced in the CGO. The balance between exploration and exploitation of solution space of optimization problem is realized, and each step is described and modeled mathematically. The performance of the CGO was evaluated on a set of 41 test functions taken from CEC2017 and CEC2022. It was then compared with eight widely recognized meta-heuristic optimization algorithms. The simulation results demonstrate that the proposed algorithm successfully achieves a balanced trade-off between exploration and exploitation, showcasing remarkable advantages when compared to seven classical algorithms. In addition, in order to further verify the effectiveness of the CGO, the CGO is applied to 8 practical engineering design problems and UAV path planning, and the results show that the CGO has strong performance in dealing with these practical optimization problems, and has a good application prospect.
\end{abstract}
\begin{document}
\author[author1]{Tai-shan Lou}
\ead{tayzan@sina.com}

\author[author1]{Guang-sheng Guan}
\corref{cor1}
\cormark[1]
\ead{guanguangsheng@163.com}

\author[author1]{Zhe-peng Yue}
\ead{1005074547@qq.com}

\author[author1]{Yu Wang}
\ead{wyu0531@163.com}

\author[author2]{Ren-long Qi}
\ead{zkyqrl@126.com}

\author[author1]{Shi-hao Tong}
\ead{tong.wen@foxmail.com}

\address[author1]{School of Electrical and Information Engineering, Zhengzhou University of Light Industry, Henan Zhengzhou 450002, China}

\address[author2]{School of Electrical Engineering, Zhengzhou University of Science and Technology, Zhengzhou 450064, China}

\cortext[cor1]{Corresponding author}

\maketitle
\section{Introduction}
The process of finding the optimal solution to a problem is commonly referred to as an optimization method\citep{lange2013optimization}. Based on the process of optimization problems, optimization algorithms can be categorized into two distinct groups. The first category consists of deterministic optimization algorithms that rely on the analytic properties of functions. These algorithms are also referred to as exact optimization algorithms. Examples of such algorithms include Newton's method\citep{more1982newton}, the mountain climbing method\citep{ChinnasamyRamachandran-518}, the branch and bound method\citep{NorkinPflug-519}, linear programming\citep{VanderbeiOthers-520} and nonlinear programming\citep{kuhn2013nonlinear}. This type of algorithm typically utilizes mathematical analytical methods to search iteratively by calculating the derivative of the objective function and utilizing gradient information. The precise optimization algorithm has achieved significant success in addressing small-scale optimization problems and has established a comprehensive mathematical optimization theory. However, many practical engineering problems are characterized by large scale, a high number of decision variables, and the presence of discrete and integer constraints. Optimization models for such problems may not be expressed analytically. Additionally, certain practical problems prioritize the need for real-time optimization, where obtaining an approximate solution is sufficient rather than aiming for the optimal solution. Hence, the second category introduces non-deterministic optimization algorithms that rely on random probability strategies.

The non-deterministic optimization algorithm is a type of approximation algorithm based on the concept of meta-heuristics\citep{FranciscoRevollar-522}. These algorithms do not depend on precise mathematical models and analytical solutions. Instead, they draw inspiration from intelligent behaviors and natural phenomena. These algorithms simulate intelligent processes to explore the solution space. While it is often not feasible to find the optimal solution to the problem in most cases. However, these algorithms exhibit high efficiency, strong timeliness, global convergence, and often display superior performance in solving practical optimization problems\citep{cavazzuti2012optimization}. Currently, two main branches of meta-heuristic algorithms, namely evolutionary strategies, and swarm intelligence strategies, are widely popular in the field.

The evolutionary algorithm is based on the natural process of evolution and simulates biological evolution to solve problems\citep{Simon-524}. The evolutionary algorithm is based on mathematical modeling of biological reproduction processes, natural selection, and Darwinian evolution. It guides the search for optimal solutions through operations of selection, crossover, and mutation based on individual fitness. Through iteration, the individual is gradually optimized, and finally the optimal solution of the problem is found. Common evolutionary algorithms include genetic algorithm\citep{Holland-525}, evolutionary programming\citep{YaoLiu-526}, and genetic programming\citep{willis1997genetic}, evolutionary mating algorithm\citep{SulaimanMustaffa-544}, coronavirus optimization algorithm\citep{SalehanDeldari-545}, among others.

The swarm intelligence algorithm is an approach that simulates group behavior and solves problems by leveraging the cooperation and interaction among individuals within the group. In swarm intelligence algorithms, individuals do not directly utilize fitness information of the problem. Instead, they adjust their behaviors by observing the actions of other individuals and adapt to the environment accordingly. Swarm intelligence algorithm has the characteristics of self-organization, local information sharing and parallel computation. Common swarm intelligence algorithms include particle swarm optimization(PSO)\citep{kennedy1995particle}, grey wolf optimizer(GWO)\citep{MirjaliliMirjalili-529}, dung beetle optimizer(DBO)\citep{XueShen-530}, snake optimization\citep{hashim2022snake}, Genghis Khan shark optimizer\citep{HuGuo-546}, and others. 

In addition to evolutionary algorithms and swarm intelligence algorithms, there are other categories within the domain of meta-heuristic optimization algorithms. Other categories of meta-heuristic optimization algorithms include the simulated annealing algorithm\citep{bertsimas1993simulated} and the snow ablation optimizer\citep{deng2023snow}, which are based on physical phenomena. In addition, there are algorithms such as subtraction optimizers based on mathematical operations and arithmetic optimization algorithms\citep{abualigah2021arithmetic}.

According to the no free lunch theorem\citep{adam2019no}, the expected performance of each algorithm is equivalent when solving all optimization problems. In other words, there does not exist a single optimization algorithm that can effectively solve all optimization problems. While an optimization algorithm may perform well in certain optimization problems, it may exhibit poor performance in others. Hence, researchers continually propose new optimization algorithms to address the growing complexity and evolving problem-solving demands. 

Therefore, a new meta-heuristic algorithm, called competitive game optimization algorithm, is proposed in this paper to solve optimization problems. The contributions of this paper are as follows:
\begin{enumerate}[a)]
\item The competitive game optimization (CGO) algorithm is derived from the evolutionary dynamics of player communities in competitive games, drawing inspiration from game mechanics and player behaviors. The mathematical model of the CGO algorithm incorporates stages that correspond to these behaviors, thus simulating their actions;
\item 41 comprehensive benchmark functions of CEC2017 and CEC2022 are used to evaluate the effectiveness and robustness of the CGO algorithm;
\item Apply the CGO to 8 practical engineering design problems. The simulation results are compared with some accepted methods;
\item The Unmanned Aerial Vehicle (UAV) path planning model is constructed, the CGO algorithm is applied to the UAV path planning, and the comparison is made with four similar algorithms;
\end{enumerate}

The rest of this article is arranged as follows. The second section describes the competitive game optimization algorithm in detail. The third section introduces and analyzes the experimental simulation results. The conclusions and some future work are presented in section 4.
\section{Competitive game optimization}
In this section, the inspiration for the CGO algorithm is derived from the game mechanics and player behavior in a military competitive gaming experience. Subsequently, a detailed presentation of the mathematical model is provided. Lastly, the pseudocode for the CGO algorithm is presented.
\subsection{Inspiration}
Once the game begins, players are deployed onto a map resembling a sandbox environment through airdrops, as illustrated in Figure 1. The players have the freedom to choose their initial parachute drop location based on the airplane's route. Random supplies, such as weapons, attachments, medicine, and protective gear, among others, are scattered throughout the residential areas in the game. The residential areas in the game map exhibit a "concentrated-dispersed" distribution. The residential areas within the urban zone are near one another, exhibiting a relatively dense distribution, while the distances between urban zones are greater, resulting in a relatively dispersed arrangement.
\begin{figure}
	\centering
 		\includegraphics[width=1\textwidth]{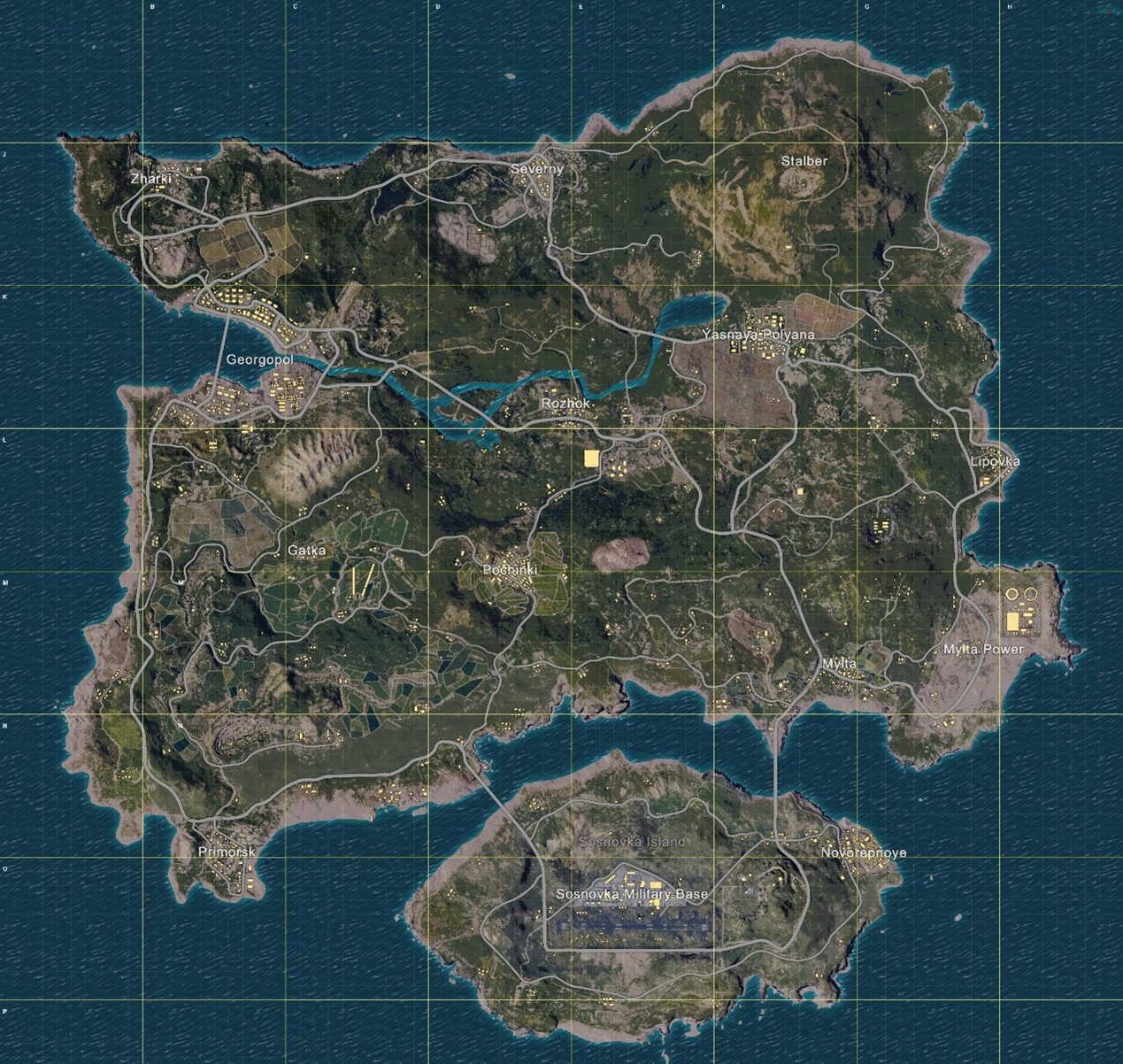}
	  \caption{Overworld of the game}\label{Fig.1}
\end{figure}
To enhance their combat effectiveness and overall score, players must explore the map to obtain a multitude of supplies. During the gameplay process, players have the potential to encounter each other while searching for supplies, resulting in battles when such situations occur. Through battles, players have access to acquire supplies and experience growth, thereby augmenting their scores and levels. Furthermore, the process and outcome of these battles may influence the player's position. Once the battle begins, the player's position becomes exposed, making them susceptible to suppressive fire and even long-range sniping if they choose to stay in the same spot. Consequently, players adopt diverse combat strategies during the combat. For example, players may strategically occupy advantageous terrain, seek cover, employ evasive maneuvering to dodge attacks, or engage in hit-and-run tactics. However, all combat strategies share the common goal of drifting towards the direction of the safe zone. Furthermore, there is a safety zone mechanism in the game, which randomly generates new, smaller safety zones within the current game process, as shown in Figure 2. The range of the safety zone will gradually decrease towards the new safety zone, and players will continuously lose health points outside of the safety zone. This mechanism encourages players to move towards the safety zone.
\begin{figure}
	\centering
 		\includegraphics[width=1\textwidth]{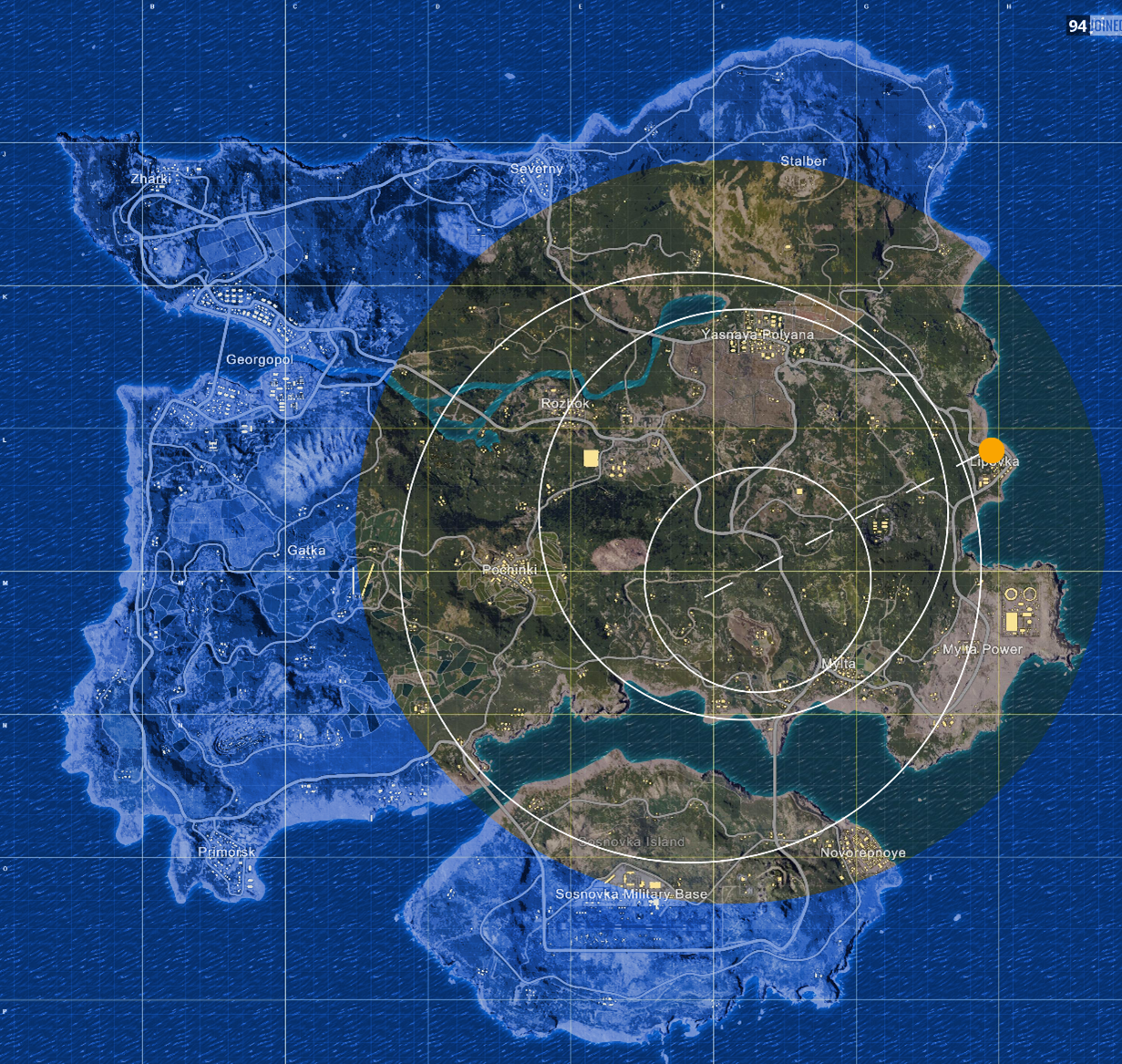}
	  \caption{Safety zone refresh chart}\label{Fig.2}
\end{figure}
\subsection{Initialization stage}
The CGO is a population-based evolution-inspired heuristic algorithm, where players are considered as members of the population for the algorithm. Every individual contains the player's location information, representing candidate solutions to the problem. In the CGO algorithm, the iteration process begins with players randomly selecting a parachute landing location. The population as a whole is represented by a matrix, following the structure of ${N}$ rows and ${Dim}$ columns, as depicted in Eq.(1).In this context, ${N}$ denotes the total count of players, while corresponds to the dimensionality of the solution space.
\begin{equation}
\begin{array}{l}
X = Lb + {r_1} \cdot (Ub - Lb)\\
 = \left[ {\begin{array}{*{20}{c}}
{{X_1}}\\
{{X_2}}\\
 \vdots \\
{{X_{N - 1}}}\\
{{X_N}}
\end{array}} \right] = {\left[ {\begin{array}{*{20}{c}}
{{x_{1,1}}}& \cdots &{{x_{1,Dim}}}\\
{{x_{2,1}}}& \cdots &{{x_{2,Dim}}}\\
 \vdots & \vdots & \vdots \\
{{x_{N - 1,1}}}& \cdots &{{x_{N - 1,Dim}}}\\
{{x_{N,1}}}& \cdots &{{x_{N,Dim}}}
\end{array}} \right]_{N \times Dim}}
\end{array}
\end{equation}
where ${Ub}$ and ${Lb}$ represent the upper bound and lower bound of the solution space, respectively. The variable ${r_1}$ denotes a random number generated within the range $\left( {0,1} \right)$.
Different candidate solutions correspond to distinct values of the objective function. These values are saved using Eq.(2).
\begin{equation}
F = {\left[ {\begin{array}{*{20}{c}}
{F({X_1})}\\
{F({X_2})}\\
 \vdots \\
{F({X_{N - 1}})}\\
{F({X_N})}
\end{array}} \right]_{N \times 1}}
\end{equation}
In metaheuristic optimization algorithms, the quality of candidate solutions is assessed through the objective function value. Consequently, the individual that corresponds to the optimal objective function value is considered the best individual. As candidate solutions undergo updates during the iterative process of the algorithm, the best individual is also updated in each iteration.
\subsection{Search phase}
\begin{figure}
	\centering
 		\includegraphics[width=1\textwidth]{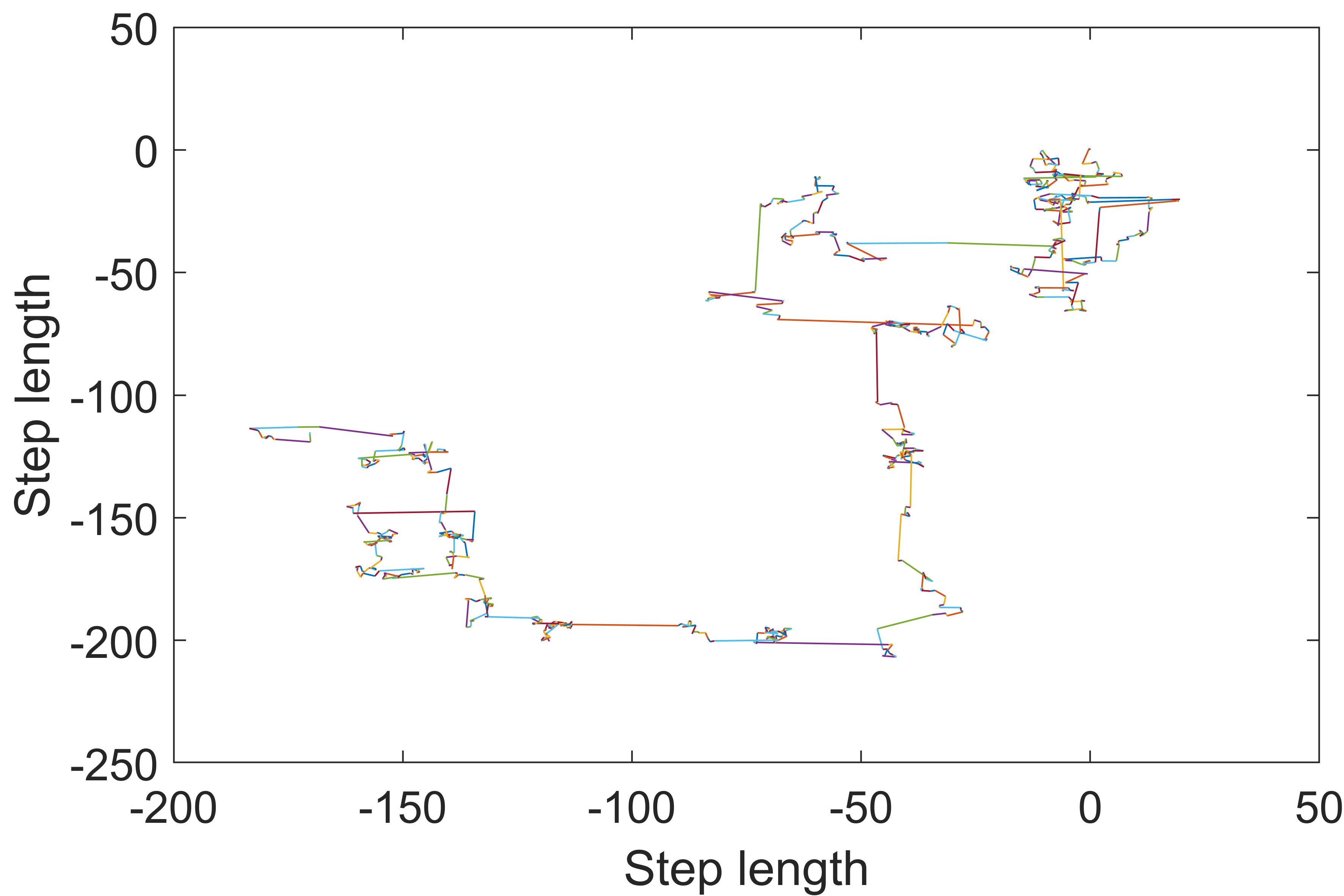}
	  \caption{Levy flight path in 2-dimensional space}\label{Fig.3}
\end{figure}
In this section, the search strategy in the CGO is described in detail. Upon completion of the parachute jump at a random location, the initiation of the iteration process and entry into the search phase are signified. As a result of the "concentrated-dispersed" distribution of resource points on the game map, players utilize a short-step intensive search strategy within individual city areas and employ a long-step search strategy during transitions between different city areas. Levy flight is a non-Gaussian stochastic process characterized by ergodicity and randomness\citep{mantegna1994fast}. The incremental advancement of Levy flight conforms to the Levy distribution, and its flight trajectory is characterized by a random walk that alternates between frequent small steps and infrequent large jumps. Therefore, the Levy flight was used to simulate the situation of players searching for supplies in a map. The trajectory diagram of Levy's flight in two-dimensional space is shown in Figure 3. The relevant mathematical representation is as follows:
\begin{equation}
S = Levy(\beta )\sim \frac{u}{{{{\left| v \right|}^{\frac{1}{\beta }}}}}
\end{equation}
\begin{equation}
u\sim N(0,{\sigma ^2})
\end{equation}
\begin{equation}
v\sim N(0,1)
\end{equation}
\begin{equation}
\sigma  = {\left[ {\frac{{\Gamma \left( {1 + \beta } \right) \cdot \sin \left( {\frac{{\pi  \cdot \beta }}{2}} \right)}}{{\Gamma \left( {\frac{{1 + \beta }}{2}} \right) \cdot \beta  \cdot {2^{\frac{{\beta  - 1}}{2}}}}}} \right]^{\frac{1}{\beta }}}
\end{equation}
where ${\beta}$ represents a numerical value between 0 and 3, and is specifically set at ${\beta = 1.5}$.
The formula for the search process is as follows:
\begin{equation}
X_i^{ts} = X_i^t + \alpha  \cdot S \cdot \left| {Xbes{t^t} - X_i^t} \right|
\end{equation}
where ${S}$ is a random step that follows the Levy distribution; ${\alpha}$ is the step scaling factor, ${\alpha = 1}$. It's the ${i}$ individual in the ${t}$ iteration. ${Xbes{t^t}}$ is the optimal individual in the ${t}$ iteration.
If the newly calculated individual enhances the value of the objective function, the search is deemed valid. Otherwise, the original individual remains unchanged. Update conditions are as follows:
\begin{equation}
X_i^{t + 1} = \left\{ {\begin{array}{*{20}{c}}
{X_i^{ts}}&{F\left( {X_i^{ts}} \right) < F\left( {X_i^t} \right)}\\
{X_i^t}&{else}
\end{array}} \right.
\end{equation}
\subsection{Battle phase}
Throughout the game, there is a probability that players will encounter other players while searching for supplies. And the probability of this encounter increases as the game progresses, because the safe zone is gradually reduced. And the rate of shrinkage of the safe zone is gradually decreasing. For the sake of simulating the real situation, the constructed encounter probability is as follows:
\begin{equation}
E = \sqrt {2 \cdot \left( {\frac{t}{{{T_{\max }}}}} \right) - {{\left( {\frac{t}{{{T_{\max }}}}} \right)}^2}}
\end{equation}
where ${t}$ is the current iteration and ${T_{\max }}$ is the maximum number of iterations.

During this phase, a random number ranging from 0 to 1 is generated and compared with the encounter probability. This determines whether the player engages in a fight during the current iteration. The player's combat behavior is described as:
\begin{equation}
{\scriptsize
\left\{ {\begin{array}{*{20}{c}}
{X_{i,j}^{ts} = {r_2} \cdot X_{i,j}^t + \left( {1 - {r_2}} \right) \cdot X_{k,j}^t + {c_1} \cdot \left( {X_{i,j}^t - X_{k,j}^t} \right) + {c_2} \cdot \left( {Xbest_j^t - X_{i,j}^t} \right)}\\
{X_{k,j}^{ts} = {r_3} \cdot X_{k,j}^t + \left( {1 - {r_3}} \right) \cdot X_{i,j}^t + {c_2} \cdot \left( {X_{k,j}^t - X_{i,j}^t} \right) + {c_1} \cdot \left( {Xbest_j^t - X_{k,j}^t} \right)}
\end{array}} \right.}
\end{equation}
where ${r_2}$ and ${r_3}$ refer to uniformly distributed random numbers within the range ${\left( {0,1} \right)}$. Likewise, ${c_1}$ and ${c_2}$ represent uniformly distributed random numbers within the range  ${\left( { - 1,1} \right)}$. ${X_{i,j}^t}$ denotes the ${j}$ dimension of the ${i}$ individual at the ${t}$ iteration, while ${Xbest_j^t}$ represents the ${j}$ dimension of the optimal individual for the current iteration. The variable ${k}$ represents random integers other than ${i}$, represents random integers other than  , chosen from the range ${\left( {1,N} \right)}$. Lastly, ${X_{k,j}^t}$ signifies the ${j}$ dimension of the individual opposing the ${i}$ individual in combat.

After the battle, calculate the objective function value of the post-battle individual. If the objective function value of the individual is improved, the position of the player in this battle is considered to have changed, that is, the individual update. Otherwise, the previous individual remains unchanged. The update conditions are as follows:
\begin{equation}
X_i^{t + 1} = \left\{ {\begin{array}{*{20}{c}}
{X_i^{ts}}&{F\left( {X_i^{ts}} \right) < F\left( {X_i^t} \right)}\\
{X_i^t}&{else}
\end{array}} \right.
\end{equation}
\subsection{Advance towards the safe zone}
Due to the safety zone mechanism in the game, players outside of the safety zone continue to lose health. Low health will cause the player's combat effectiveness to decrease and their points to decrease. This mechanic encourages the player to move towards the safe zone as it shrinks. In this study, individuals with low objective function values were considered players in the non-safe zone. Therefore, in each iteration, all individuals are sorted by fitness value, and the twenty percent individuals with the worst fitness value will move towards the position of the best individual. The player's actions toward the safe zone are described as:
\begin{equation}
X_i^{t + 1} = Xbes{t^t} + chy \cdot \left( {Xbes{t^t} - X_i^t} \right)
\end{equation}
\begin{equation}
chy = \tan \left[ {\pi  \cdot \left( {{r_4} - 0.5} \right)} \right]
\end{equation}
where ${r_4}$ denote a random number within the range of ${\left( {0,1} \right)}$.

The population is updated according to Eq.(3) to Eq.(13) for individual updates as the player searches for supplies and battles and moves toward the safe zone. This process is repeated, recording the best individuals for the current iteration until the last iteration is completed. When the running of the CGO is complete, the optimal player individual obtained during the algorithm iteration is output as the best solution. The entire process of the CGO algorithm is summarized in Algorithm 1.
\begin{algorithm}[!h]
    \caption{Competitive game optimizer (CGO)}
    \label{alg:AOS}
    \renewcommand{\algorithmicrequire}{\textbf{Input:}}
    \renewcommand{\algorithmicensure}{\textbf{Output:}}
    \begin{algorithmic}[1]
        \REQUIRE maximum number of iterations ${T_{\max }}$, population number ${N}$, search for upper and lower bounds ${Ub}$ and ${Lb}$; objective function ${F\left(  *  \right)}$.  
        \ENSURE best individual    
        \STATE   Initialization: the player group ${X_i}\left( {i = 1,2, \cdots ,N} \right)$, $t = 0$, ${N_1} = 0.8*N$
        \STATE Fitness evaluation
        \STATE Record the current best individual
        \WHILE{$t < {T_{\max }}$}
            \STATE Calculate the probability $E$ through Eq.(9).
            \STATE Calculate the Levy random step size $S$ through Eq.(3).
            \FOR{$i \leftarrow {N_1} + 1\begin{array}{*{20}{c}}{}\end{array}to\begin{array}{*{20}{c}}{}\end{array}N$}
                \STATE The position of the inferior twenty percent individual’s is updated by Eq.(12)
            \ENDFOR
            \FOR{every individual}
                \STATE Calculate every individual’s new position through 
                Eq.(7).
                \STATE Update every individual’s position through 
                Eq.(8).
            \ENDFOR
            \FOR{every individual}
                \IF {$rand < E$}
                    \STATE Calculate every individual’s new position through Eq.(10).
                    \STATE Update every individual’s position through Eq.(11).
                \ENDIF
            \ENDFOR
            \STATE Fitness evaluation
            \STATE Update best individual
            \STATE $t = t + 1$
        \ENDWHILE
        \RETURN best individual
    \end{algorithmic}
\end{algorithm}
\section{Experimental results and analysis}
In this section, the performance of the CGO is verified through a total of 41 test functions for CEC2017 and CEC2022, as well as 8 real-world engineering design problems. The whole experiment consists of the following three parts: (1) benchmark functions and experimental settings; (2) experiments on the CEC2017 and CEC2022 test suite; and (3) experimental research on 8 real-world constrained optimization problems.
\subsection{Benchmark functions and experimental settings}
The 30-dimensional benchmark problems in 29 CEC2017 test suites and 20-dimensional benchmark problems in 12 CEC2022 test suites are studied experimentally. According to their characteristics, 41 unconstrained benchmark problems are classified, as shown in Table 1.
\begin{table*}
\caption{Benchmark problems are classified}
\label{tab_fwdc}
 \begin{tabular*}{\tblwidth}{@{}LLLL@{}}
\toprule 
Benchmark problems & Category name \\ 
\midrule 
$\rm CEC2017_1$, $\rm CEC2017_3$, $\rm CEC2022_1$ & unimodal problems \\
$\rm CEC2017_4$-$\rm CEC2017_{10}$, $\rm CEC2022_2$-$\rm CEC2022_5$ & simple multimodal problems \\
$\rm CEC2017_{11}$-$\rm CEC2017_{20}$, $\rm CEC2022_6$-$\rm CEC2022_8$ & hybrid problems \\
$\rm CEC2017_{21}$- $\rm CEC2017_{30}$, $\rm CEC2022_9$-$\rm CEC2022_{12}$ & composition problems \\
\bottomrule 
\end{tabular*}
\end{table*}
Table 1 presents the utilization of unimodal problems to assess algorithm exploitation capabilities, given their possession of a singular global optimum. Conversely, simple multimodal problems pose greater difficulties than single-modal problems due to their abundance of local optimal solutions. Additionally, hybrid and combined problems can effectively simulate real search spaces by encompassing distinct characteristics within different regions. Utilizing these benchmark functions aids in reflecting the potential performance of an algorithm when confronted with a real optimization problem. For this study, termination conditions consisted of 50 search agents and a total of 1000 iterations. In order to compare with the CGO, seven widely recognized optimization algorithms, namely the GWO, beluga whale optimization (BWO)\citep{ZhongLi-541}, the DBO, sparrow search algorithm (SSA)\citep{XueShen-540}, the PSO, golden jackal optimization (GJO)\citep{ChopraAnsari-539}, and slime mould algorithm (SMA)\citep{li2020slime}, were chosen. To ensure fairness, the parameter settings for each algorithm remained consistent.
\subsection{Experiments on the CEC2017 and CEC2022 test suite}
In this section, the CGO is compared to the GWO, the BWO, the DBO, the SSA, the PSO, the GJO, and the SMA by using a total of 41 test suites CEC2017 and CEC2022. Table 2 and Table 3 summarize the simulation results of 8 algorithms on CEC2017 and CEC2022 respectively. Based on optimal values, standard deviations, and mean values, the comments are as follows:

For unimodal problems $\rm CEC2017_1$ and $\rm CEC2017_3$, as well as $\rm CEC2022_1$, the CGO showed strong competitiveness. Even though the SSA achieved the best result on $\rm CEC2017_1$, it can be seen from the data in the table that the result of the CGO is much better than other algorithms except the SSA, and the result is almost the same as that of the SSA. In $\rm CEC2017_3$, the global optimal solution is directly searched, while other algorithms cannot. The SMA also provided the best result on $\rm CEC2022_1$, but in terms of standard deviation, it was surpassed by the CGO.

For simple multimodal problems $\rm CEC2017_4$-$\rm CEC2017_{10}$ and $\rm CEC2022_2$-$\rm CEC2022_5$, the CGO outperforms other algorithms on $\rm CEC2017_4$-$\rm CEC2017_6$ and $\rm CEC2022_2$-$\rm CEC2022_4$. On other test suites, although they are inferior to other algorithms in terms of optimal value, they have different degrees of exceedance in terms of their mean and standard deviation.

For hybrid problems CEC2017$_{11}$-CEC2017$_{20}$ and CEC 2022$_6$-CEC2022$_8$, the CGO achieved the best results on $\rm CEC2017_{12}$-$\rm CEC2017_{15}$, $\rm CEC2017_{18}$, $\rm CEC2017_{20}$ and CEC 2022$_7$-$\rm CEC2022_8$, respectively. On $\rm CEC2017_{11}$, $\rm CEC2017_{17}$, and $\rm CEC2017_{19}$, there is little difference between the results of the CGO and the results of the algorithm that obtained the optimal value.

For composition problems $\rm CEC2017_{21}$-$\rm CEC2017_{30}$ and $\rm CEC2022_9$-$\rm CEC2022_{12}$, the CGO achieved the best results on all CEC2017 test suites except $\rm CEC2017_{27}$ and $\rm CEC2017_{30}$. On $\rm CEC2022_9$ and $\rm CEC2022_{12}$, the CGO is inferior to the PSO in terms of optimal values. On $\rm CEC2022_{10}$ and $\rm CEC2022_{11}$, the CGO is inferior to the SMA in terms of optimal value. However, the CGO outperforms other comparison algorithms in terms of standard deviations and means on individual test suites.

{
\tiny
\begin{longtable}{cccccccccc}
	\caption{Experimental results obtained by 8 algorithms on CEC2017 test suite}	\\ 
 \hline
     &   & SMA & GWO & BWO & DBO & SSA & PSO & GJO & CGO \bigstrut\\
			\hline 
			\endfirsthead
			
			\multicolumn{3}{c}%
			{{\bfseries  -- continued from previous page}} \\
			\hline 	
      &   & SMA & GWO & BWO & DBO & SSA & PSO & GJO & CGO \bigstrut\\
                \hline 
			\endhead
			
			\hline \multicolumn{3}{l}{{Continued on next page}} \\ 
			\endfoot
			
			\hline
			\endlastfoot		
    \hline
    $\rm CEC2007_1$ & best  & 9.45E+02 & 8.34E+07 & 3.97E+10 & 1.01E+04 & \textbf{1.00E+02} & 5.68E+04 & 5.07E+09 & 1.01E+02 \bigstrut[t]\\
          & std   & 6.67E+03 & 1.39E+09 & 3.89E+09 & 1.46E+07 & \textbf{3.14E+03} & 1.17E+05 & 3.75E+09 & 4.35E+03 \\
          & mean  & 8.16E+03 & 1.90E+09 & 5.01E+10 & 1.24E+07 & \textbf{2.69E+03} & 2.36E+05 & 1.02E+10 & 4.08E+03 \\
    $\rm CEC2017_3$ & best  & 3.78E+02 & 1.83E+04 & 5.21E+04 & 4.31E+04 & 2.08E+04 & 2.21E+03 & 2.64E+04 & \textbf{3.00E+02} \\
          & std   & 7.82E+02 & 1.25E+04 & 5.51E+03 & 1.37E+04 & 4.55E+03 & 2.92E+03 & 1.07E+04 & \textbf{5.50E+02} \\
          & mean  & 9.51E+02 & 4.45E+04 & 7.58E+04 & 6.78E+04 & 2.70E+04 & 6.28E+03 & 4.85E+04 & \textbf{5.37E+02} \\
    $\rm CEC2017_4$ & best  & 4.71E+02 & 4.93E+02 & 9.16E+03 & 4.67E+02 & 4.04E+02 & 4.21E+02 & 6.41E+02 & \textbf{4.00E+02} \\
          & std   & \textbf{1.56E+01} & 1.26E+02 & 1.30E+03 & 9.45E+01 & 2.74E+01 & 2.60E+01 & 9.29E+02 & 3.33E+01 \\
          & mean  & 4.95E+02 & 6.22E+02 & 1.14E+04 & 5.77E+02 & 4.93E+02 & \textbf{4.78E+02} & 1.32E+03 & 4.84E+02 \\
    $\rm CEC2017_5$ & best  & 5.56E+02 & 5.53E+02 & 8.63E+02 & 6.38E+02 & 6.68E+02 & 6.09E+02 & 6.31E+02 & \textbf{5.50E+02} \\
          & std   & 2.59E+01 & 3.72E+01 & \textbf{2.14E+01} & 4.90E+01 & 5.09E+01 & 2.70E+01 & 5.16E+01 & 2.60E+01 \\
          & mean  & 6.01E+02 & 6.04E+02 & 9.09E+02 & 7.19E+02 & 7.59E+02 & 6.70E+02 & 7.20E+02 & \textbf{5.88E+02} \\
    $\rm CEC2017_6$ & best  & 6.01E+02 & 6.03E+02 & 6.70E+02 & 6.20E+02 & 6.20E+02 & 6.21E+02 & 6.20E+02 & \textbf{6.00E+02} \\
          & std   & 4.57E+00 & 4.01E+00 & 6.35E+00 & 1.18E+01 & 1.20E+01 & 6.68E+00 & 7.40E+00 & \textbf{7.41E-01} \\
          & mean  & 6.06E+02 & 6.09E+02 & 6.86E+02 & 6.39E+02 & 6.47E+02 & 6.40E+02 & 6.34E+02 & \textbf{6.00E+02} \\
    $\rm CEC2017_7$ & best  & \textbf{7.81E+02} & 8.04E+02 & 1.26E+03 & 8.32E+02 & 9.01E+02 & 8.06E+02 & 9.10E+02 & 7.99E+02 \\
          & std   & 3.30E+01 & 4.08E+01 & 3.82E+01 & 8.25E+01 & 1.08E+02 & \textbf{2.78E+01} & 5.87E+01 & 3.94E+01 \\
          & mean  & \textbf{8.52E+02} & 8.63E+02 & 1.37E+03 & 9.43E+02 & 1.21E+03 & 8.53E+02 & 1.02E+03 & 8.78E+02 \\
    $\rm CEC2017_8$ & best  & 8.72E+02 & \textbf{8.50E+02} & 1.09E+03 & 9.10E+02 & 9.33E+02 & 8.86E+02 & 9.07E+02 & 8.53E+02 \\
          & std   & 2.41E+01 & 4.13E+01 & \textbf{1.95E+01} & 4.81E+01 & 2.51E+01 & 2.15E+01 & 2.98E+01 & 2.52E+01 \\
          & mean  & 9.18E+02 & 8.97E+02 & 1.13E+03 & 1.01E+03 & 9.85E+02 & 9.22E+02 & 9.58E+02 & \textbf{8.88E+02} \\
    $\rm CEC2017_9$ & best  & \textbf{9.79E+02} & 1.00E+03 & 7.93E+03 & 1.40E+03 & 3.34E+03 & 3.29E+03 & 2.77E+03 & 1.16E+03 \\
          & std   & 1.50E+03 & 4.88E+02 & 8.68E+02 & 2.16E+03 & 4.85E+02 & 1.17E+03 & 1.50E+03 & \textbf{4.25E+02} \\
          & mean  & 3.16E+03 & 1.69E+03 & 1.04E+04 & 5.36E+03 & 5.18E+03 & 5.00E+03 & 4.86E+03 & \textbf{1.68E+03} \\
    $\rm CEC2017_{10}$ & best  & \textbf{2.57E+03} & 3.58E+03 & 7.92E+03 & 4.66E+03 & 3.94E+03 & 3.32E+03 & 4.00E+03 & 3.88E+03 \\
          & std   & 7.22E+02 & 1.16E+03 & \textbf{3.01E+02} & 7.47E+02 & 6.83E+02 & 7.68E+02 & 1.61E+03 & 6.24E+02 \\
          & mean  & \textbf{4.38E+03} & 4.81E+03 & 8.51E+03 & 5.88E+03 & 5.38E+03 & 4.80E+03 & 6.39E+03 & 5.13E+03 \\
    $\rm CEC2017_{11}$ & best  & 1.21E+03 & 1.30E+03 & 4.58E+03 & 1.21E+03 & 1.16E+03 & \textbf{1.14E+03} & 1.53E+03 & 1.16E+03 \\
          & std   & 5.28E+01 & 9.23E+02 & 1.02E+03 & 1.25E+02 & 5.53E+01 & \textbf{3.32E+01} & 1.56E+03 & 6.04E+01 \\
          & mean  & 1.28E+03 & 2.01E+03 & 6.96E+03 & 1.47E+03 & 1.26E+03 & \textbf{1.21E+03} & 2.98E+03 & 1.24E+03 \\
    $\rm CEC2017_{12}$ & best  & 2.11E+05 & 2.30E+06 & 6.02E+09 & 1.47E+06 & 9.75E+04 & 9.58E+04 & 7.10E+07 & \textbf{9.17E+03} \\
          & std   & 2.02E+06 & 7.50E+07 & 1.74E+09 & 2.62E+07 & 3.83E+05 & 7.59E+05 & 5.21E+08 & \textbf{6.22E+04} \\
          & mean  & 2.99E+06 & 6.15E+07 & 9.27E+09 & 1.95E+07 & 5.44E+05 & 9.64E+05 & 6.82E+08 & \textbf{7.98E+04} \\
    $\rm CEC2017_{13}$ & best  & 7.33E+03 & 2.64E+04 & 1.95E+09 & 2.13E+04 & 3.01E+03 & 2.15E+03 & 2.03E+05 & \textbf{1.59E+03} \\
          & std   & 2.38E+04 & 9.66E+06 & 1.91E+09 & 4.63E+06 & 1.95E+04 & \textbf{1.93E+04} & 3.68E+08 & 2.11E+04 \\
          & mean  & 2.91E+04 & 4.04E+06 & 5.28E+09 & 2.62E+06 & \textbf{1.99E+04} & 1.85E+04 & 1.82E+08 & 2.14E+04 \\
    $\rm CEC2017_{14}$ & best  & 1.06E+04 & 8.98E+03 & 2.16E+05 & 1.00E+04 & 3.05E+03 & 2.05E+03 & 3.53E+04 & \textbf{1.56E+03} \\
          & std   & 4.73E+04 & 5.04E+05 & 1.52E+06 & 9.34E+04 & 2.65E+04 & \textbf{1.71E+04} & 6.22E+05 & 2.21E+04 \\
          & mean  & 7.00E+04 & 4.95E+05 & 2.64E+06 & 1.09E+05 & 3.47E+04 & 1.69E+04 & 5.88E+05 & \textbf{8.84E+03} \\
    $\rm CEC2017_{15}$ & best  & 2.24E+03 & 8.37E+03 & 7.92E+07 & 9.45E+03 & 1.83E+03 & 1.83E+03 & 2.69E+04 & \textbf{1.77E+03} \\
          & std   & 1.40E+04 & 1.08E+06 & 1.08E+08 & 9.00E+04 & 1.09E+04 & \textbf{6.59E+03} & 1.85E+06 & 1.24E+04 \\
          & mean  & 2.28E+04 & 6.36E+05 & 2.31E+08 & 7.83E+04 & \textbf{1.04E+04} & 8.18E+03 & 1.51E+06 & 1.44E+04 \\
    $\rm CEC2017_{16}$ & best  & \textbf{1.88E+03} & 1.92E+03 & 4.18E+03 & 2.06E+03 & 2.32E+03 & 2.20E+03 & 2.19E+03 & 2.01E+03 \\
          & std   & 3.40E+02 & 2.77E+02 & 3.93E+02 & 4.74E+02 & 4.09E+02 & \textbf{2.66E+02} & 3.56E+02 & 2.86E+02 \\
          & mean  & 2.59E+03 & 2.48E+03 & 5.29E+03 & 3.26E+03 & 2.97E+03 & 2.71E+03 & 2.88E+03 & \textbf{2.47E+03} \\
    $\rm CEC2017_{17}$ & best  & 1.87E+03 & 1.82E+03 & 3.05E+03 & 1.87E+03 & 1.97E+03 & \textbf{1.79E+03} & 1.82E+03 & 1.80E+03 \\
          & std   & 2.20E+02 & \textbf{1.92E+02} & 4.31E+02 & 2.80E+02 & 2.84E+02 & 2.54E+02 & 2.08E+02 & 2.78E+02 \\
          & mean  & 2.30E+03 & \textbf{2.08E+03} & 3.81E+03 & 2.47E+03 & 2.44E+03 & 2.25E+03 & 2.18E+03 & 2.18E+03 \\
    $\rm CEC2017_{18}$ & best  & 2.02E+05 & 6.37E+04 & 6.27E+06 & 3.04E+04 & 4.29E+04 & 4.80E+04 & 4.83E+04 & \textbf{4.83E+03} \\
          & std   & 1.32E+06 & 1.83E+06 & 1.69E+07 & 1.22E+06 & 2.93E+05 & 3.31E+05 & 1.84E+06 & \textbf{3.05E+04} \\
          & mean  & 1.41E+06 & 1.34E+06 & 2.67E+07 & 8.30E+05 & 2.69E+05 & 4.27E+05 & 1.66E+06 & \textbf{3.73E+04} \\
    $\rm CEC2017_{19}$ & best  & 2.04E+03 & 5.60E+03 & 6.52E+07 & 2.10E+03 & 1.99E+03 & \textbf{1.96E+03} & 2.75E+04 & 1.97E+03 \\
          & std   & 2.16E+04 & 7.03E+05 & 1.49E+08 & 9.61E+05 & 1.12E+04 & \textbf{5.39E+03} & 3.00E+07 & 1.77E+04 \\
          & mean  & 2.73E+04 & 7.19E+05 & 2.94E+08 & 4.81E+05 & 8.97E+03 & \textbf{6.62E+03} & 1.18E+07 & 2.03E+04 \\
    $\rm CEC2017_{20}$ & best  & 2.22E+03 & 2.16E+03 & 2.57E+03 & 2.17E+03 & 2.33E+03 & 2.18E+03 & 2.23E+03 & \textbf{2.10E+03} \\
          & std   & 1.78E+02 & 1.63E+02 & \textbf{1.37E+02} & 2.18E+02 & 2.25E+02 & 2.51E+02 & 1.71E+02 & 2.41E+02 \\
          & mean  & 2.53E+03 & \textbf{2.43E+03} & 2.91E+03 & 2.68E+03 & 2.74E+03 & 2.60E+03 & 2.55E+03 & 2.49E+03 \\
    $\rm CEC2017_{21}$ & best  & 2.37E+03 & 2.36E+03 & 2.57E+03 & 2.42E+03 & 2.41E+03 & 2.41E+03 & 2.41E+03 & \textbf{2.33E+03} \\
          & std   & 3.23E+01 & 3.05E+01 & 4.04E+01 & 4.51E+01 & 5.77E+01 & 3.14E+01 & 2.84E+01 & \textbf{2.38E+01} \\
          & mean  & \textbf{2.42E+03} & 2.39E+03 & 2.70E+03 & 2.51E+03 & 2.51E+03 & 2.47E+03 & 2.47E+03 & 2.38E+03 \\
    $\rm CEC2017_{22}$ & best  & 2.30E+03 & 2.39E+03 & 7.30E+03 & \textbf{2.30E+03} & \textbf{2.30E+03} & \textbf{2.30E+03} & 2.67E+03 & 2.30E+03 \\
          & std   & 1.11E+03 & 1.97E+03 & 5.35E+02 & 2.10E+03 & 1.75E+03 & 1.80E+03 & 2.41E+03 & 1.58E+03 \\
          & mean  & 5.66E+03 & 5.08E+03 & 8.25E+03 & 5.62E+03 & 6.35E+03 & \textbf{3.45E+03} & 5.47E+03 & 6.30E+03 \\
    $\rm CEC2017_{23}$ & best  & 2.71E+03 & 2.71E+03 & 3.19E+03 & 2.84E+03 & 2.81E+03 & 2.85E+03 & 2.80E+03 & \textbf{2.69E+03} \\
          & std   & 2.73E+01 & 5.41E+01 & 5.96E+01 & 5.29E+01 & 7.67E+01 & 1.06E+02 & 3.97E+01 & \textbf{2.42E+01} \\
          & mean  & 2.75E+03 & 2.77E+03 & 3.29E+03 & 2.93E+03 & 2.93E+03 & 3.08E+03 & 2.87E+03 & \textbf{2.74E+03} \\
    $\rm CEC2017_{24}$ & best  & 2.89E+03 & 2.88E+03 & 3.33E+03 & 2.96E+03 & 2.95E+03 & 3.03E+03 & 2.96E+03 & \textbf{2.87E+03} \\
          & std   & \textbf{3.26E+01} & 5.03E+01 & 6.98E+01 & 6.94E+01 & 9.43E+01 & 8.76E+01 & 6.60E+01 & 3.30E+01 \\
          & mean  & 2.94E+03 & 2.95E+03 & 3.52E+03 & 3.08E+03 & 3.10E+03 & 3.17E+03 & 3.06E+03 & \textbf{2.92E+03} \\
    $\rm CEC2017_{25}$ & best  & \textbf{2.88E+03} & 2.93E+03 & 4.11E+03 & 2.89E+03 & \textbf{2.88E+03} & \textbf{2.88E+03} & 3.00E+03 & \textbf{2.88E+03} \\
          & std   & \textbf{4.40E+00} & 4.01E+01 & 1.10E+02 & 4.41E+01 & 1.72E+01 & 1.48E+01 & 1.27E+02 & 1.75E+01 \\
          & mean  & \textbf{2.89E+03} & 2.99E+03 & 4.34E+03 & 2.93E+03 & 2.90E+03 & \textbf{2.89E+03} & 3.17E+03 & 2.90E+03 \\
    $\rm CEC2017_{26}$ & best  & 4.40E+03 & 4.37E+03 & 9.11E+03 & 4.83E+03 & 2.90E+03 & 2.81E+03 & 5.06E+03 & \textbf{2.80E+03} \\
          & std   & \textbf{2.67E+02} & 3.55E+02 & 5.51E+02 & 6.28E+02 & 1.22E+03 & 2.00E+03 & 5.01E+02 & 4.38E+02 \\
          & mean  & 4.85E+03 & 4.81E+03 & 1.04E+04 & 6.42E+03 & 6.29E+03 & 5.80E+03 & 5.81E+03 & \textbf{4.76E+03} \\
    $\rm CEC2017_{27}$ & best  & 3.19E+03 & 3.22E+03 & 3.65E+03 & 3.22E+03 & 3.22E+03 & \textbf{3.16E+03} & 3.24E+03 & 3.20E+03 \\
          & std   & 1.78E+01 & 1.93E+01 & 1.42E+02 & 4.86E+01 & 3.40E+01 & 1.56E+02 & 6.26E+01 & \textbf{1.69E+01} \\
          & mean  & \textbf{3.22E+03} & 3.25E+03 & 3.90E+03 & 3.29E+03 & 3.27E+03 & 3.31E+03 & 3.34E+03 & 3.23E+03 \\
    $\rm CEC2017_{28}$ & best  & 3.21E+03 & 3.31E+03 & 5.73E+03 & 3.25E+03 & 3.17E+03 & 3.19E+03 & 3.49E+03 & \textbf{3.10E+03} \\
          & std   & 3.88E+01 & 4.52E+01 & 2.74E+02 & 6.16E+01 & \textbf{2.38E+01} & 2.57E+01 & 2.98E+02 & 4.64E+01 \\
          & mean  & 3.25E+03 & 3.40E+03 & 6.26E+03 & 3.36E+03 & \textbf{3.22E+03} & 3.23E+03 & 3.82E+03 & \textbf{3.22E+03} \\
    $\rm CEC2017_{29}$ & best  & 3.48E+03 & 3.57E+03 & 5.46E+03 & 3.51E+03 & 3.71E+03 & 3.59E+03 & 3.64E+03 & \textbf{3.43E+03} \\
          & std   & 2.18E+02 & \textbf{1.76E+02} & 6.49E+02 & 2.52E+02 & 2.05E+02 & 3.13E+02 & 2.48E+02 & 2.27E+02 \\
          & mean  & 3.89E+03 & 3.84E+03 & 6.54E+03 & 4.13E+03 & 4.15E+03 & 4.09E+03 & 4.15E+03 & \textbf{3.81E+03} \\
    $\rm CEC2017_{30}$ & best  & 6.85E+03 & 1.46E+06 & 2.38E+08 & 9.54E+03 & \textbf{5.02E+03} & \textbf{1.03E+04} & 6.02E+06 & 5.99E+03 \\
          & std   & 9.34E+03 & 4.73E+06 & 2.62E+08 & 7.63E+06 & 5.89E+03 & 4.24E+04 & 2.26E+07 & \textbf{4.13E+03} \\
          & mean  & 2.31E+04 & 6.55E+06 & 7.88E+08 & 2.53E+06 & 1.35E+04 & 4.76E+04 & 3.86E+07 & \textbf{1.07E+04} \bigstrut[b]\\
        \label{table9}
\end{longtable}
}

\begin{table*}[!t]
  \centering
  \caption{Experimental results obtained by 8 algorithms on CEC2022 test suite}
    \resizebox{\textwidth}{!}{
    \begin{tabular}{cccccccccc}
    \hline
      &   & SMA & GWO & BWO & DBO & SSA & PSO & GJO & CGO \bigstrut\\
    \hline
    $\rm CEC2022_1$ & best & \textbf{3.00E+02} & 2.42E+03 & 4.18E+04 & 6.36E+03 & 3.11E+02 & 3.04E+02 & 4.95E+03 & \textbf{3.00E+02} \bigstrut[t]\\
        & std & 2.29E-01 & 5.84E+03 & 1.63E+04 & 8.19E+03 & 3.06E+02 & 3.14E+00 & 4.45E+03 & \textbf{1.05E-08} \\
        & mean & \textbf{3.00E+02} & 9.86E+03 & 6.44E+04 & 1.43E+04 & 5.77E+02 & 3.08E+02 & 1.28E+04 & \textbf{3.00E+02} \\
    $\rm CEC2022_2$ & best & 4.23E+02 & 4.45E+02 & 1.40E+03 & 4.09E+02 & \textbf{4.00E+02} & 4.15E+02 & 4.91E+02 & \textbf{4.00E+02} \\
        & std & 2.37E+01 & 2.44E+01 & 2.50E+02 & 3.68E+01 & 2.53E+01 & 2.71E+01 & 7.49E+01 & \textbf{2.06E+01} \\
        & mean & 4.57E+02 & 4.77E+02 & 2.01E+03 & 4.65E+02 & 4.46E+02 & \textbf{4.39E+02} & 5.93E+02 & 4.43E+02 \\
    $\rm CEC2022_3$ & best & \textbf{6.00E+02} & 6.01E+02 & 6.64E+02 & 6.01E+02 & 6.10E+02 & 6.18E+02 & 6.09E+02 & \textbf{6.00E+02} \\
        & std & 5.74E-01 & 3.04E+00 & 5.68E+00 & 8.97E+00 & 1.09E+01 & 8.46E+00 & 1.04E+01 & \textbf{1.55E-02} \\
        & mean & 6.01E+02 & 6.04E+02 & 6.76E+02 & 6.20E+02 & 6.29E+02 & 6.33E+02 & 6.22E+02 & \textbf{6.00E+02} \\
    $\rm CEC2022_4$ & best & 8.34E+02 & 8.20E+02 & 9.38E+02 & 8.60E+02 & 8.67E+02 & 8.31E+02 & 8.46E+02 & \textbf{8.19E+02} \\
        & std & 2.35E+01 & 1.75E+01 & \textbf{1.18E+01} & 3.03E+01 & 1.78E+01 & 2.12E+01 & 2.40E+01 & 1.91E+01 \\
        & mean & 8.67E+02 & \textbf{8.45E+02} & 9.64E+02 & 9.08E+02 & 8.98E+02 & 8.62E+02 & 8.92E+02 & 8.47E+02 \\
    $\rm CEC2022_5$ & best & 9.42E+02 & 9.09E+02 & 2.83E+03 & 9.69E+02 & 1.99E+03 & \textbf{9.02E+02} & 1.16E+03 & 9.30E+02 \\
        & std & 3.59E+02 & 1.99E+02 & 2.68E+02 & 4.88E+02 & 1.83E+02 & 4.95E+02 & 3.70E+02 & \textbf{1.60E+02} \\
        & mean & 1.41E+03 & 1.10E+03 & 3.46E+03 & 1.64E+03 & 2.47E+03 & 1.86E+03 & 1.65E+03 & \textbf{1.09E+03} \\
    $\rm CEC2022_6$ & best & 2.35E+03 & 2.46E+03 & 1.92E+08 & 1.95E+03 & \textbf{1.84E+03} & 1.89E+03 & 6.05E+03 & 2.01E+03 \\
        & std & 6.82E+03 & 1.49E+06 & 3.91E+08 & 4.12E+05 & 6.64E+03 & \textbf{4.52E+03} & 1.76E+07 & 7.36E+03 \\
        & mean & 2.14E+04 & 3.25E+05 & 9.97E+08 & 1.51E+05 & 9.06E+03 & \textbf{5.02E+03} & 1.24E+07 & 9.69E+03 \\
    $\rm CEC2022_7$ & best & \textbf{2.02E+03} & \textbf{2.02E+03} & 2.16E+03 & 2.05E+03 & 2.04E+03 & 2.05E+03 & 2.04E+03 & \textbf{2.02E+03} \\
        & std & 4.15E+01 & 4.06E+01 & \textbf{2.28E+01} & 3.91E+01 & 6.10E+01 & 4.49E+01 & 4.36E+01 & 4.06E+01 \\
        & mean & 2.08E+03 & \textbf{2.07E+03} & 2.21E+03 & 2.11E+03 & 2.13E+03 & 2.11E+03 & 2.12E+03 & \textbf{2.07E+03} \\
    $\rm CEC2022_8$ & best & \textbf{2.22E+03} & \textbf{2.22E+03} & 2.25E+03 & \textbf{2.22E+03} & \textbf{2.22E+03} & \textbf{2.22E+03} & 2.23E+03 & \textbf{2.22E+03} \\
        & std & 4.34E+01 & 5.61E+01 & \textbf{1.18E+01} & 6.18E+01 & 6.93E+01 & 7.89E+01 & 5.21E+01 & 3.56E+01 \\
        & mean & \textbf{2.24E+03} & 2.27E+03 & 2.27E+03 & 2.28E+03 & 2.29E+03 & 2.28E+03 & 2.25E+03 & \textbf{2.24E+03} \\
    $\rm CEC2022_9$ & best & 2.48E+03 & 2.48E+03 & 2.78E+03 & 2.48E+03 & 2.48E+03 & \textbf{2.47E+03} & 2.52E+03 & 2.48E+03 \\
        & std & 9.02E-02 & 1.73E+01 & 1.11E+02 & 4.04E+01 & 4.06E-03 & 4.95E-03 & 3.74E+01 & \textbf{2.70E-05} \\
        & mean & 2.48E+03 & 2.51E+03 & 2.95E+03 & 2.51E+03 & 2.48E+03 & \textbf{2.47E+03} & 2.58E+03 & 2.48E+03 \\
    $\rm CEC2022_{10}$ & best & \textbf{2.44E+03} & 2.50E+03 & 2.62E+03 & 2.50E+03 & 2.50E+03 & 2.50E+03 & 2.50E+03 & 2.50E+03 \\
        & std & \textbf{2.84E+02} & 8.13E+02 & 4.57E+02 & 8.97E+02 & 5.62E+02 & 8.85E+02 & 1.48E+03 & 7.82E+02 \\
        & mean & \textbf{2.88E+03} & 3.23E+03 & 3.14E+03 & 3.08E+03 & 3.71E+03 & 3.74E+03 & 3.59E+03 & 3.72E+03 \\
    $\rm CEC2022_{11}$ & best & \textbf{2.60E+03} & 3.08E+03 & 5.25E+03 & 2.90E+03 & \textbf{2.60E+03} & 2.90E+03 & 3.92E+03 & 2.90E+03 \\
        & std & 1.51E+02 & 4.20E+02 & 6.36E+02 & 2.54E+01 & 7.74E+01 & 4.69E-01 & 5.74E+02 & \textbf{9.52E-13} \\
        & mean & 2.97E+03 & 3.58E+03 & 7.84E+03 & 2.91E+03 & 2.92E+03 & \textbf{2.90E+03} & 5.00E+03 & \textbf{2.90E+03} \\
    $\rm CEC2022_{12}$ & best & 2.94E+03 & 2.94E+03 & 3.11E+03 & 2.95E+03 & 2.95E+03 & \textbf{2.89E+03} & 2.96E+03 & 2.94E+03 \\
        & std & \textbf{8.53E+00} & 2.28E+01 & 8.85E+01 & 4.72E+01 & 7.04E+01 & 1.37E+02 & 3.94E+01 & 1.18E+01 \\
        & mean & \textbf{2.95E+03} & 2.97E+03 & 3.28E+03 & 3.01E+03 & 3.00E+03 & 3.03E+03 & 3.01E+03 & \textbf{2.95E+03} \bigstrut[b]\\
    \hline
    \end{tabular}%
    }
  \label{tab:addlabel}%
\end{table*}%

\begin{figure}
	\centering
 		\includegraphics[width=0.8\textwidth]{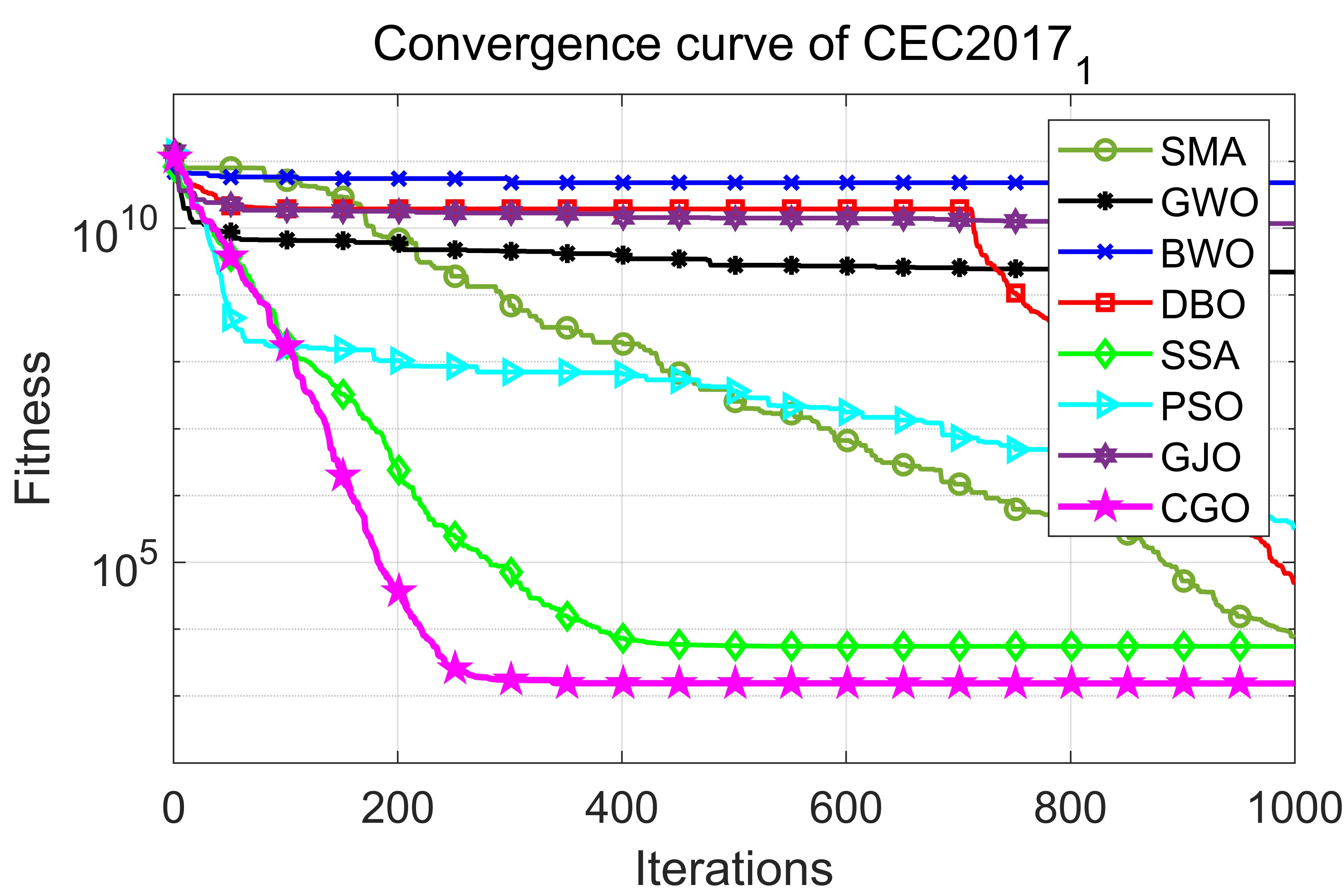}
	  \caption{Convergence curves of 8 algorithms in $\rm CEC2017_1$}\label{Fig.4}
\end{figure}
\begin{figure}
	\centering
 		\includegraphics[width=0.8\textwidth]{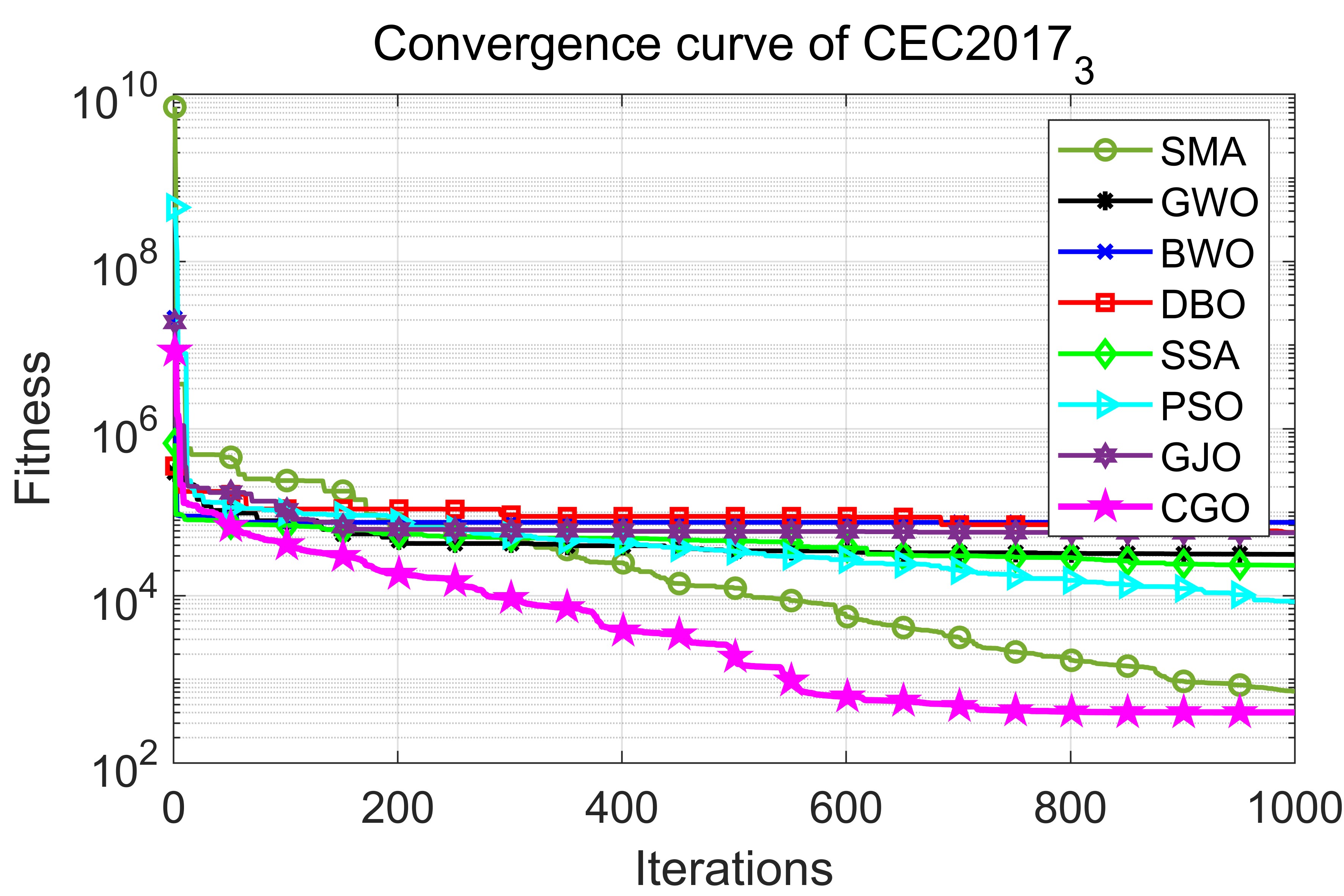}
	  \caption{Convergence curves of 8 algorithms in $\rm CEC2017_3$}\label{Fig.5}
\end{figure}
\begin{figure}
	\centering
 		\includegraphics[width=0.8\textwidth]{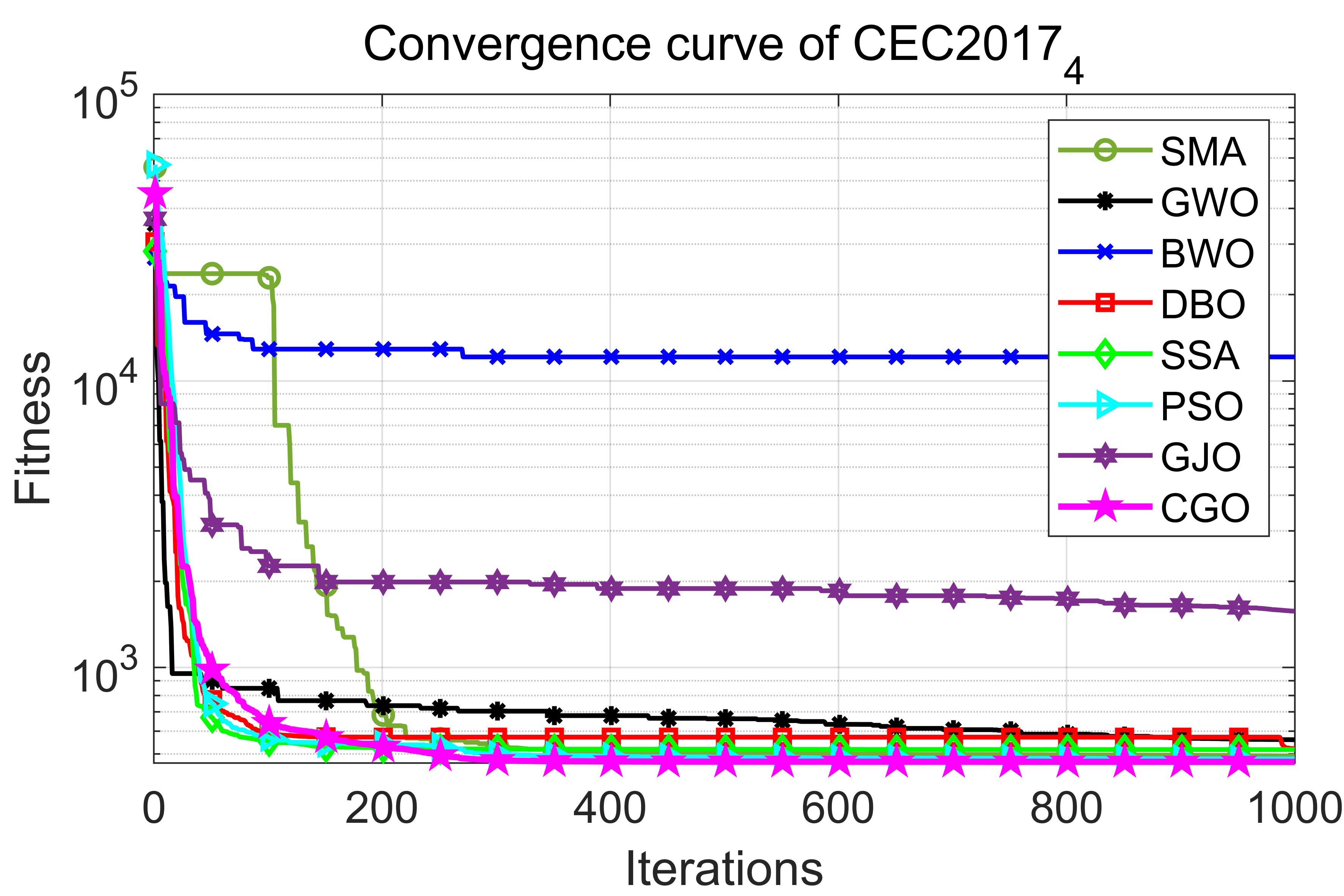}
	  \caption{Convergence curves of 8 algorithms in $\rm CEC2017_4$}\label{Fig.6}
\end{figure}
\begin{figure}
	\centering
 		\includegraphics[width=0.8\textwidth]{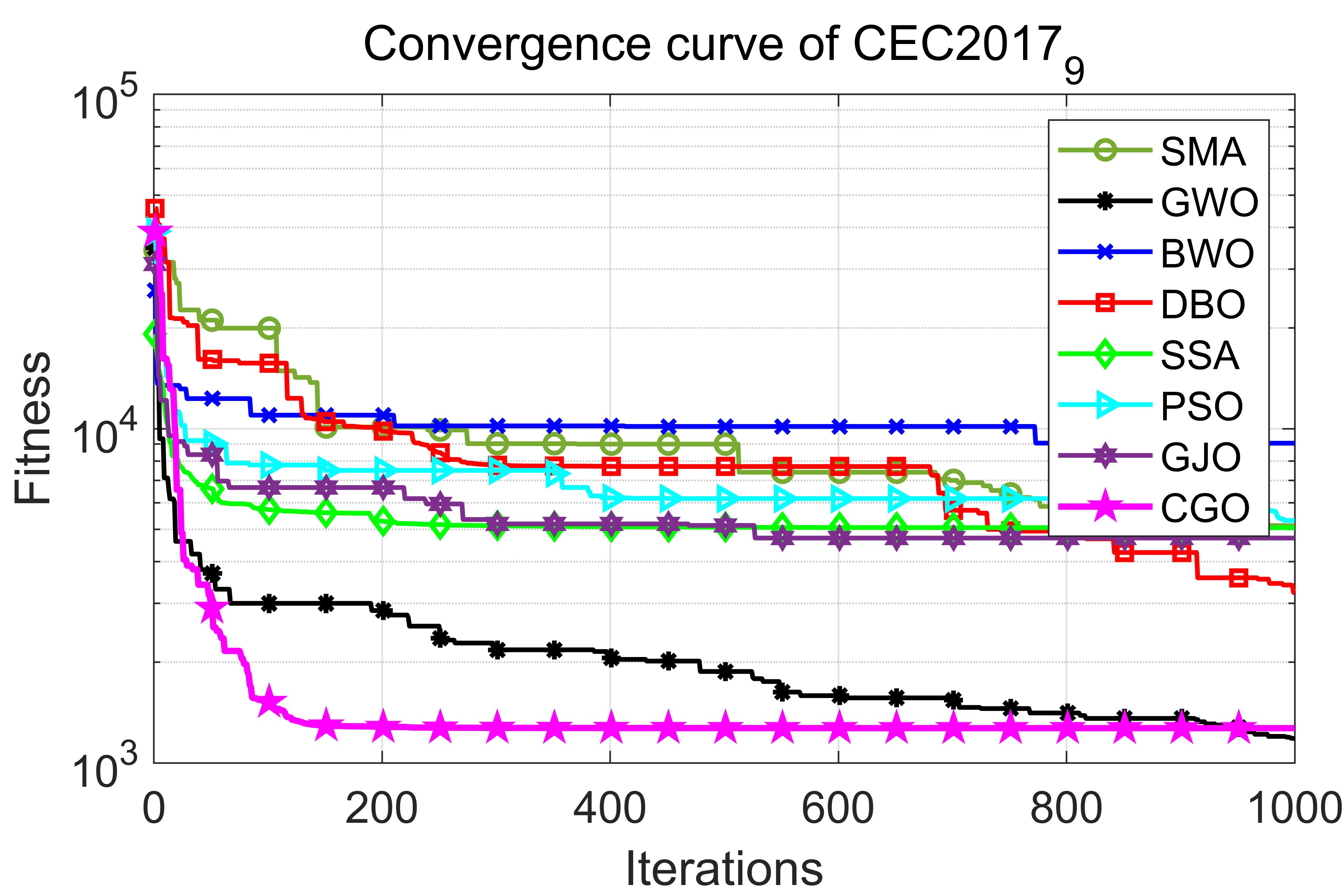}
	  \caption{Convergence curves of 8 algorithms in $\rm CEC2017_9$}\label{Fig.7}
\end{figure}
\begin{figure}
	\centering
 		\includegraphics[width=0.8\textwidth]{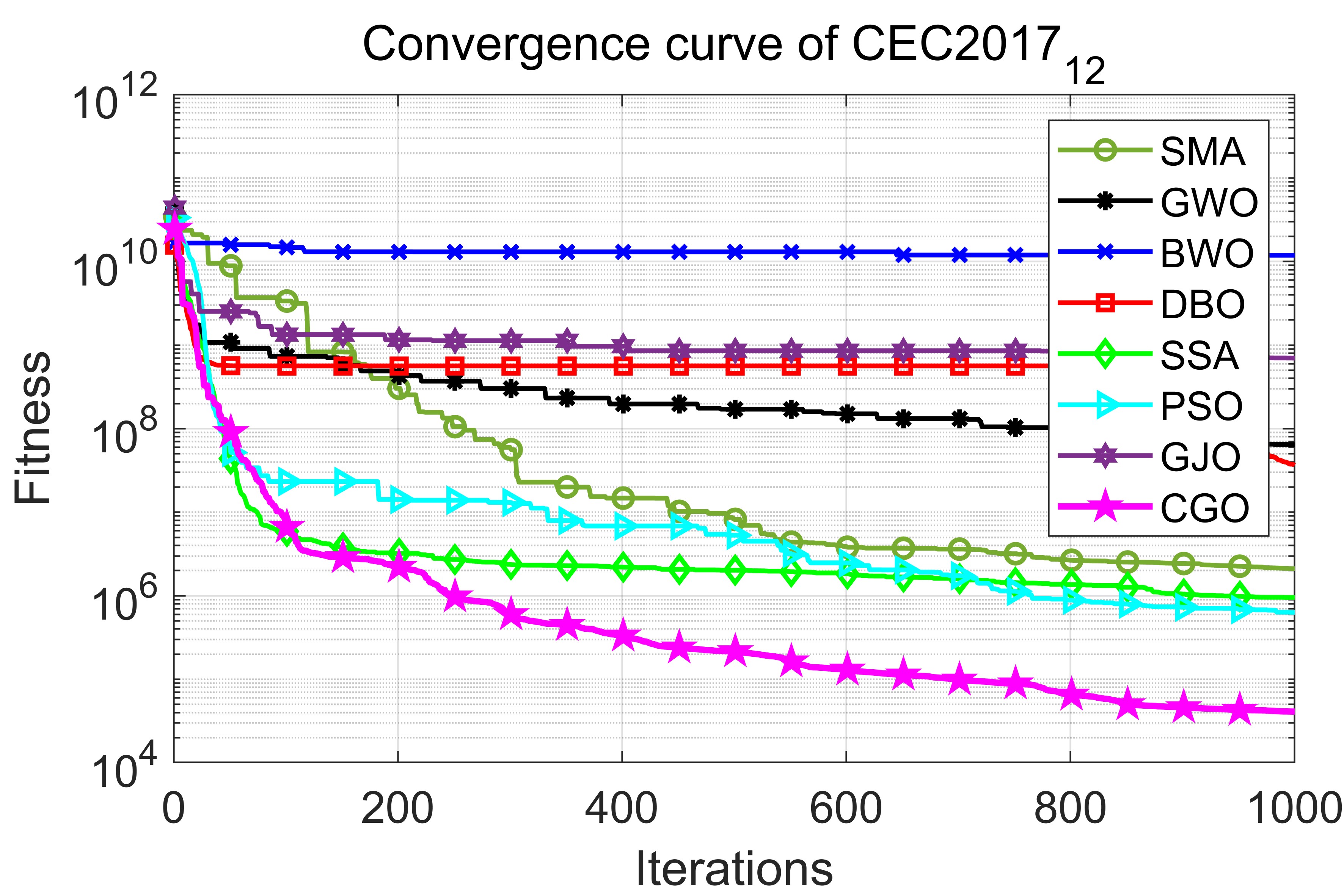}
	  \caption{Convergence curves of 8 algorithms in $\rm CEC2017_{12}$}\label{Fig.8}
\end{figure}
\begin{figure}
	\centering
 		\includegraphics[width=0.8\textwidth]{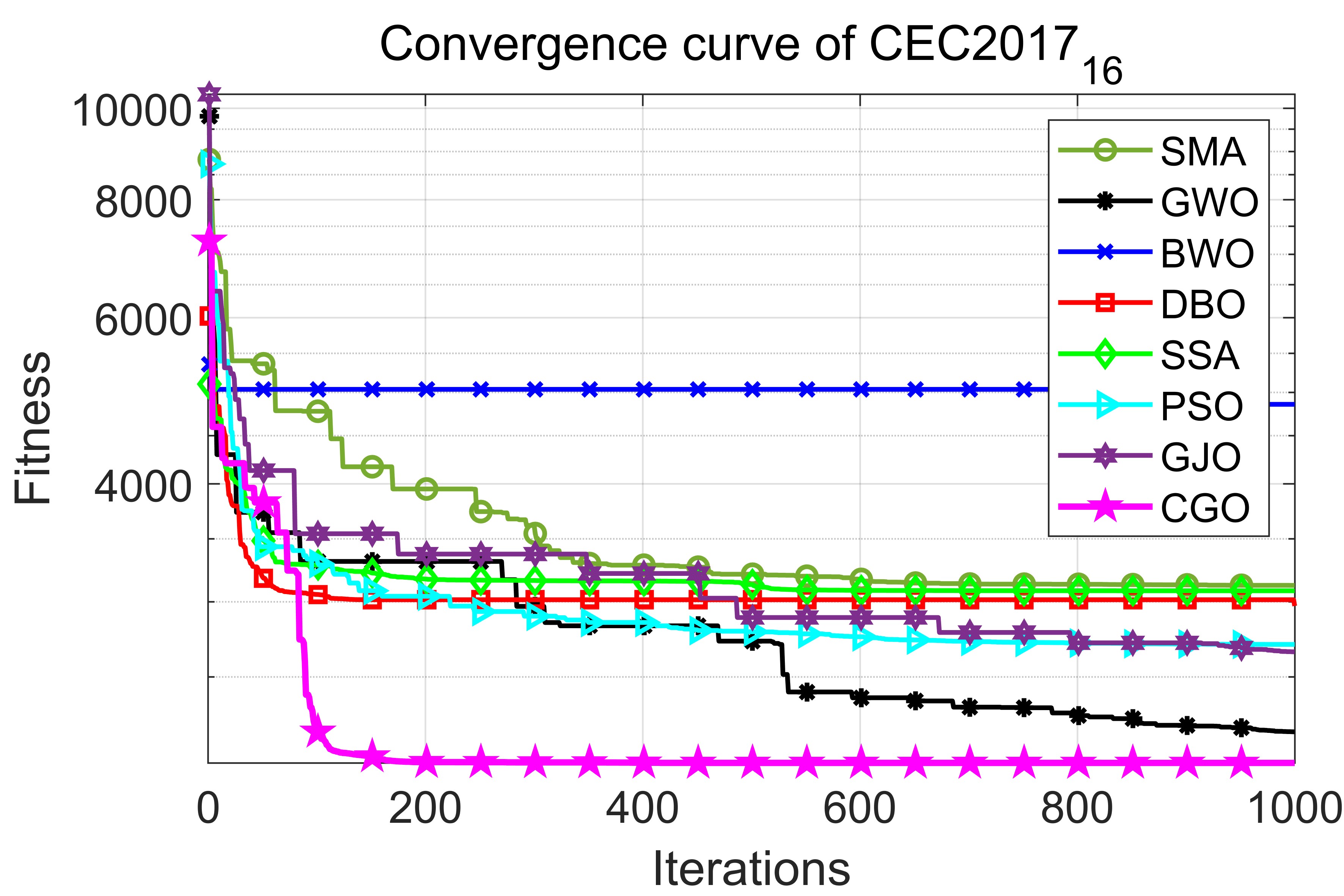}
	  \caption{Convergence curves of 8 algorithms in $\rm CEC2017_{16}$}\label{Fig.9}
\end{figure}
\begin{figure}
	\centering
 		\includegraphics[width=0.8\textwidth]{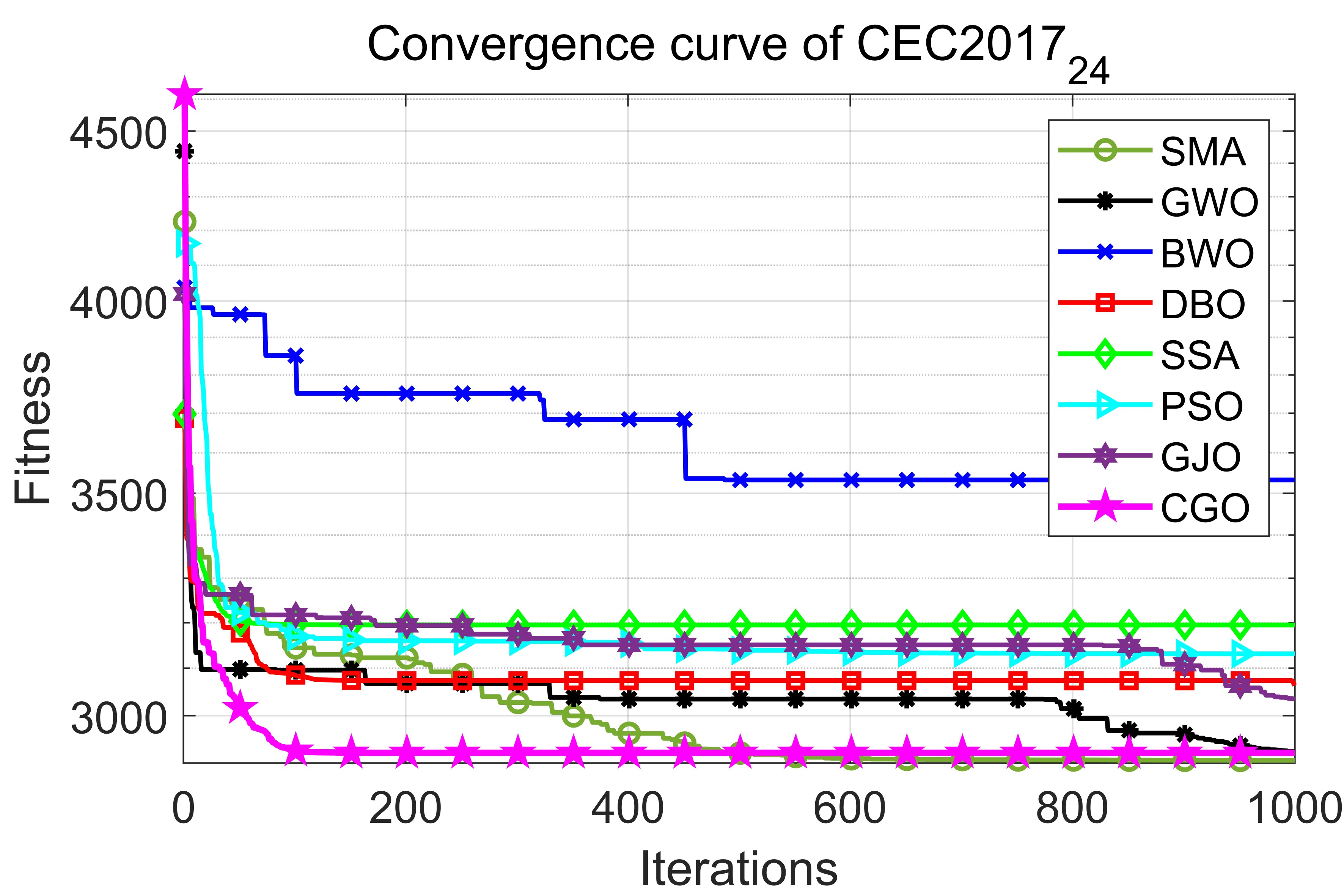}
	  \caption{Convergence curves of 8 algorithms in $\rm CEC2017_{24}$}\label{Fig.10}
\end{figure}
\begin{figure}
	\centering
 		\includegraphics[width=0.8\textwidth]{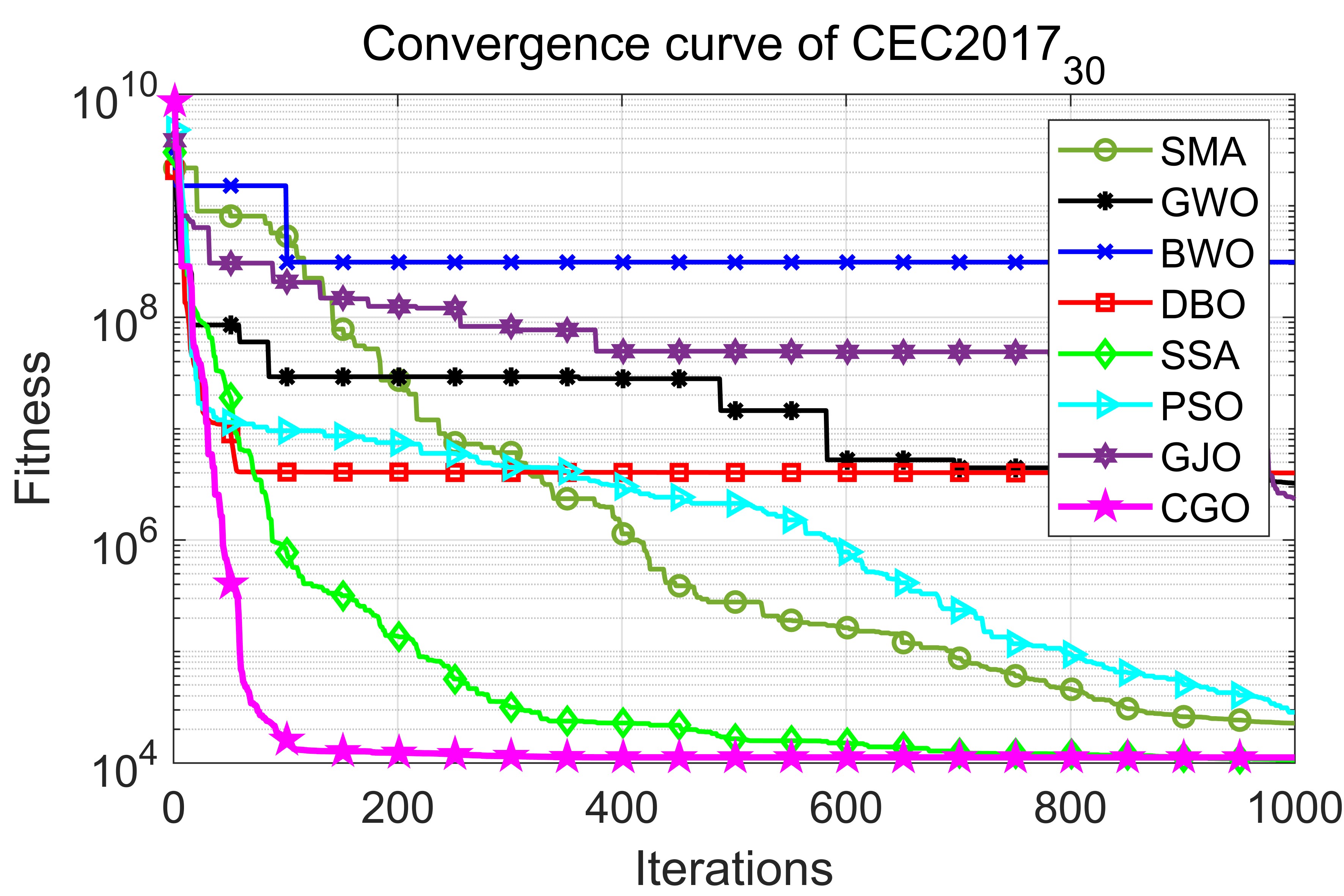}
	  \caption{Convergence curves of 8 algorithms in $\rm CEC2017_{30}$}\label{Fig.11}
\end{figure}
\begin{figure}
	\centering
 		\includegraphics[width=0.8\textwidth]{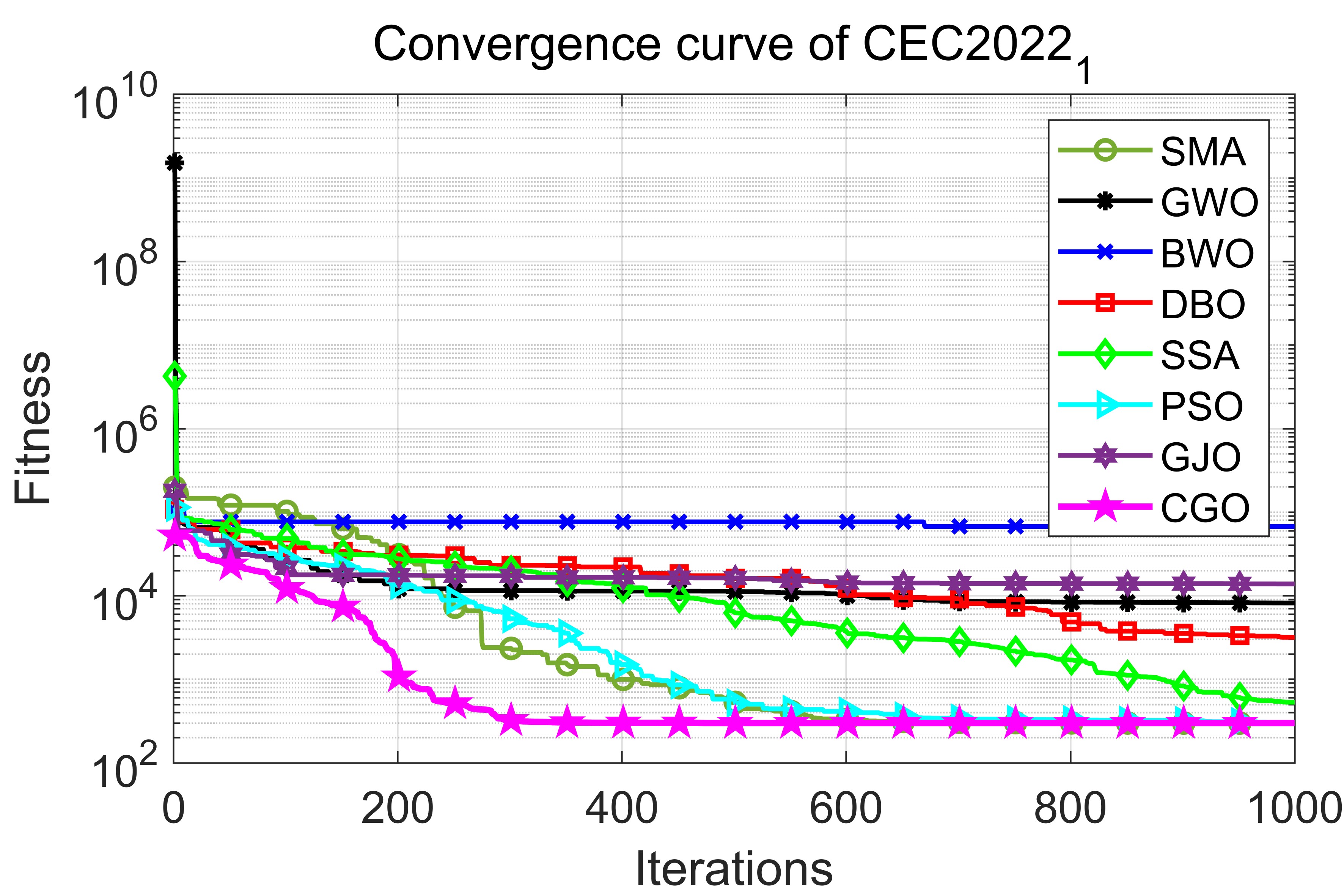}
	  \caption{Convergence curves of 8 algorithms in $\rm CEC2022_{1}$}\label{Fig.12}
\end{figure}
\begin{figure}
	\centering
 		\includegraphics[width=0.8\textwidth]{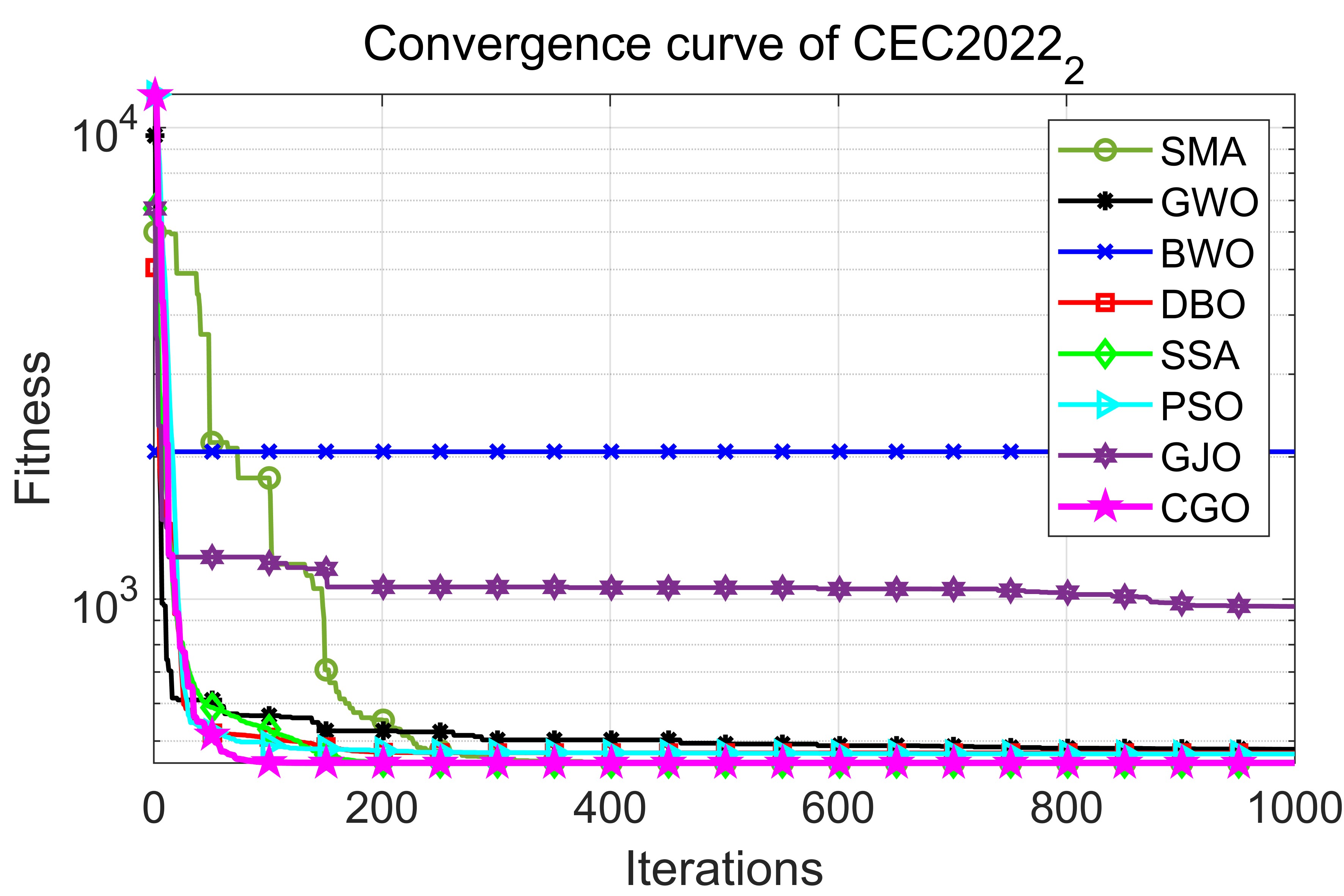}
	  \caption{Convergence curves of 8 algorithms in $\rm CEC2022_{2}$}\label{Fig.13}
\end{figure}
\begin{figure}
	\centering
 		\includegraphics[width=0.8\textwidth]{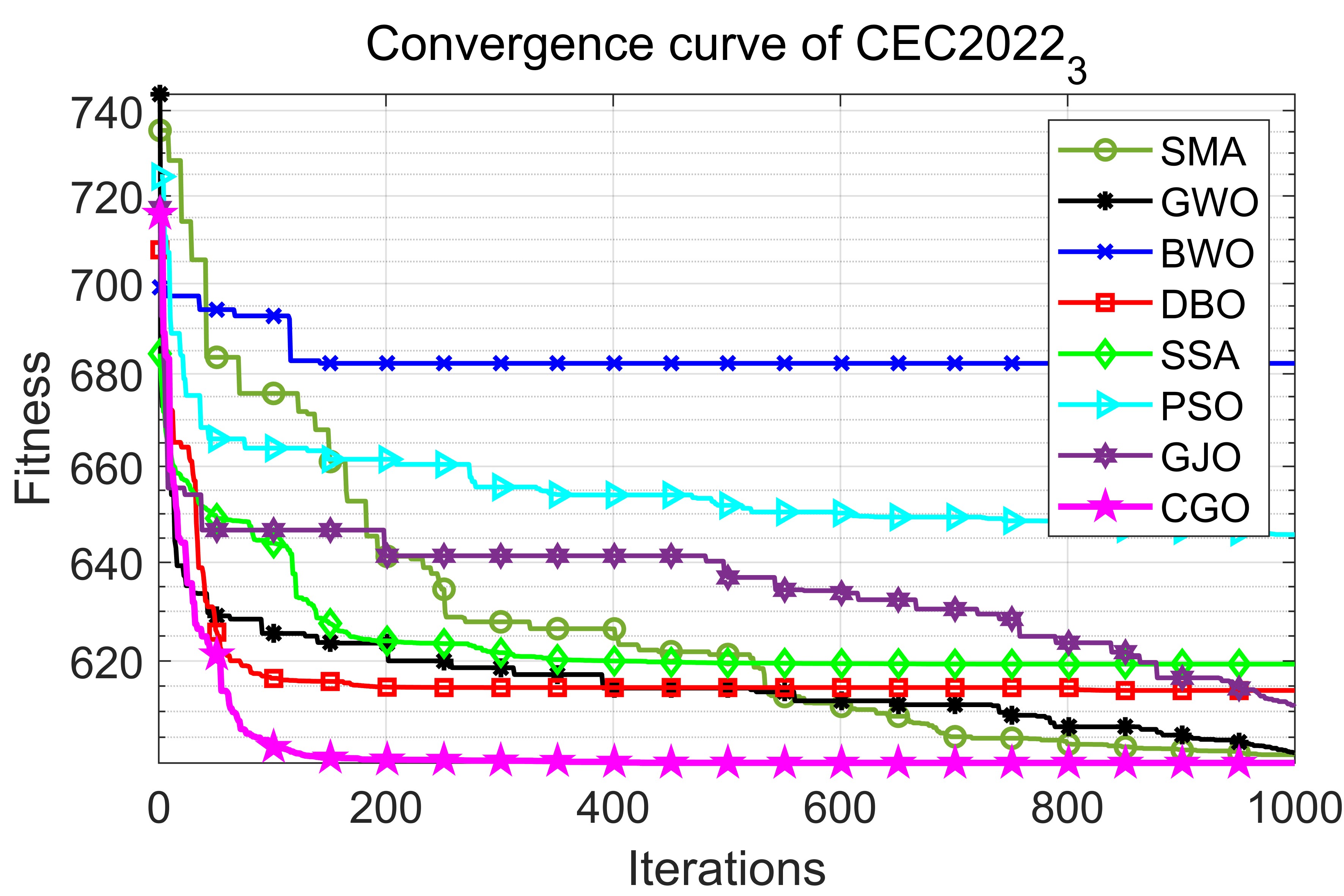}
	  \caption{Convergence curves of 8 algorithms in $\rm CEC2022_{3}$}\label{Fig.14}
\end{figure}
\begin{figure}
	\centering
 		\includegraphics[width=0.8\textwidth]{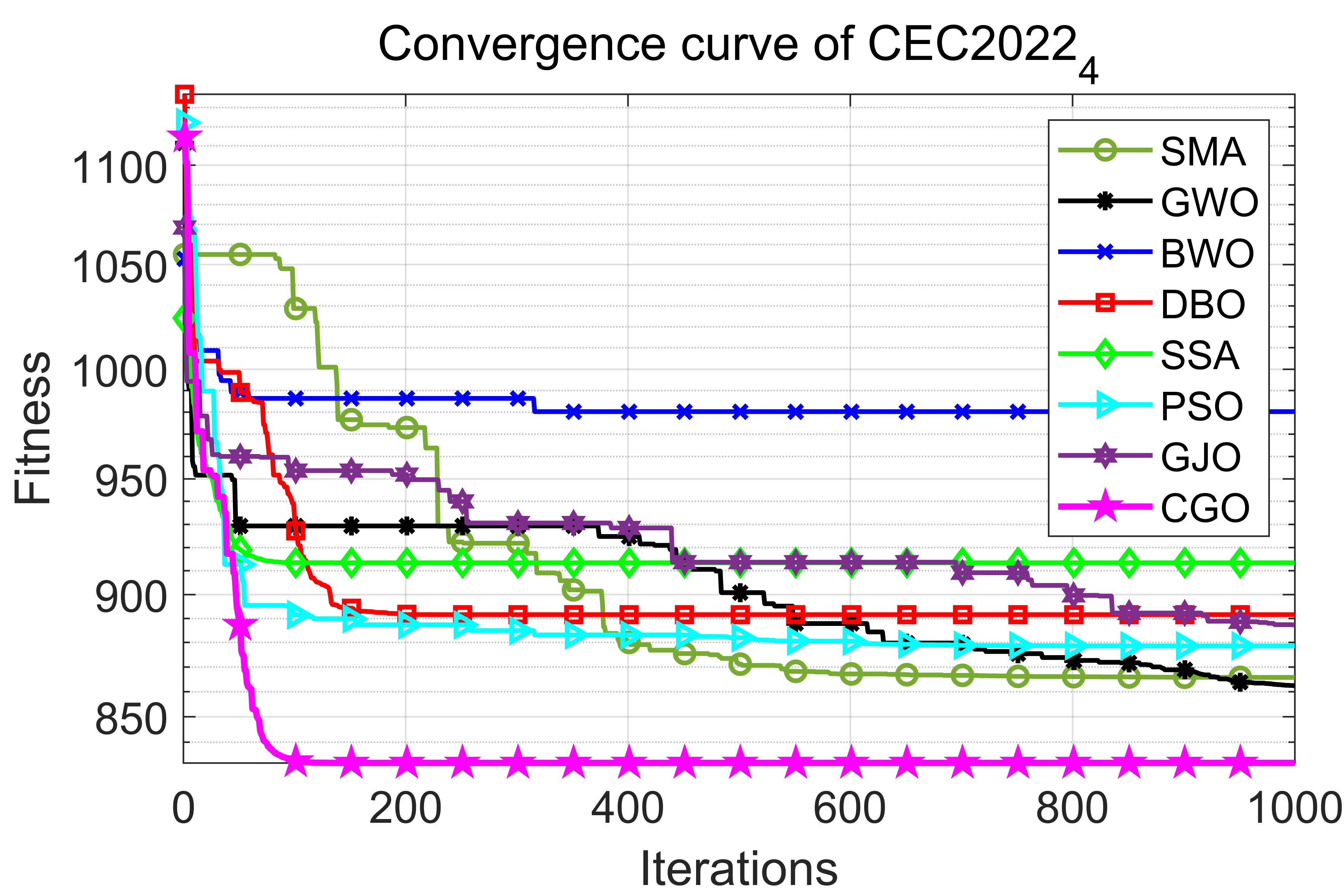}
	  \caption{Convergence curves of 8 algorithms in $\rm CEC2022_{4}$}\label{Fig.15}
\end{figure}
\begin{figure}
	\centering
 		\includegraphics[width=0.8\textwidth]{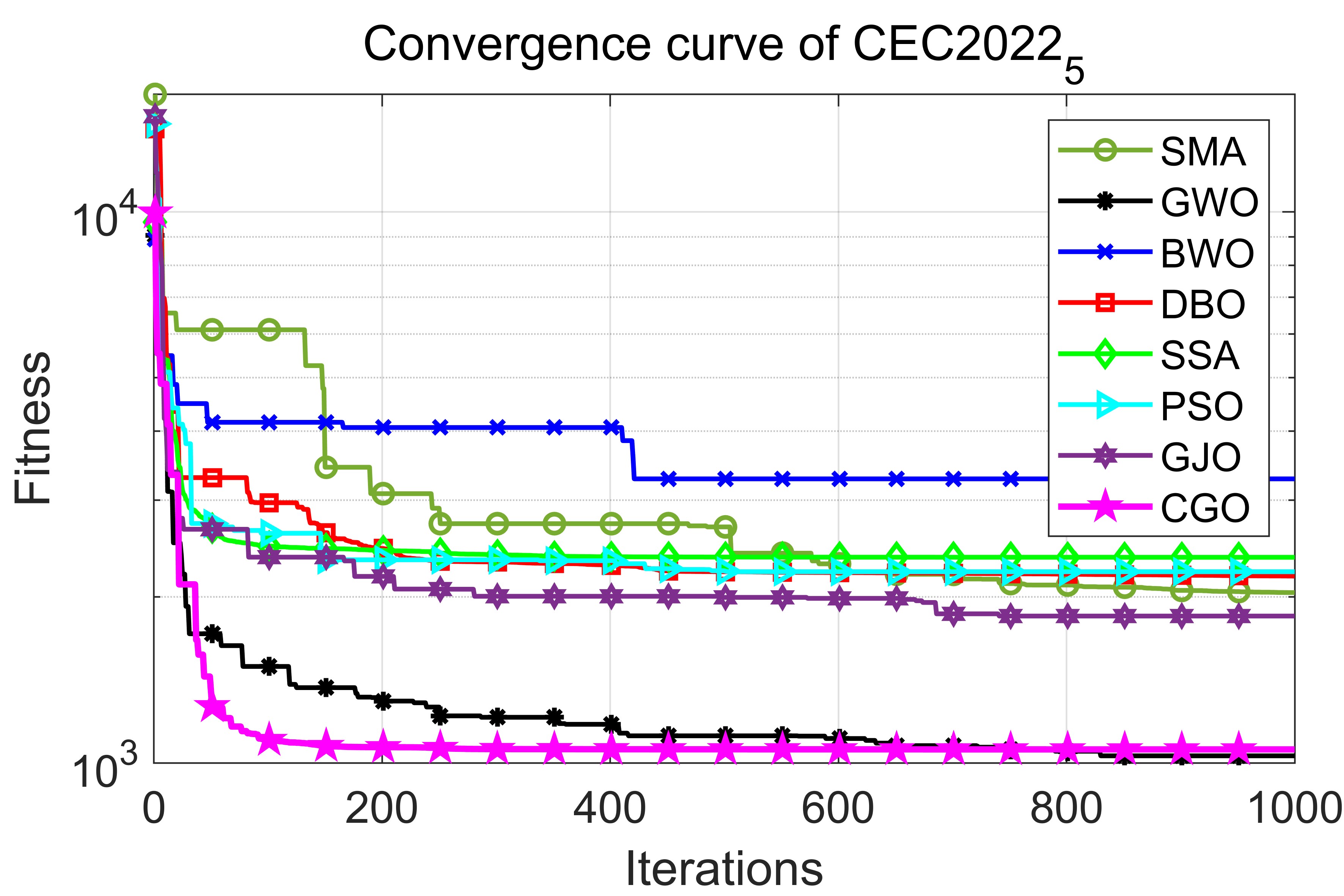}
	  \caption{Convergence curves of 8 algorithms in $\rm CEC2022_{5}$}\label{Fig.16}
\end{figure}
\begin{figure}
	\centering
 		\includegraphics[width=0.8\textwidth]{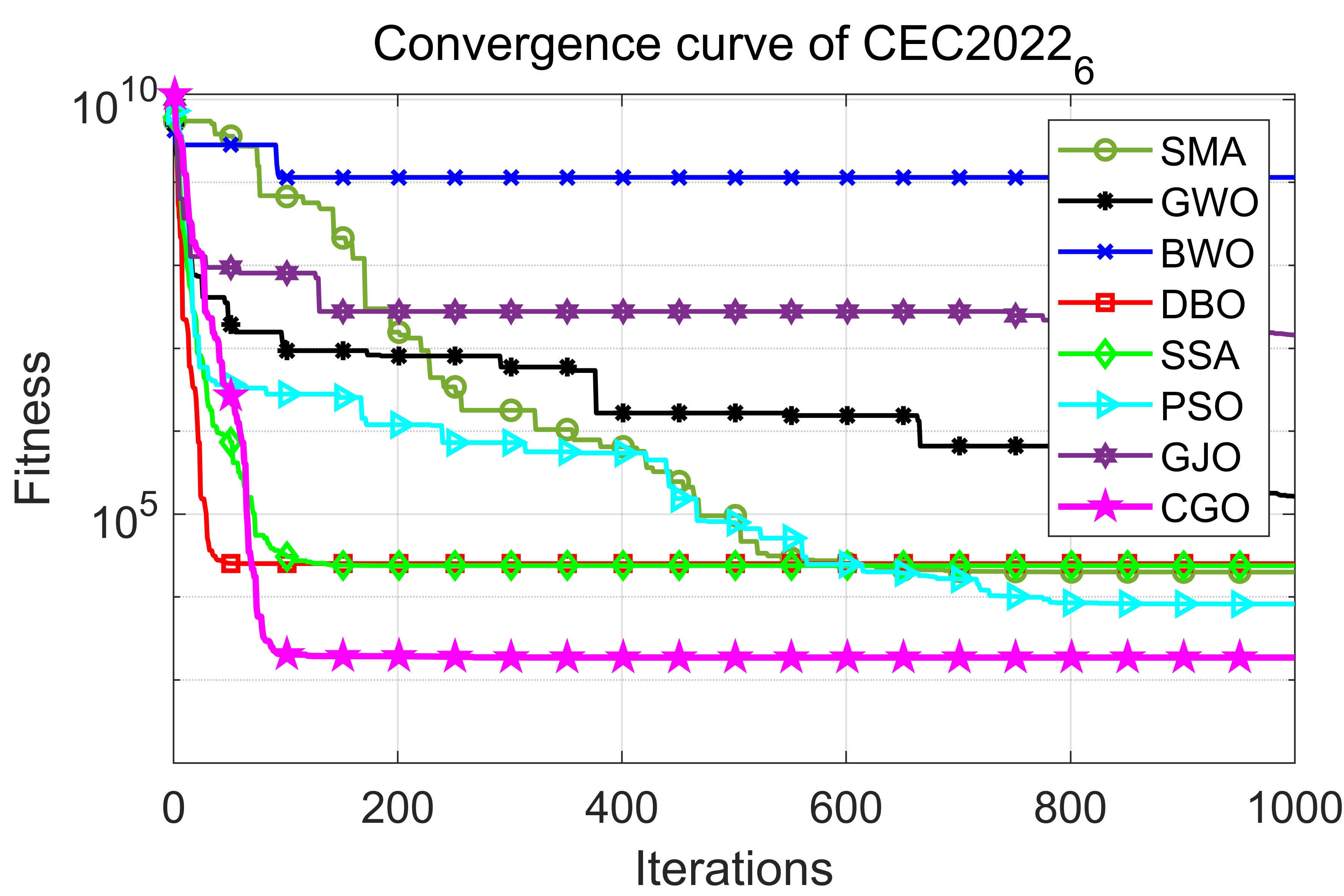}
	  \caption{Convergence curves of 8 algorithms in $\rm CEC2022_{6}$}\label{Fig.17}
\end{figure}
\begin{figure}
	\centering
 		\includegraphics[width=0.8\textwidth]{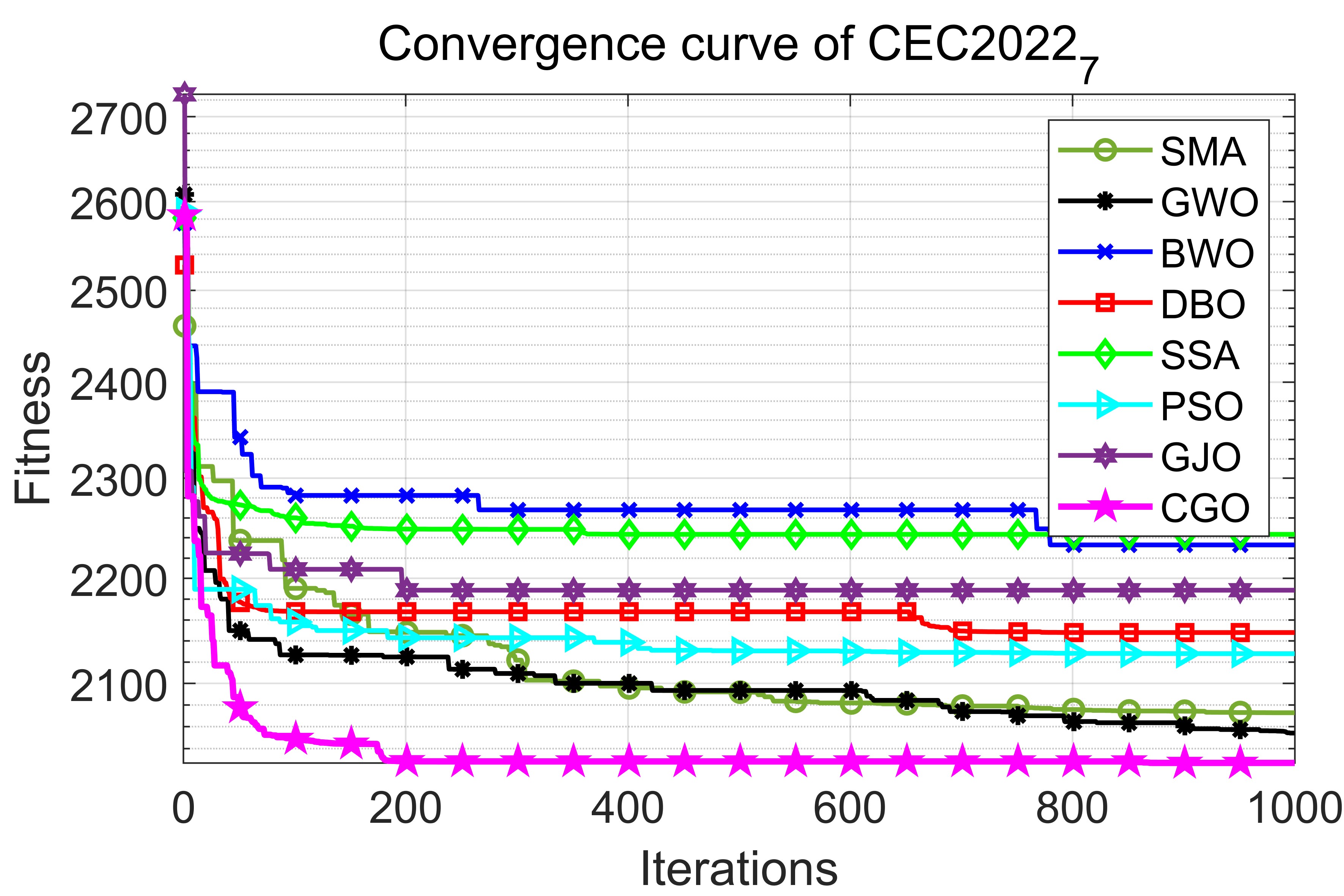}
	  \caption{Convergence curves of 8 algorithms in $\rm CEC2022_{7}$}\label{Fig.18}
\end{figure}
\begin{figure}
	\centering
 		\includegraphics[width=0.8\textwidth]{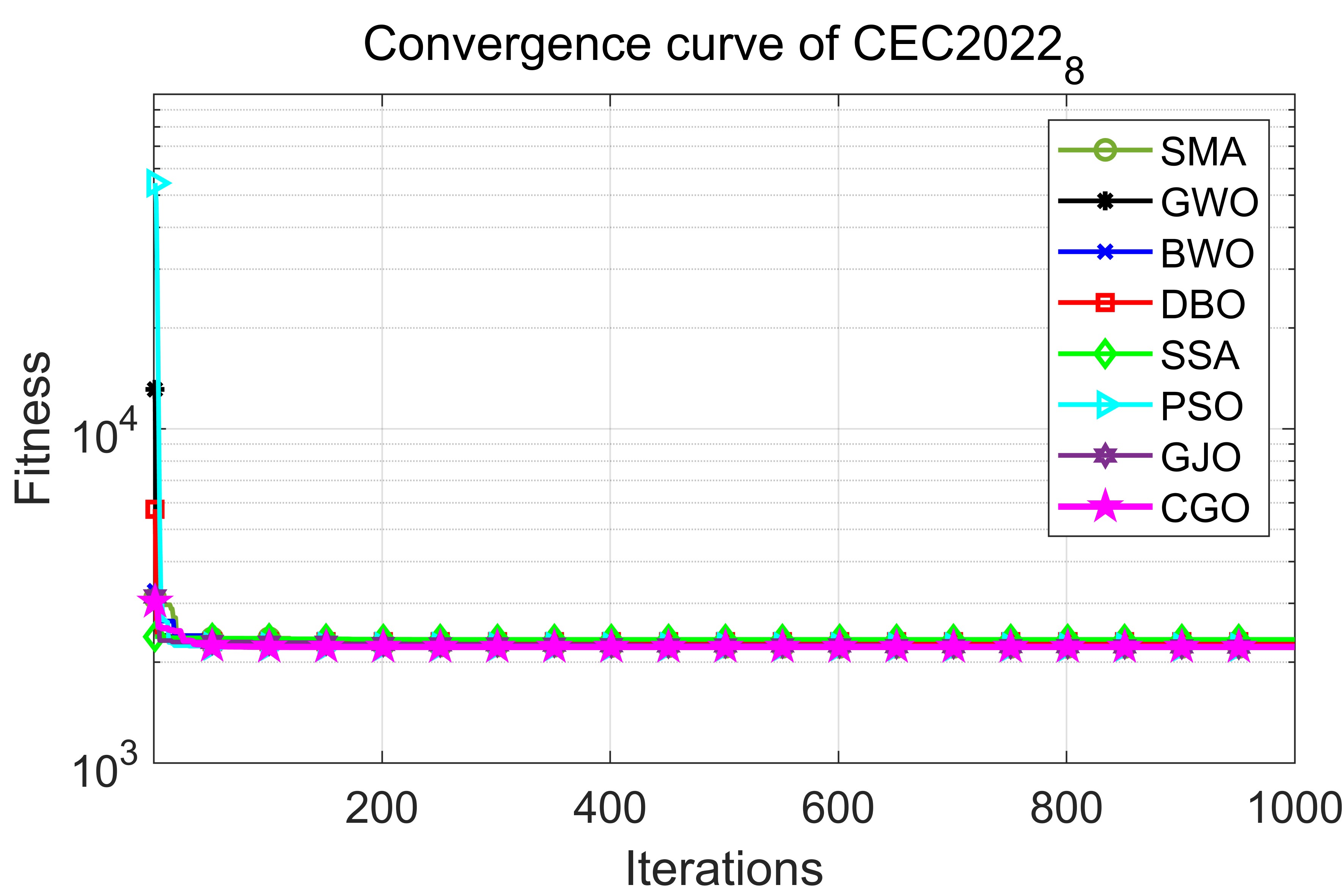}
	  \caption{Convergence curves of 8 algorithms in $\rm CEC2022_{8}$}\label{Fig.19}
\end{figure}
\begin{figure}
	\centering
 		\includegraphics[width=0.8\textwidth]{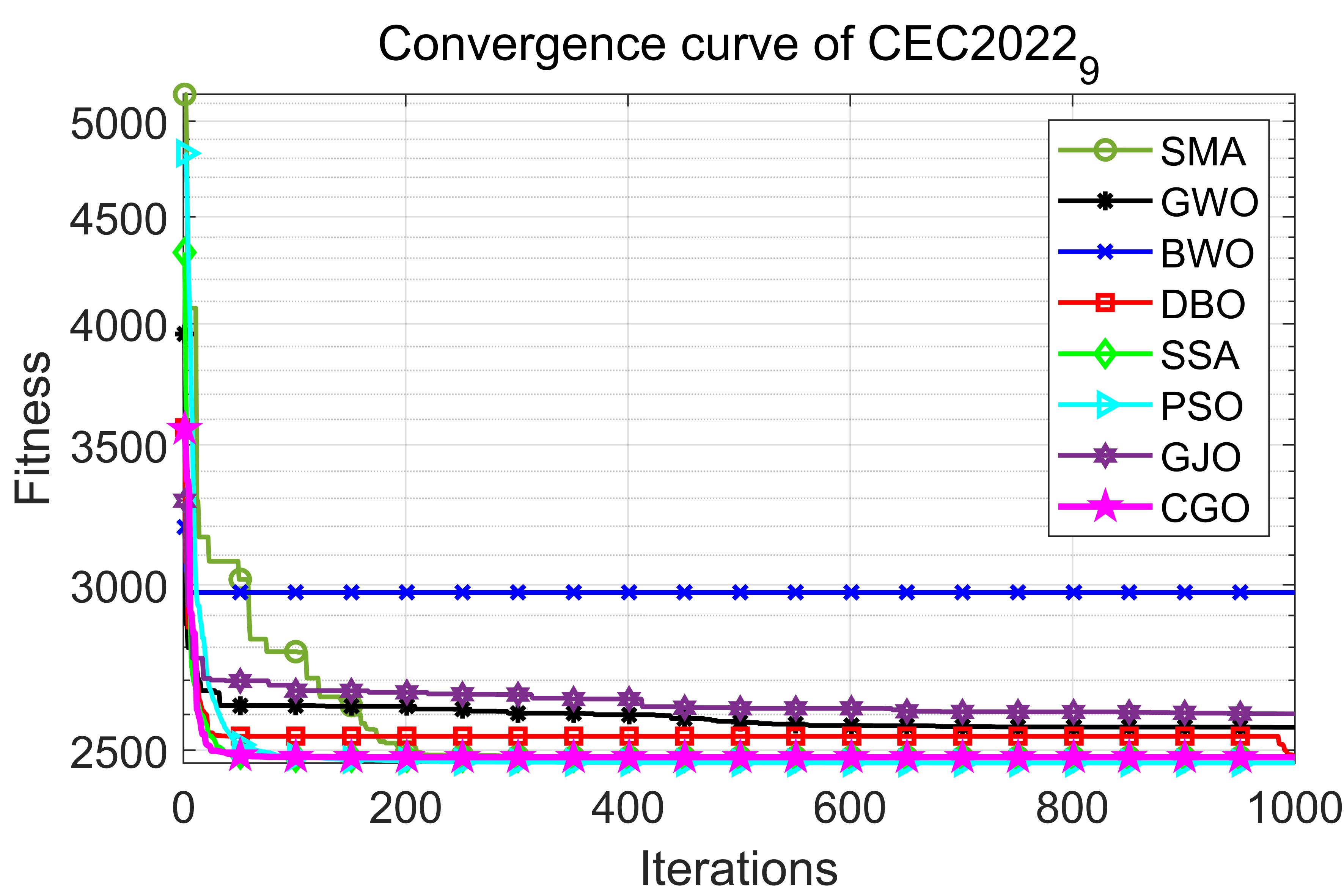}
	  \caption{Convergence curves of 8 algorithms in $\rm CEC2022_{9}$}\label{Fig.20}
\end{figure}
\begin{figure}
	\centering
 		\includegraphics[width=0.8\textwidth]{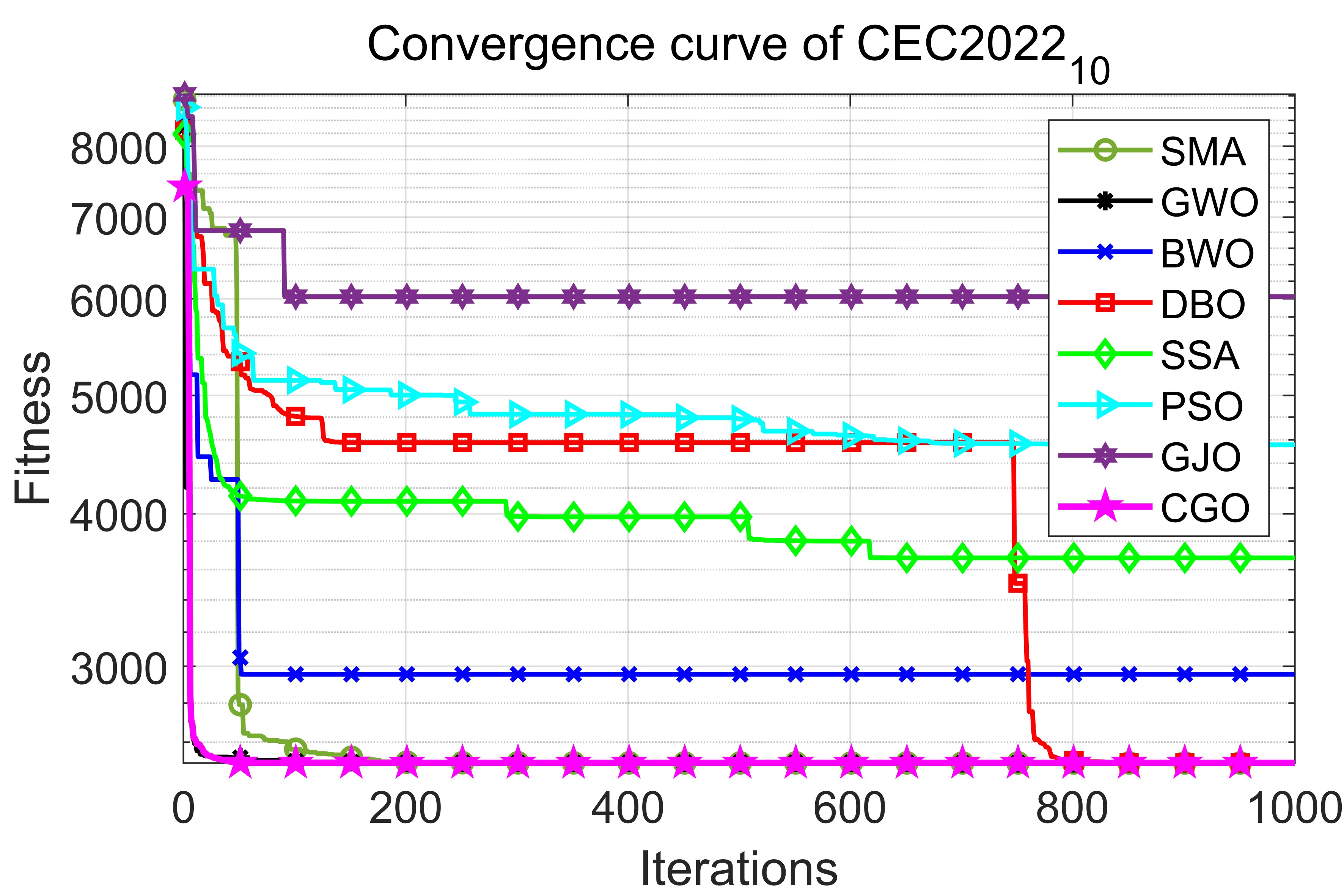}
	  \caption{Convergence curves of 8 algorithms in $\rm CEC2022_{10}$}\label{Fig.21}
\end{figure}
\begin{figure}
	\centering
 		\includegraphics[width=0.8\textwidth]{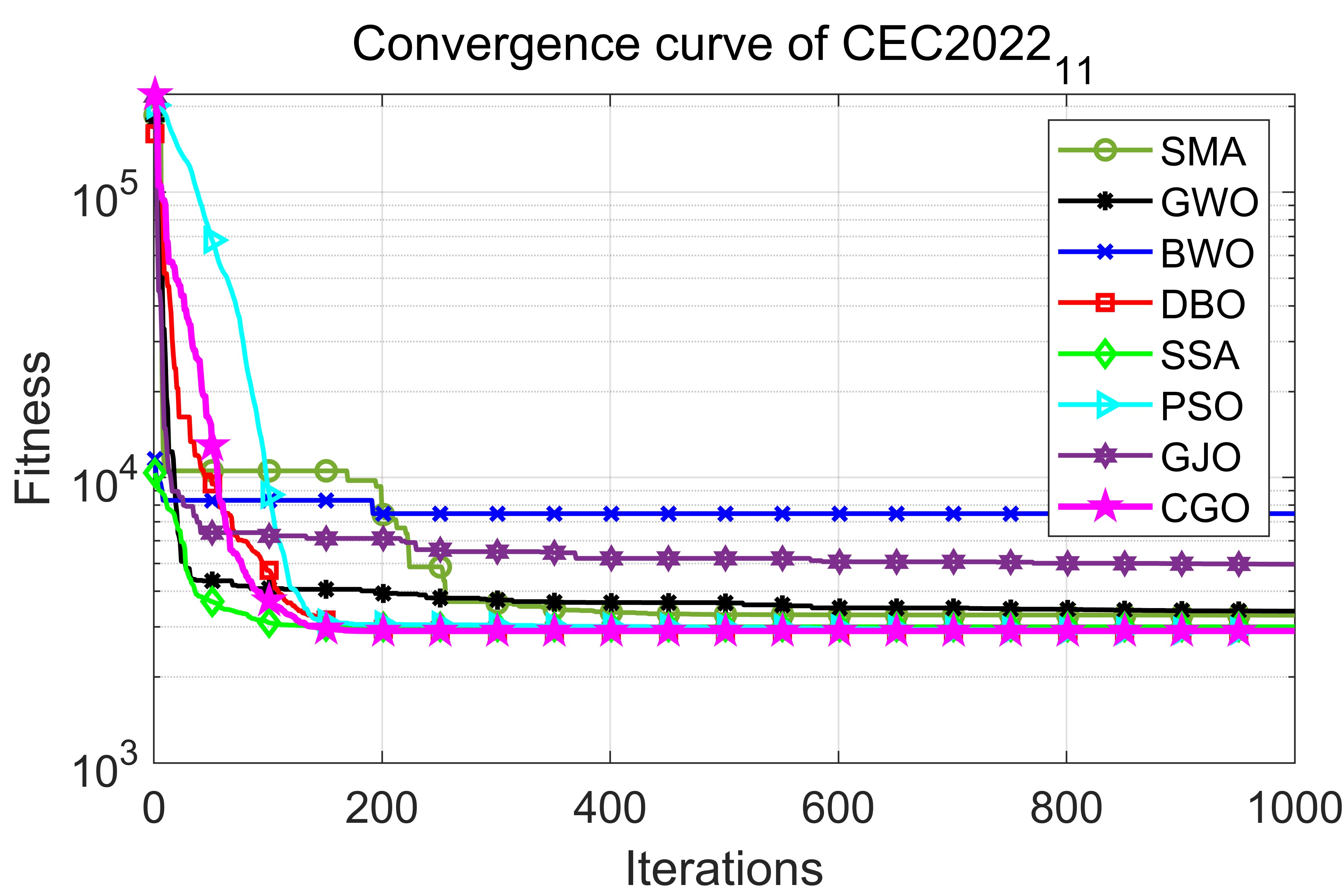}
	  \caption{Convergence curves of 8 algorithms in $\rm CEC2022_{11}$}\label{Fig.22}
\end{figure}
\begin{figure}
	\centering
 		\includegraphics[width=0.8\textwidth]{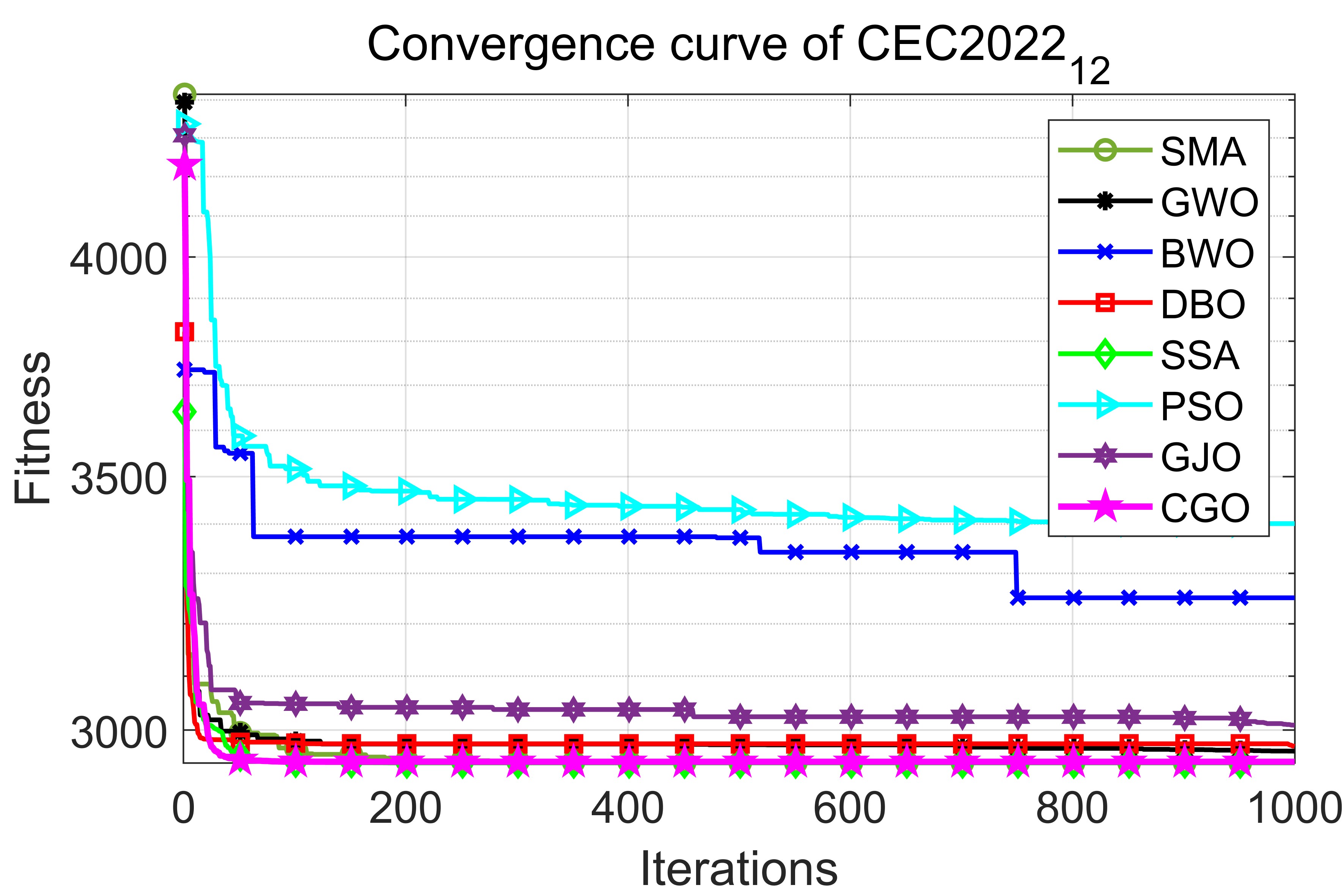}
	  \caption{Convergence curves of 8 algorithms in $\rm CEC2022_{12}$}\label{Fig.23}
\end{figure}

Figure 4-23 shows the convergence curves obtained by 8 algorithms for 8 CEC2017 and all CEC2022 benchmark problems. As can be seen from the figure, the CGO algorithm achieves a high level of convergence speed and accuracy, and is superior to other comparison algorithms in most cases.

The Wilcoxon test\citep{derrac2011practical} is conducted on these 8 algorithms based on CEC2017 and CEC2022 test suites. As shown in the test results in Tables 4 and 5, in most cases, the average obtained is less than 0.05. This shows that the developed the CGO algorithm is statistically superior to other comparison algorithms.
\begin{table*}[!t]
  \centering
  \caption{Comparison outcomes attained by Wilcoxon’s test on the CEC2017 test suite}
    \resizebox{\textwidth}{!}{
    \begin{tabular}{cccccccc}
    \hline
        & SMA & GWO & BWO & DBO & SSA & PSO & GJO \bigstrut\\
    \hline
    $\rm CEC2017_1$ & 3.85E-03 & 3.02E-11 & 3.02E-11 & 3.69E-11 & 1.37E-01 & 3.02E-11 & 3.02E-11 \bigstrut[t]\\
    $\rm CEC2017_3$ & 1.02E-05 & 3.02E-11 & 3.02E-11 & 3.02E-11 & 3.02E-11 & 3.69E-11 & 3.02E-11 \\
    $\rm CEC2017_4$ & 5.37E-02 & 9.92E-11 & 3.02E-11 & 3.26E-07 & 3.26E-01 & 1.33E-01 & 3.02E-11 \\
    $\rm CEC2017_5$ & 2.61E-02 & 4.84E-02 & 3.02E-11 & 6.07E-11 & 3.02E-11 & 1.96E-10 & 5.49E-11 \\
    $\rm CEC2017_6$ & 1.33E-10 & 3.02E-11 & 3.02E-11 & 3.02E-11 & 3.02E-11 & 3.02E-11 & 3.02E-11 \\
    $\rm CEC2017_7$ & 4.43E-03 & 9.63E-02 & 3.02E-11 & 1.00E-03 & 6.70E-11 & 5.83E-03 & 1.33E-10 \\
    $\rm CEC2017_8$ & 2.96E-05 & 7.62E-01 & 3.02E-11 & 7.39E-11 & 4.98E-11 & 1.09E-05 & 5.57E-10 \\
    $\rm CEC2017_9$ & 2.49E-06 & 9.35E-01 & 3.02E-11 & 1.17E-09 & 3.02E-11 & 3.02E-11 & 3.34E-11 \\
    $\rm CEC2017_{10}$ & 8.66E-05 & 1.52E-03 & 3.02E-11 & 1.58E-04 & 1.33E-01 & 1.54E-01 & 1.52E-03 \\
    $\rm CEC2017_{11}$ & 4.03E-03 & 9.92E-11 & 3.02E-11 & 3.50E-09 & 7.24E-02 & 4.51E-02 & 3.02E-11 \\
    $\rm CEC2017_{12}$ & 4.50E-11 & 3.02E-11 & 3.02E-11 & 3.02E-11 & 5.57E-10 & 8.15E-11 & 3.02E-11 \\
    $\rm CEC2017_{13}$ & 3.03E-02 & 1.29E-09 & 3.02E-11 & 5.46E-09 & 6.52E-01 & 9.71E-01 & 3.02E-11 \\
    $\rm CEC2017_{14}$ & 5.57E-10 & 1.46E-10 & 3.02E-11 & 3.16E-10 & 7.09E-08 & 2.39E-04 & 8.15E-11 \\
    $\rm CEC2017_{15}$ & 1.44E-02 & 1.20E-08 & 3.02E-11 & 1.36E-07 & 1.41E-01 & 6.79E-02 & 8.15E-11 \\
    $\rm CEC2017_{16}$ & 1.45E-01 & 6.84E-01 & 3.02E-11 & 2.83E-08 & 1.02E-05 & 2.50E-03 & 3.83E-05 \\
    $\rm CEC2017_{17}$ & 5.37E-02 & 2.01E-01 & 3.02E-11 & 6.20E-04 & 1.30E-03 & 3.04E-01 & 8.65E-01 \\
    $\rm CEC2017_{18}$ & 3.02E-11 & 4.98E-11 & 3.02E-11 & 2.92E-09 & 8.89E-10 & 1.46E-10 & 7.39E-11 \\
    $\rm CEC2017_{19}$ & 1.96E-01 & 2.67E-09 & 3.02E-11 & 1.64E-05 & 4.86E-03 & 7.70E-04 & 6.70E-11 \\
    $\rm CEC2017_{20}$ & 4.29E-01 & 3.11E-01 & 1.20E-08 & 1.86E-03 & 2.39E-04 & 9.33E-02 & 2.52E-01 \\
    $\rm CEC2017_{21}$ & 2.43E-05 & 2.52E-01 & 3.02E-11 & 4.08E-11 & 6.07E-11 & 4.50E-11 & 4.50E-11 \\
    $\rm CEC2017_{22}$ & 2.62E-03 & 2.16E-03 & 8.48E-09 & 3.63E-01 & 3.87E-01 & 1.53E-05 & 1.45E-01 \\
    $\rm CEC2017_{23}$ & 3.27E-02 & 3.92E-02 & 3.02E-11 & 3.02E-11 & 3.02E-11 & 3.02E-11 & 3.02E-11 \\
    $\rm CEC2017_{24}$ & 3.85E-03 & 1.91E-02 & 3.02E-11 & 4.98E-11 & 9.92E-11 & 3.02E-11 & 1.33E-10 \\
    $\rm CEC2017_{25}$ & 3.27E-02 & 1.09E-10 & 3.02E-11 & 4.08E-05 & 4.83E-01 & 8.66E-05 & 3.02E-11 \\
    $\rm CEC2017_{26}$ & 8.77E-01 & 3.18E-01 & 3.02E-11 & 1.33E-10 & 2.02E-08 & 5.32E-03 & 5.49E-11 \\
    $\rm CEC2017_{27}$ & 6.15E-02 & 1.06E-03 & 3.02E-11 & 5.60E-07 & 3.09E-06 & 2.46E-01 & 9.92E-11 \\
    $\rm CEC2017_{28}$ & 7.70E-04 & 4.08E-11 & 3.02E-11 & 1.96E-10 & 1.26E-01 & 8.65E-01 & 3.02E-11 \\
    $\rm CEC2017_{29}$ & 4.29E-01 & 6.41E-01 & 3.02E-11 & 5.61E-05 & 2.88E-06 & 2.27E-03 & 3.16E-05 \\
    $\rm CEC2017_{30}$ & 6.53E-08 & 3.02E-11 & 3.02E-11 & 3.82E-10 & 6.15E-02 & 8.89E-10 & 3.02E-11 \bigstrut[b]\\
    \hline
    \end{tabular}%
    }
  \label{tab:addlabel}%
\end{table*}%
\begin{table*}[!t]
  \centering
  \caption{Comparison outcomes attained by Wilcoxon’s test on the CEC2022 test suite}
  \resizebox{\textwidth}{!}{
    \begin{tabular}{cccccccc}
    \hline
        & SMA & GWO & BWO & DBO & SSA & PSO & GJO \bigstrut\\
    \hline
    $\rm CEC2022_1$ & 3.02E-11 & 3.02E-11 & 3.02E-11 & 3.02E-11 & 3.02E-11 & 3.02E-11 & 3.02E-11 \bigstrut[t]\\
    $\rm CEC2022_2$ & 1.68E-04 & 7.03E-07 & 3.01E-11 & 7.42E-05 & 2.41E-02 & 9.82E-01 & 3.01E-11 \\
    $\rm CEC2022_3$ & 2.72E-11 & 2.72E-11 & 2.72E-11 & 2.72E-11 & 2.72E-11 & 2.72E-11 & 2.72E-11 \\
    $\rm CEC2022_4$ & 5.56E-04 & 8.77E-01 & 3.02E-11 & 8.10E-10 & 2.15E-10 & 4.86E-03 & 9.26E-09 \\
    $\rm CEC2022_5$ & 8.15E-05 & 5.69E-01 & 3.02E-11 & 4.31E-08 & 3.02E-11 & 2.44E-09 & 5.00E-09 \\
    $\rm CEC2022_6$ & 7.04E-07 & 7.62E-03 & 3.02E-11 & 1.67E-01 & 7.84E-01 & 1.60E-03 & 3.20E-09 \\
    $\rm CEC2022_7$ & 1.71E-01 & 9.00E-01 & 4.50E-11 & 3.99E-04 & 2.77E-05 & 3.99E-04 & 1.87E-05 \\
    $\rm CEC2022_8$ & 5.01E-02 & 3.77E-04 & 1.01E-08 & 1.87E-05 & 3.01E-04 & 7.62E-03 & 6.55E-04 \\
    $\rm CEC2022_9$ & 2.23E-11 & 2.23E-11 & 2.23E-11 & 2.23E-11 & 4.67E-10 & 2.23E-11 & 2.23E-11 \\
    $\rm CEC2022_{10}$ & 9.79E-05 & 3.50E-03 & 6.97E-03 & 4.51E-02 & 4.83E-01 & 8.65E-01 & 7.28E-01 \\
    $\rm CEC2022_{11}$ & 2.26E-10 & 1.08E-11 & 1.08E-11 & 8.75E-07 & 2.26E-10 & 1.08E-11 & 1.08E-11 \\
    $\rm CEC2022_{12}$ & 6.52E-01 & 4.35E-05 & 3.02E-11 & 7.12E-09 & 3.82E-09 & 1.22E-01 & 1.33E-10 \bigstrut[b]\\
    \hline
    \end{tabular}%
    }
  \label{tab:addlabel}%
\end{table*}%
\subsection{Experimental research on 8 real-world constrained optimization problems}
In this section, eight real-world engineering constrained optimization problems in different engineering domains are used to verify the performance of the proposed CGO algorithm. Table 6 briefly describes these challenging problems, and mathematical expressions for them can be found in the literature\citep{KumarWu-542}. In the experiment, each algorithm uses 50 individuals, runs 30 times independently, and the total iteration is 50 times as the termination standard.

Figure 24-31 shows the convergence curves of the objective functions of 8 algorithms on 8 practical engineering design problems. As can be seen from the figure, the CGO is ahead of other comparison algorithms in F2, F3, F6, and F7 problems. For F4, F5 and F8, the optimal values of all algorithms are almost the same, but the convergence speed of the CGO is ahead of most of the comparison algorithms.
\begin{table*}[htbp]
\caption{Real-world engineering constrained optimization problems. $d$ is the total number of decision variables of the problem, $g$ is the number of inequality constraints and $h$ is the number of equality constraints, $f\left(  *  \right)$ is best known feasible objective function value.}
\label{tab_fwdc}
\begin{tabular*}{\tblwidth}{@{}LLLLLL@{}}
\toprule 
No. & Name & $d$ & $g$ & $h$ & $f\left(  *  \right)$ \\ 
\midrule 
F1 & Tension/compression spring design & 3 & 3 & 0 & 1.2665233E-02 \\
F2 & Pressure vessel design & 4 & 4 & 0 & 5.8853328E+03 \\
F3 & Three-bar truss design problem & 2 & 3 & 0 & 2.6389584E+02 \\
F4 & Welded beam design & 4 & 5 & 0 & 1.6702177E+00 \\
F5 & Weight Minimization of a Speed Reducer & 7 & 11 & 0 & 2.9944245E+03 \\
F6 & Gear train design Problem & 4 & 1 & 1 & 0.0000000E+00 \\
F7 & Cantilever beam design problem & 5 & 1 & 0 & 1.3395842E+00 \\
F8 & Planetary gear train design optimization problem & 9 & 10 & 1 & 2.3576871E-01 \\
\bottomrule 
\end{tabular*}
\end{table*}
\begin{table*}[!t]
  \centering
  \caption{Experimental results of 8 algorithms on 8 practical engineering design optimization problems}
    \resizebox{\textwidth}{!}{
    \begin{tabular}{cccccccccc}
    \hline
        &     & SMA & GWO & BWO & DBO & SSA & PSO & GJO & CGO \bigstrut\\
    \hline
    F1 & best & 1.27E-02 & 1.28E-02 & 1.32E-02 & 1.27E-02 & 1.27E-02 & 1.27E-02 & 1.28E-02 & \textbf{1.27E-02} \bigstrut[t]\\
        & std & 2.21E-03 & 7.45E-04 & 7.33E-03 & 2.76E-03 & 1.91E-03 & 2.01E-03 & 1.60E-03 & \textbf{1.61E-04} \\
        & mean & 1.49E-02 & 1.35E-02 & 1.96E-02 & 1.58E-02 & 1.42E-02 & 1.42E-02 & 1.46E-02 & \textbf{1.28E-02} \\
    F2 & best & 5.89E+03 & 5.96E+03 & 8.66E+03 & 5.94E+03 & 6.10E+03 & 6.23E+03 & 6.21E+03 & \textbf{5.89E+03} \\
        & std & 4.90E+02 & 3.88E+02 & 2.81E+04 & 5.51E+02 & 5.18E+02 & \textbf{3.73E+02} & 5.75E+02 & 5.51E+02 \\
        & mean & \textbf{6.32E+03} & 6.45E+03 & 3.28E+04 & 6.63E+03 & 6.79E+03 & 6.66E+03 & 7.00E+03 & 6.34E+03 \\
    F3 & best & 2.64E+02 & \textbf{2.64E+02} & 2.64E+02 & \textbf{2.64E+02} & \textbf{2.64E+02} & 2.64E+02 & 2.64E+02 & \textbf{2.64E+02} \\
        & std & 3.56E+00 & 9.28E-02 & 1.70E+00 & 7.40E-02 & 1.13E-01 & 8.52E-02 & 1.10E-01 & \textbf{1.00E-03} \\
        & mean & 2.70E+02 & 2.64E+02 & 2.66E+02 & 2.64E+02 & 2.64E+02 & 2.64E+02 & 2.64E+02 & \textbf{2.64E+02} \\
    F4 & best & 1.69E+00 & 1.71E+00 & 2.01E+00 & 1.70E+00 & 1.72E+00 & 1.72E+00 & 1.72E+00 & \textbf{1.69E+00} \\
        & std & 4.72E-01 & \textbf{8.46E-03} & 3.57E+96 & 2.11E-01 & 3.23E-01 & 7.76E-02 & 2.49E-02 & 3.75E-02 \\
        & mean & 1.91E+00 & 1.72E+00 & 6.52E+95 & 1.88E+00 & 2.03E+00 & 1.83E+00 & 1.76E+00 & \textbf{1.70E+00} \\
    F5 & best & 2.99E+03 & 3.03E+03 & 3.05E+03 & 2.99E+03 & 3.00E+03 & 3.01E+03 & 3.04E+03 & \textbf{2.99E+03} \\
        & std & \textbf{1.75E+00} & 1.30E+01 & 4.85E+01 & 3.31E+01 & 1.09E+01 & 7.17E+01 & 3.65E+01 & 7.48E+00 \\
        & mean & \textbf{3.00E+03} & 3.05E+03 & 3.14E+03 & 3.03E+03 & 3.00E+03 & 3.09E+03 & 3.11E+03 & 3.00E+03 \\
    F6 & best & 9.92E-10 & 2.31E-11 & 8.89E-10 & 8.89E-10 & 9.92E-10 & 2.31E-11 & 2.31E-11 & \textbf{2.70E-12} \\
        & std & 3.99E-08 & 1.77E-09 & 3.42E-07 & 1.61E-07 & 3.48E-07 & 4.58E-09 & 8.26E-09 & \textbf{1.02E-09} \\
        & mean & 3.49E-08 & 1.91E-09 & 1.42E-07 & 4.70E-08 & 1.00E-07 & 3.21E-09 & 5.76E-09 & \textbf{8.07E-10} \\
    F7 & best & 1.34E+00 & 1.34E+00 & 1.47E+00 & 1.34E+00 & 1.34E+00 & 1.34E+00 & 1.34E+00 & \textbf{1.34E+00} \\
        & std & 3.14E-03 & 1.10E-03 & 2.63E-01 & 8.85E-03 & 1.12E-02 & 1.14E-02 & 2.67E-03 & \textbf{4.28E-04} \\
        & mean & 1.34E+00 & 1.34E+00 & 1.82E+00 & 1.35E+00 & 1.35E+00 & 1.36E+00 & 1.35E+00 & \textbf{1.34E+00} \\
    F8 & best & \textbf{2.35E-01} & 2.37E-01 & 3.47E-01 & 2.40E-01 & 2.48E-01 & 2.37E-01 & 2.40E-01 & \textbf{2.35E-01} \\
        & std & 1.96E-01 & 8.78E-02 & 4.10E+101 & 1.64E+101 & 4.76E+102 & 4.38E-02 & 3.35E-01 & \textbf{2.17E-02} \\
        & mean & 2.92E-01 & 2.78E-01 & 2.37E+101 & 3.00E+100 & 2.75E+102 & 2.59E-01 & 4.76E-01 & \textbf{2.48E-01} \bigstrut[b]\\
    \hline
    \end{tabular}%
    }
  \label{tab:addlabel}%
\end{table*}%
\begin{figure}
	\centering
 		\includegraphics[width=0.8\textwidth]{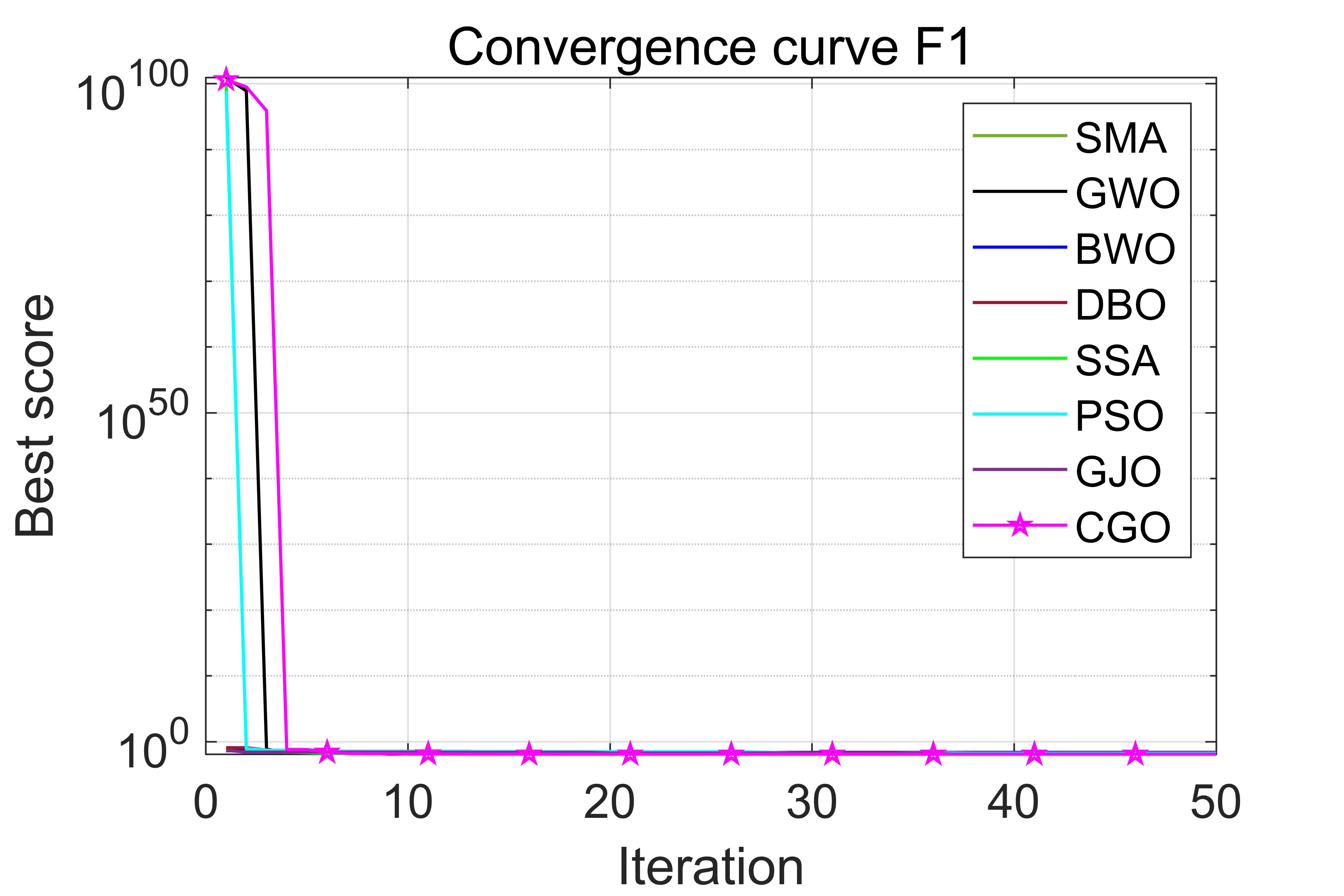}
	  \caption{Convergence curves of 8 algorithms on F1 engineering problems}\label{Fig.24}
\end{figure}
\begin{figure}
	\centering
 		\includegraphics[width=0.8\textwidth]{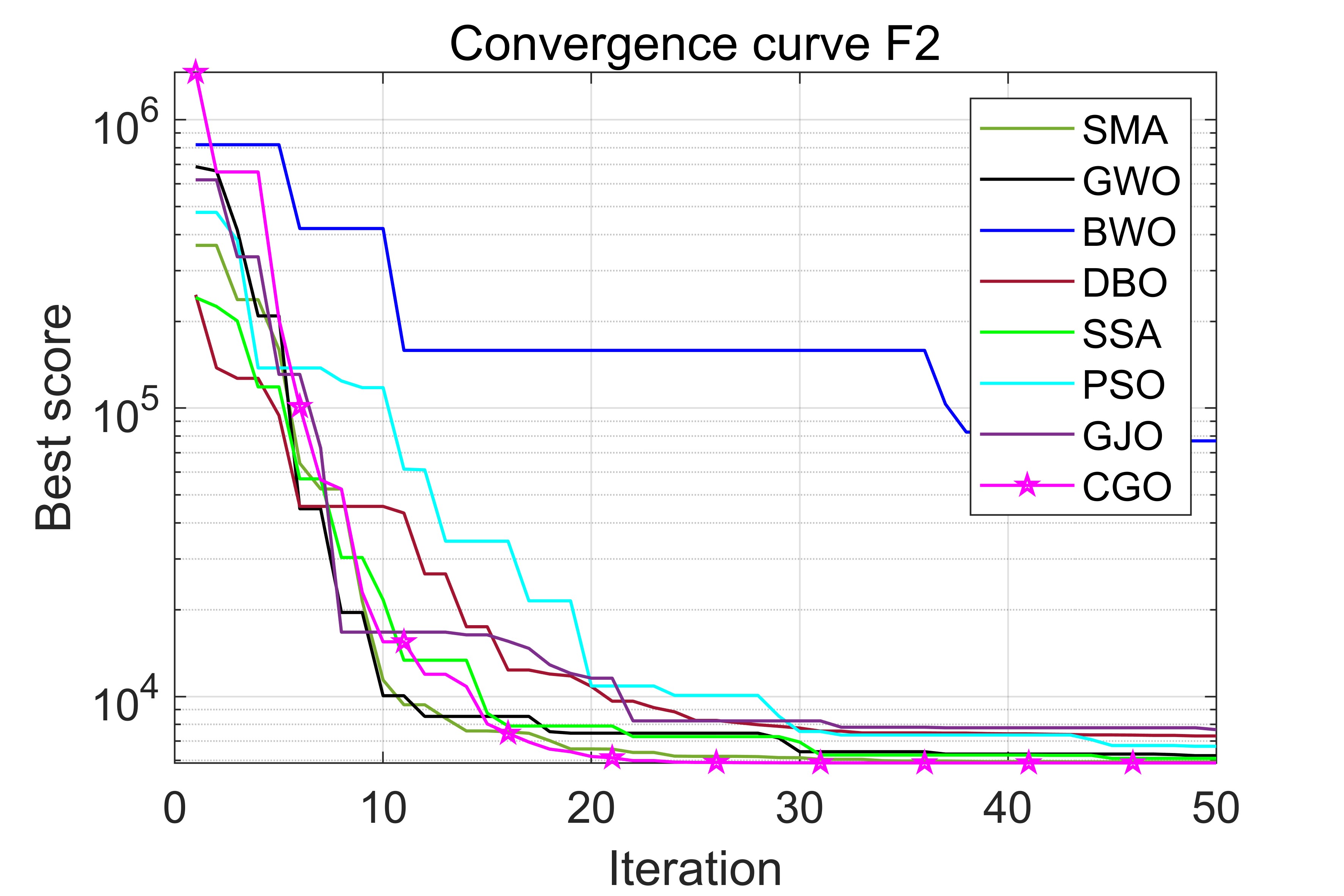}
	  \caption{Convergence curves of 8 algorithms on F2 engineering problems}\label{Fig.25}
\end{figure}
\begin{figure}
	\centering
 		\includegraphics[width=0.8\textwidth]{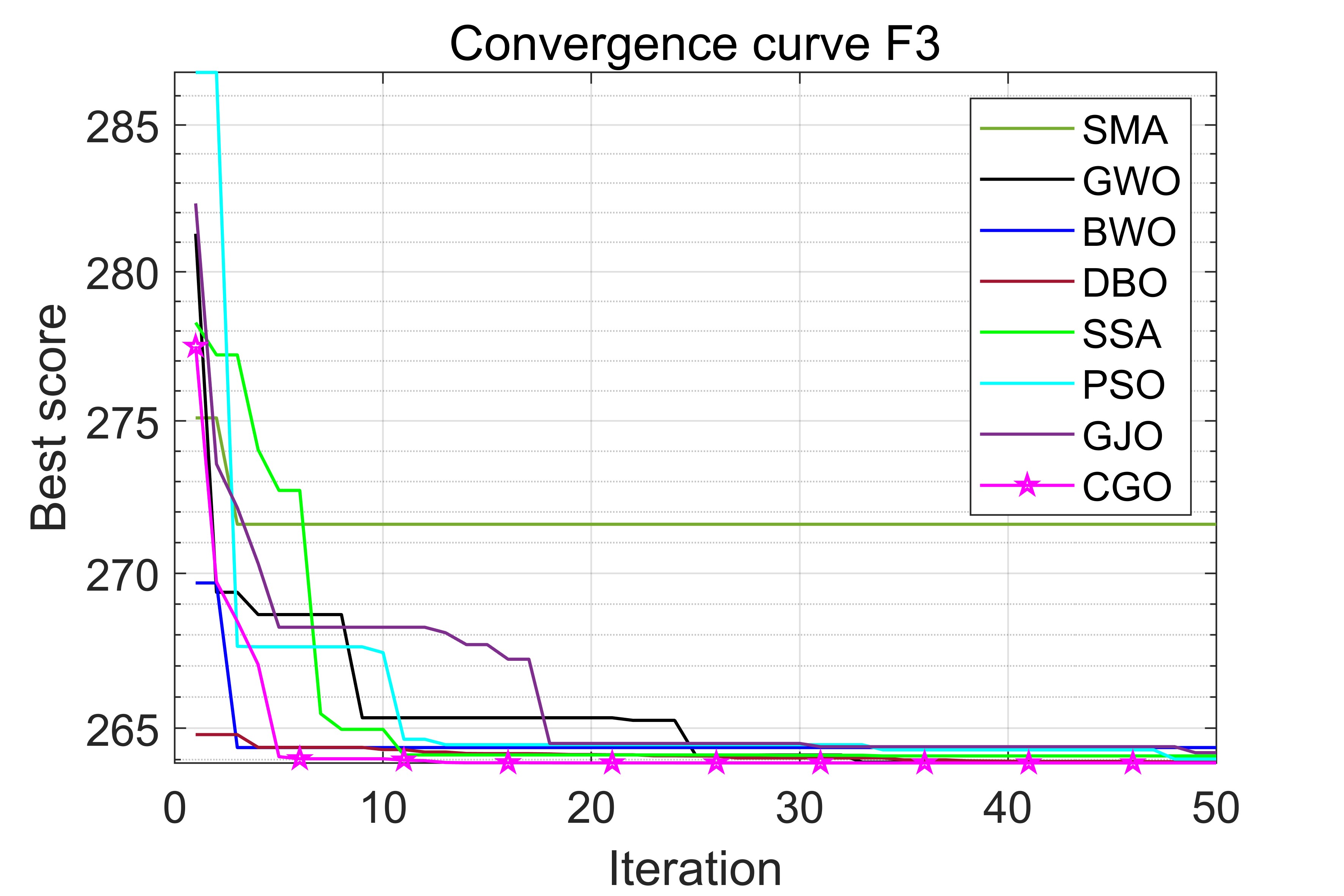}
	  \caption{Convergence curves of 8 algorithms on F3 engineering problems}\label{Fig.26}
\end{figure}
\begin{figure}
	\centering
 		\includegraphics[width=0.8\textwidth]{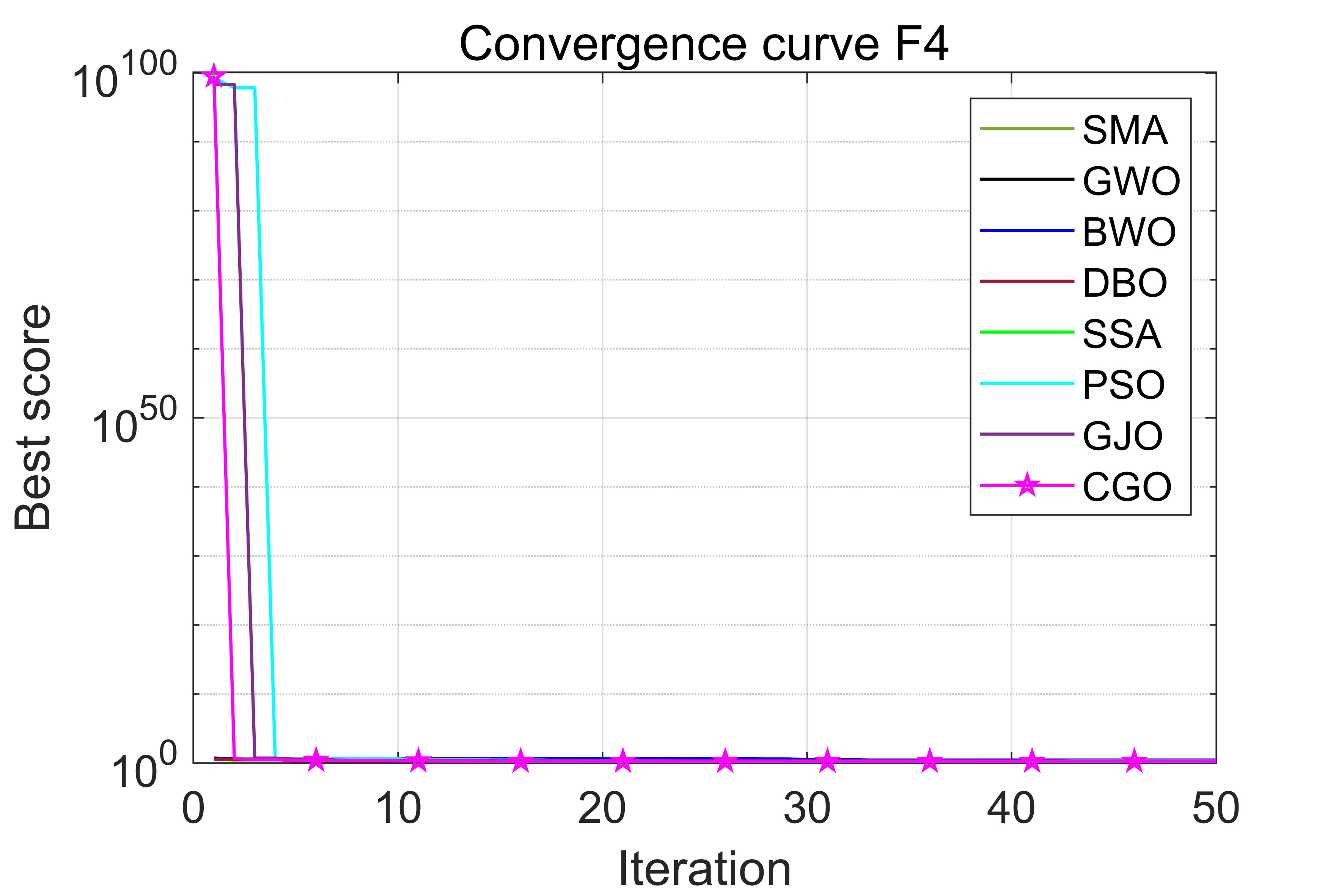}
	  \caption{Convergence curves of 8 algorithms on F4 engineering problems}\label{Fig.27}
\end{figure}
\begin{figure}
	\centering
 		\includegraphics[width=0.8\textwidth]{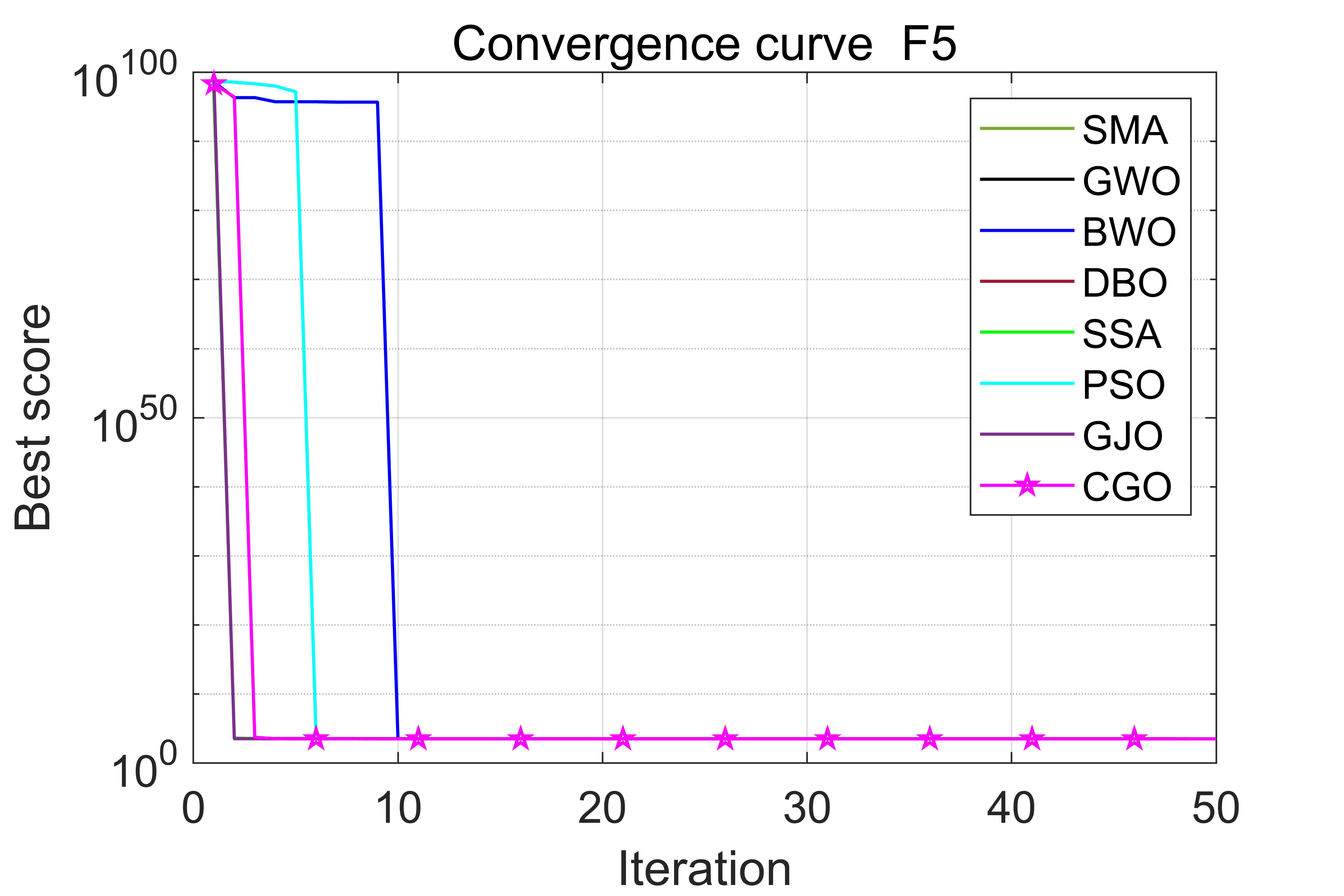}
	  \caption{Convergence curves of 8 algorithms on F5 engineering problems}\label{Fig.28}
\end{figure}
\begin{figure}
	\centering
 		\includegraphics[width=0.8\textwidth]{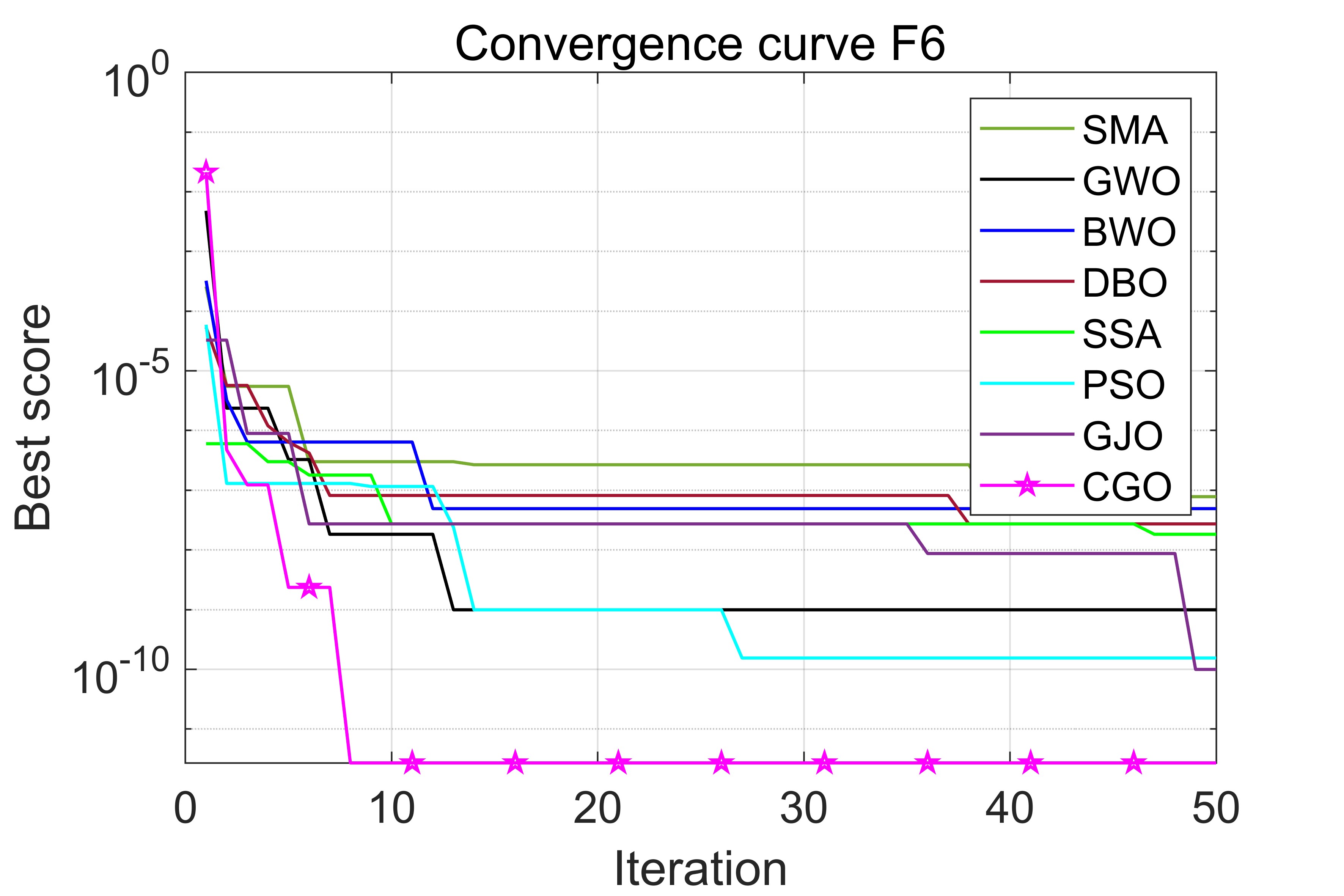}
	  \caption{Convergence curves of 8 algorithms on F6 engineering problems}\label{Fig.29}
\end{figure}
\begin{figure}
	\centering
 		\includegraphics[width=0.8\textwidth]{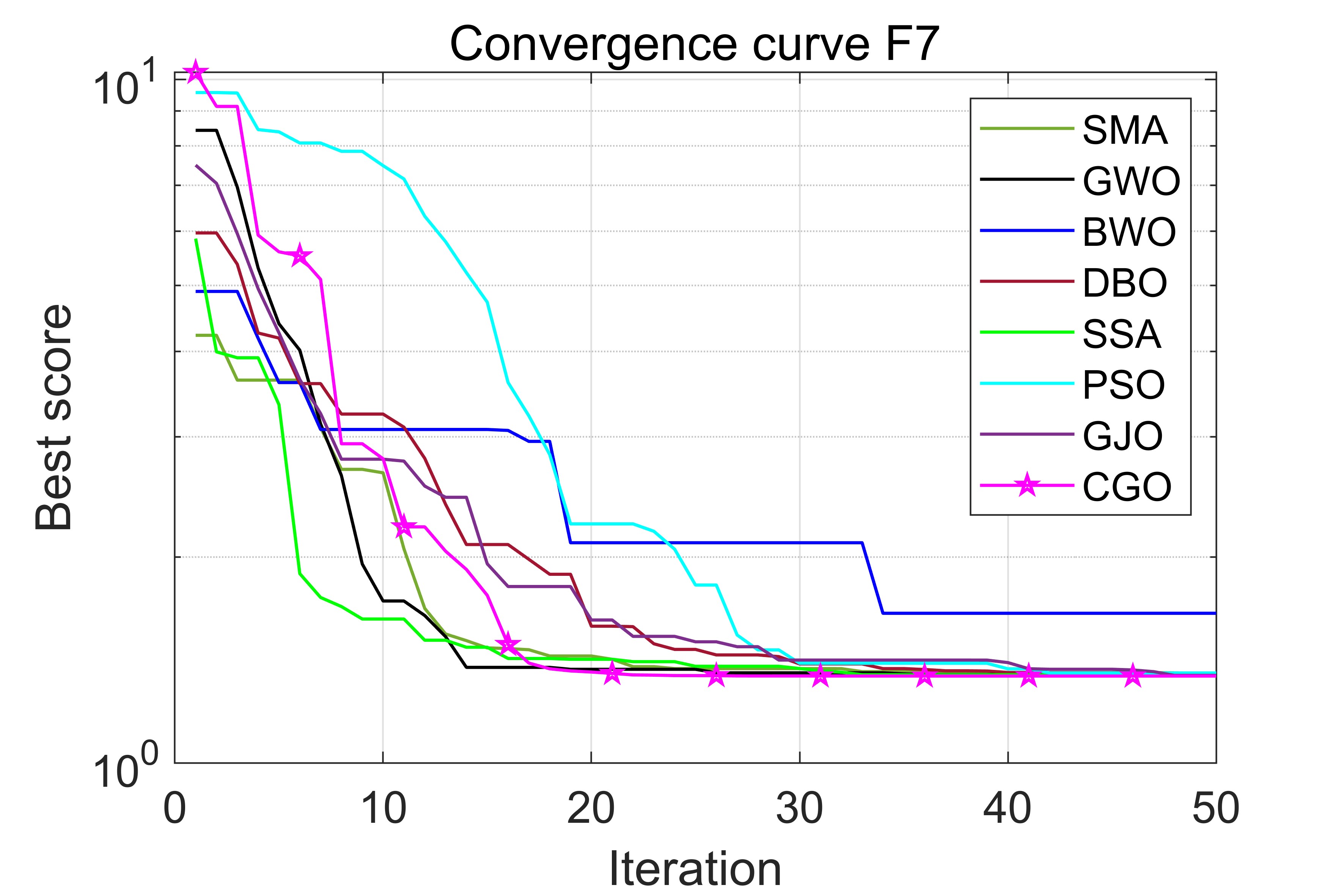}
	  \caption{Convergence curves of 8 algorithms on F7 engineering problems}\label{Fig.30}
\end{figure}
\begin{figure}
	\centering
 		\includegraphics[width=0.8\textwidth]{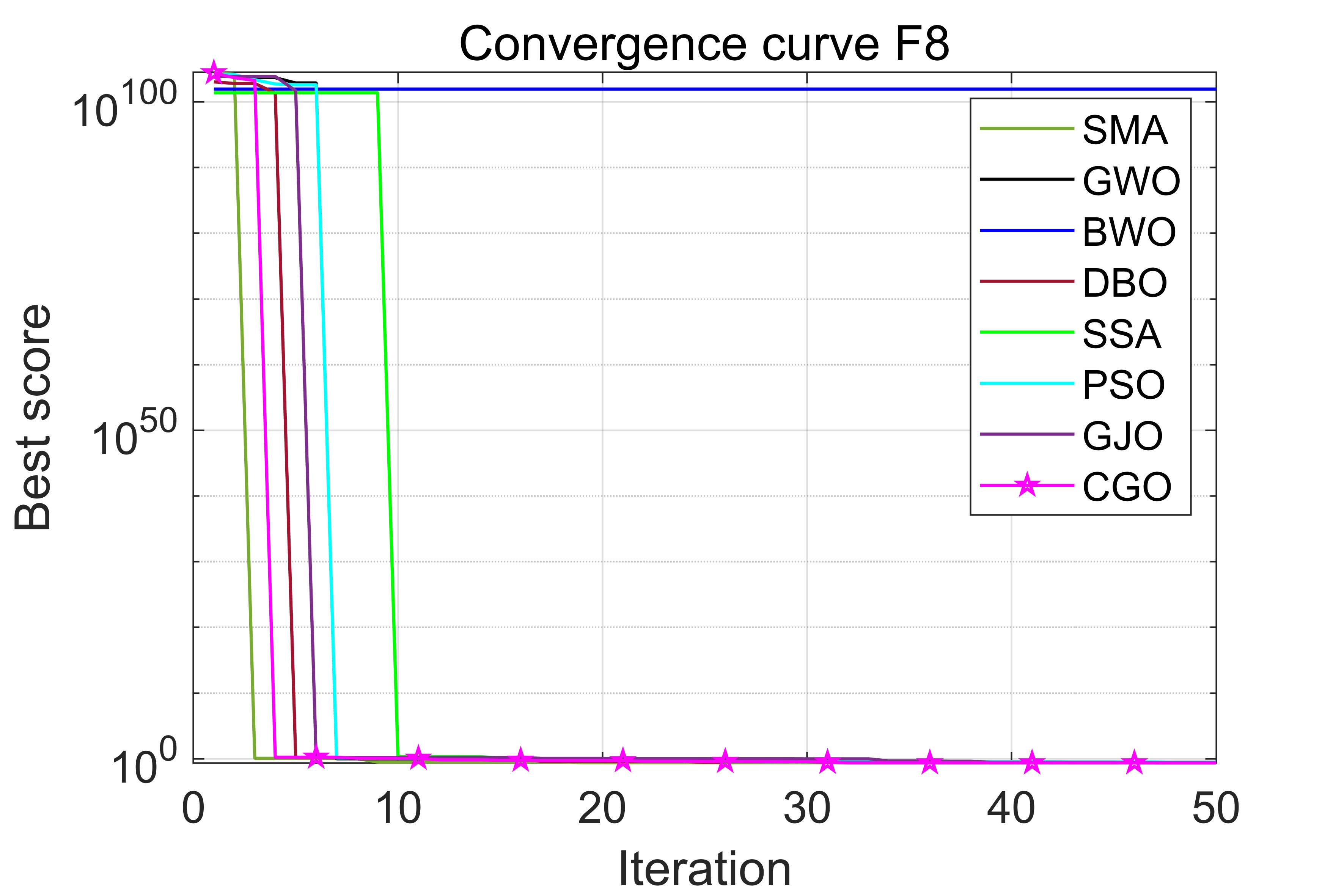}
	  \caption{Convergence curves of 8 algorithms on F8 engineering problems}\label{Fig.31}
\end{figure}

Table 7 describes the test results of 30 independent runs of 8 algorithms on 8 practical engineering design problems. It can be seen from the results that the proposed the CGO algorithm has strong competitiveness in practical engineering design problems. For F3 and F8 problems, although some algorithms also find the optimal value, the CGO algorithm is still ahead of other comparison algorithms in terms of standard deviation and mean value.
\subsection{Application to UAV flight path planning}
In this section, a more meaningful real-world engineering application is used to evaluate the advantages of the CGO over other recognized competitors. In the field of unmanned aerial vehicles, path planning is a key technical problem. In the working environment of UAV, it is particularly important to plan an optimal path from the starting point to the target point and around obstacles. The optimal path reduces drone flight time and energy consumption\cite{PhungHa-597}. In this study, the UAV flight path planning problem is transformed into an optimization problem, and the distance cost, obstacle cost and height cost are constructed as the objective functions of the optimization problem, and the optimal flight path of the UAV is obtained by minimizing the cost function. Therefore, the CGO algorithm and other five algorithms are used in this study to plan the flight path of UAVs.

In the whole flight process of the UAV, the flight path is composed of $N$ track points, and the track points correspond to the coordinate ${P_{ij}} = \left( {{x_{ij}},{y_{ij}},{z_{ij}}} \right)$ in the map. The cost function of UAV path length is as follows:
\begin{equation}
{F_1}\left( {{X_i}} \right) = \sum\limits_{j = 1}^{n - 1} {\left\| {\overrightarrow {{P_{ij}}{P_{i,j + 1}}} } \right\|}
\end{equation}
where $\left\| {\overrightarrow {{P_{ij}}{P_{i,j + 1}}} } \right\|$ represents the Euclidean distance between the $j$-th track point and the $j + 1$-th track point.

During the flight of UAV, when the distance ${d_k}$ between any path segment $\overrightarrow {{P_{ij}}{P_{i,j + 1}}} $ and the obstacle center ${C_k}$ is less than the radius ${R_k}$ of the obstacle, collision will occur. Set the safety width $S$.If the distance between the UAV and the obstacle is within the flight width, it indicates that the UAV has the risk of collision. Therefore, the flight path needs to keep a distance from the obstacle, and the obstacle cost function is constructed as follows:

\begin{equation}
\left\{ {\begin{array}{*{20}{c}}
{{F_2}\left( {{X_i}} \right) = \sum\limits_{j = 1}^{n - 1} {\sum\limits_{k = 1}^K {{T_k}\left( {\overrightarrow {{P_{ij}}{P_{i,j + 1}}} } \right).} } }\\
{{T_k}\left( {\overrightarrow {{P_{ij}}{P_{i,j + 1}}} } \right) = \left\{ {\begin{array}{*{20}{c}}
0&{{d_k} > S + {R_k}}\\
{S + {R_k} - {d_k},}&{{R_k} < {d_k} \le S + {R_k}}\\
\infty &{{d_k} < {R_k}}
\end{array}} \right.}
\end{array}} \right.
\end{equation}
where $K$ represents the number of obstacles and $n$ represents the number of track points. \[{T_k}\left( {\overrightarrow {{P_{ij}}{P_{i,j + 1}}} } \right)\] represents the distance of the track section between the $j$-th track point and the $j + 1$-th track point from the center of the $k$ obstacle.

The cost function of constructing height according to the height change information is as follows:
\begin{equation}
\left\{ {\begin{array}{*{20}{c}}
{{F_3}\left( {{X_i}} \right) = \sum\limits_{j = 1}^n {{H_{ij}}} }\\
{{H_{ij}} = \left\{ {\begin{array}{*{20}{c}}
{\left| {{h_{ij}} - \frac{{{h_{\min }} + {h_{\max }}}}{2}} \right|,}&{{h_{\min }} \le {h_{ij}} \le {h_{\max }}}\\
\infty &{otherwise}
\end{array}} \right.}
\end{array}} \right.
\end{equation}
where ${h_{\max }}$ and ${h_{\min }}$ are the maximum and minimum height of the flight path of the UAV, and ${h_{ij}}$ is the distance from the $j$-th track point of the $i$-th track to the ground.

The total cost function of UAV path planning is as follows:
\begin{equation}
\left\{ {\begin{array}{*{20}{c}}
{F\left( {{X_i}} \right) = \sum\limits_{l = 1}^3 {{\omega _l} \cdot {F_l}\left( {{X_i}} \right)} }\\
{\sum\limits_{l = 1}^3 {{\omega _l} = 1} }
\end{array}} \right.
\end{equation}
where ${\omega _l}$ is the weight coefficient.

The scenarios used for evaluation are based on real digital elevation model (DEM) maps derived from LiDAR sensors\cite{Australia-596}. The cylinders are used to represent obstacles, generating two scenarios based on the location and number of different obstacles. For comparison, all algorithms use the same parameters: the population number is set to 50, the maximum number of iterations is 300, the number of track points is chosen to be 50, and all algorithms are run 20 times to find the mean and standard deviation.

\begin{figure}
	\centering
 		\includegraphics[width=0.8\textwidth]{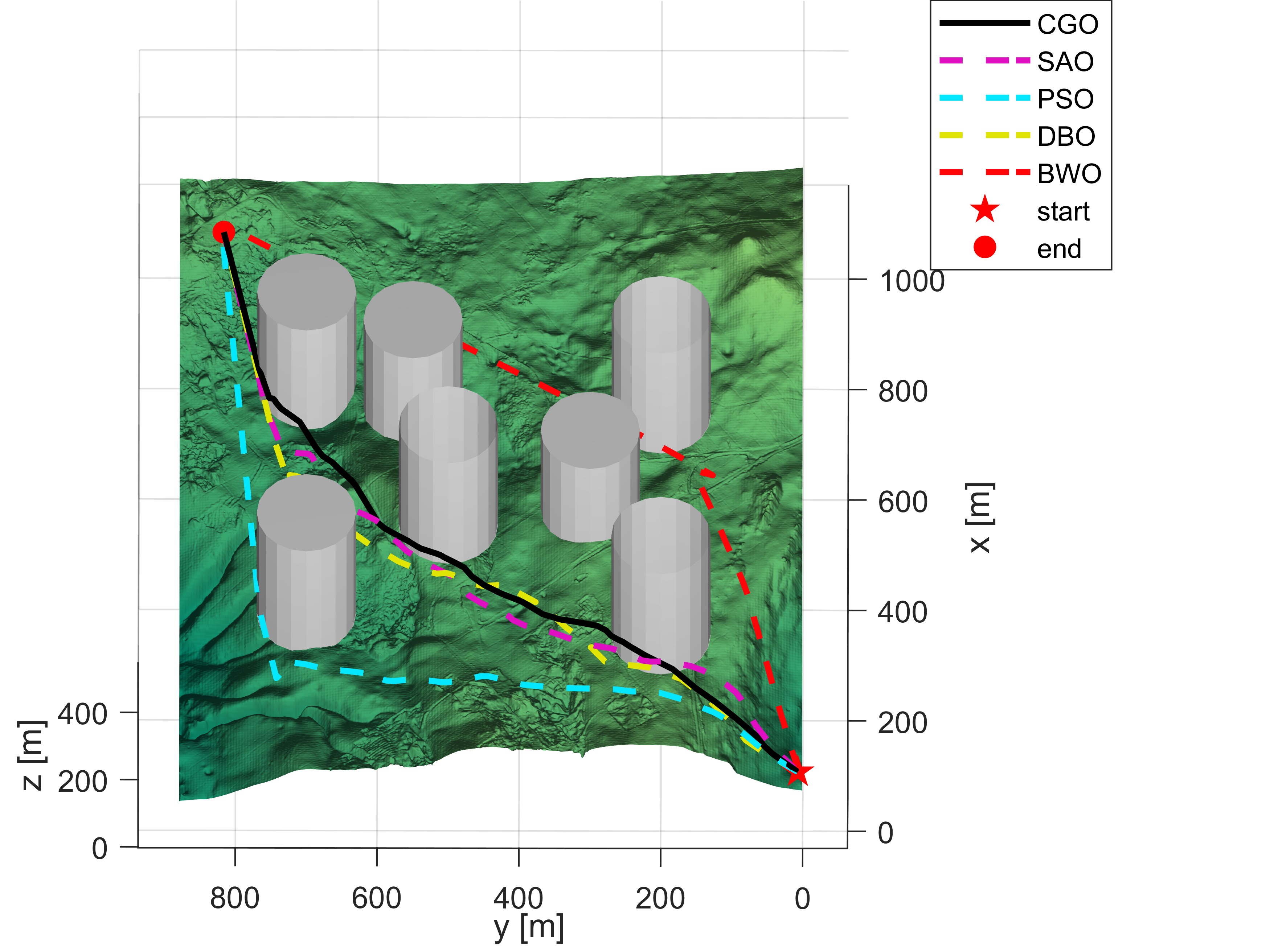}
	  \caption{3D view of the drone path planned by five algorithms under seven obstacles}\label{Fig.32}
\end{figure}

\begin{figure}
	\centering
 		\includegraphics[width=0.8\textwidth]{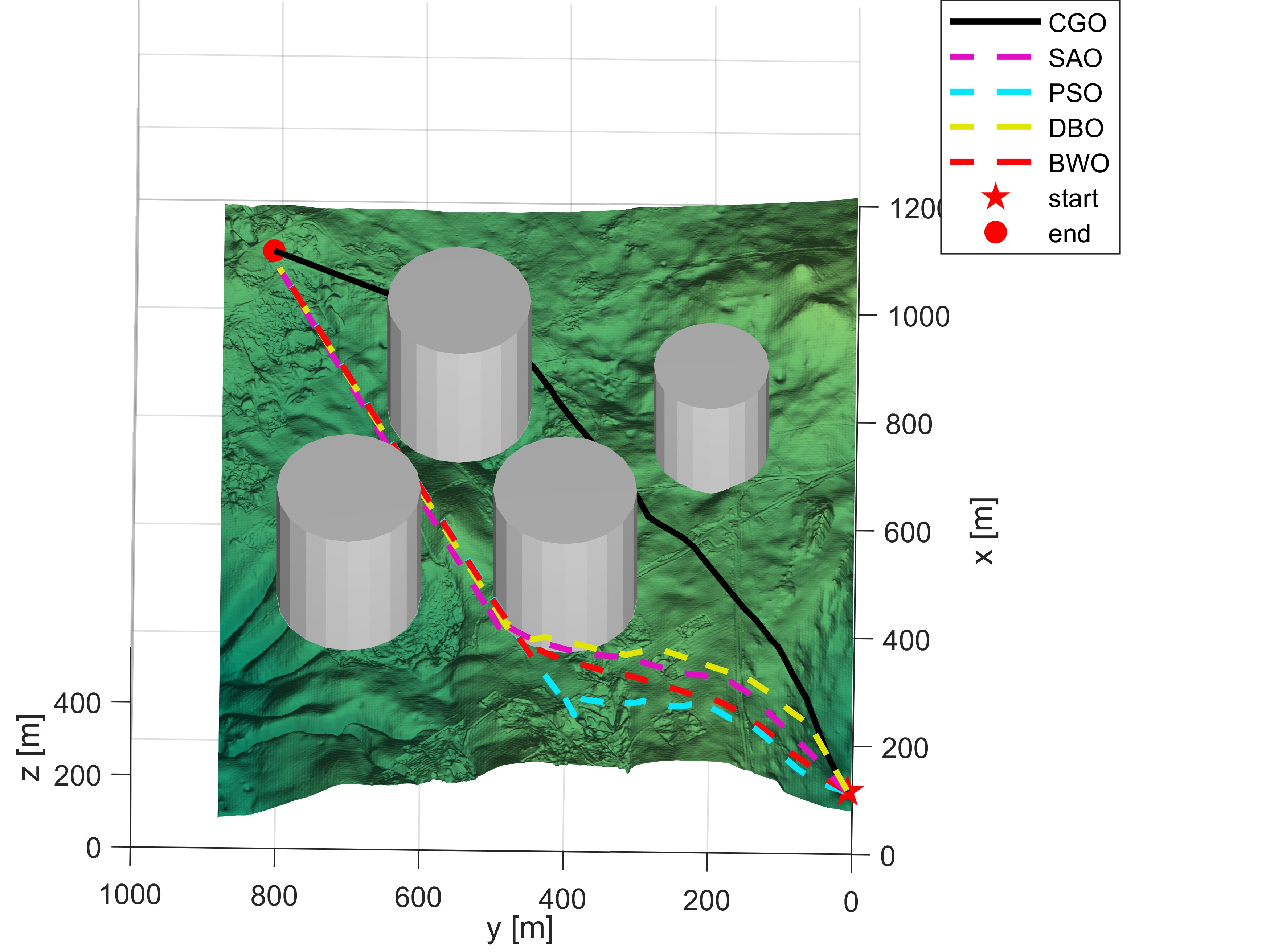}
	  \caption{3D view of the drone path planned by five algorithms under four obstacles}\label{Fig.33}
\end{figure}

\begin{figure}
	\centering
 		\includegraphics[width=0.8\textwidth]{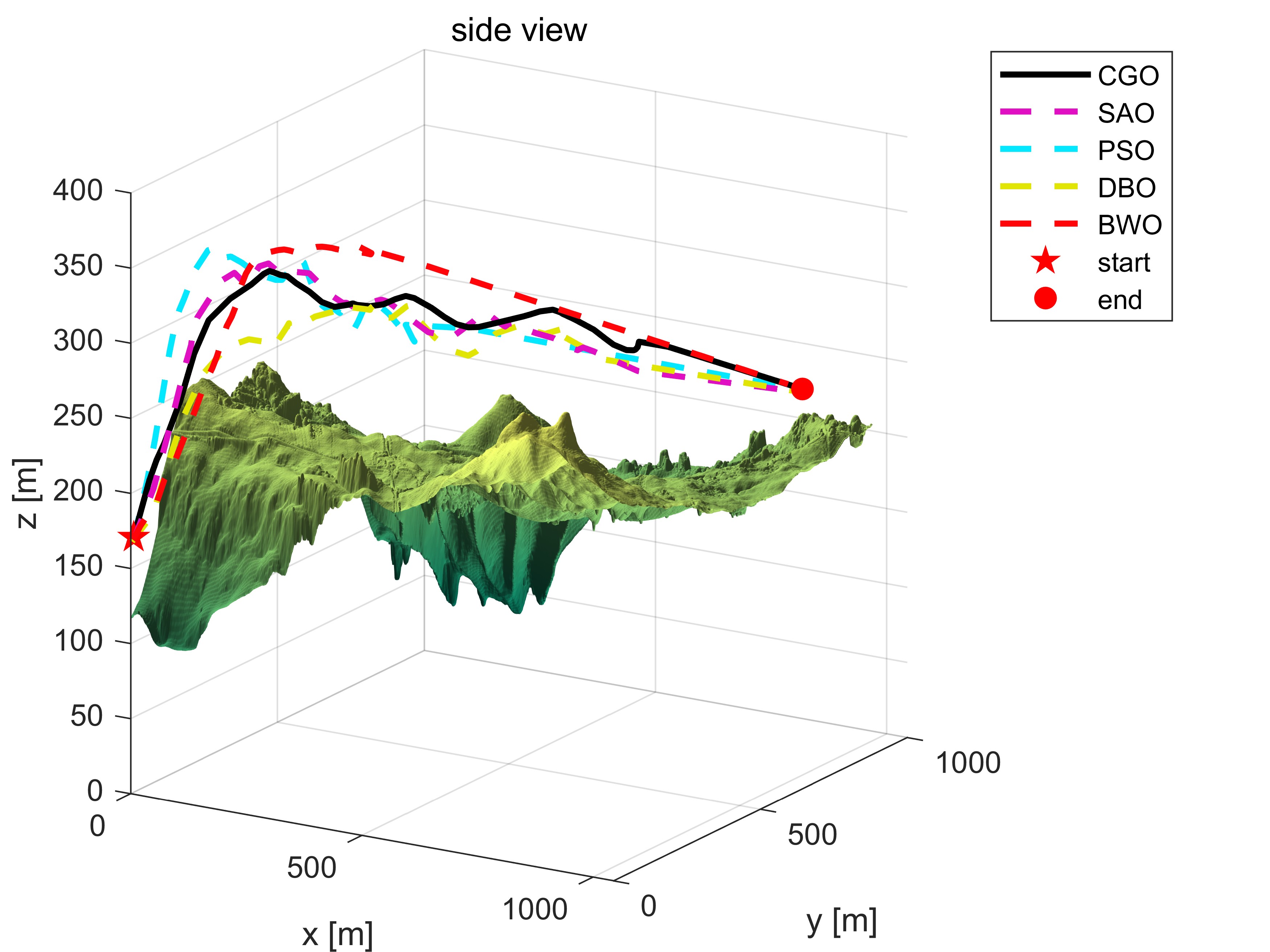}
	  \caption{Side view of the drone path planned by five algorithms under seven obstacles}\label{Fig.34}
\end{figure}

\begin{figure}
	\centering
 		\includegraphics[width=0.8\textwidth]{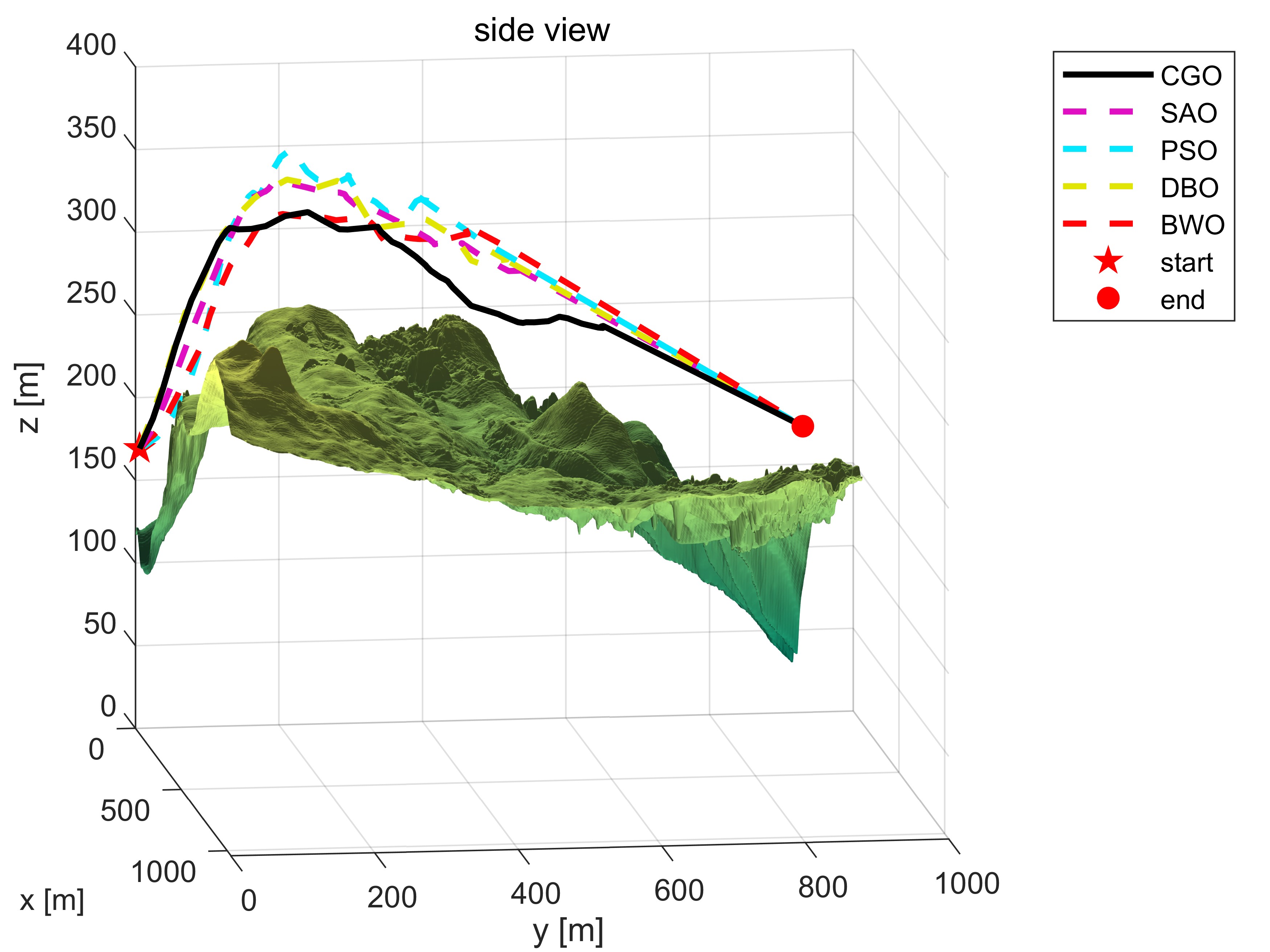}
	  \caption{Side view of the drone path planned by five algorithms under four obstacles}\label{Fig.35}
\end{figure}

\begin{figure}
	\centering
 		\includegraphics[width=0.8\textwidth]{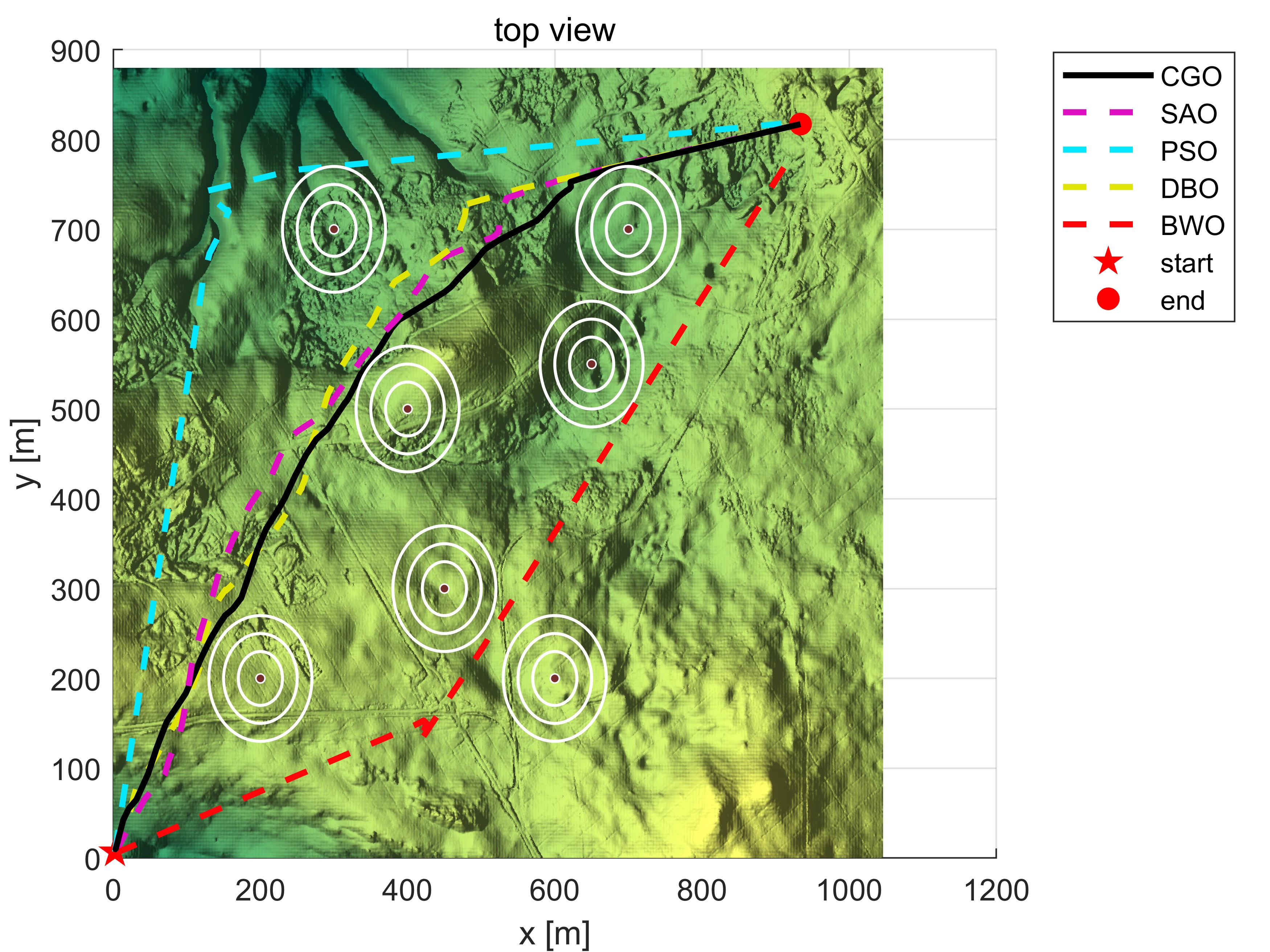}
	  \caption{Top view of UAV path planned by five algorithms under seven obstacles}\label{Fig.36}
\end{figure}

\begin{figure}
	\centering
 		\includegraphics[width=0.8\textwidth]{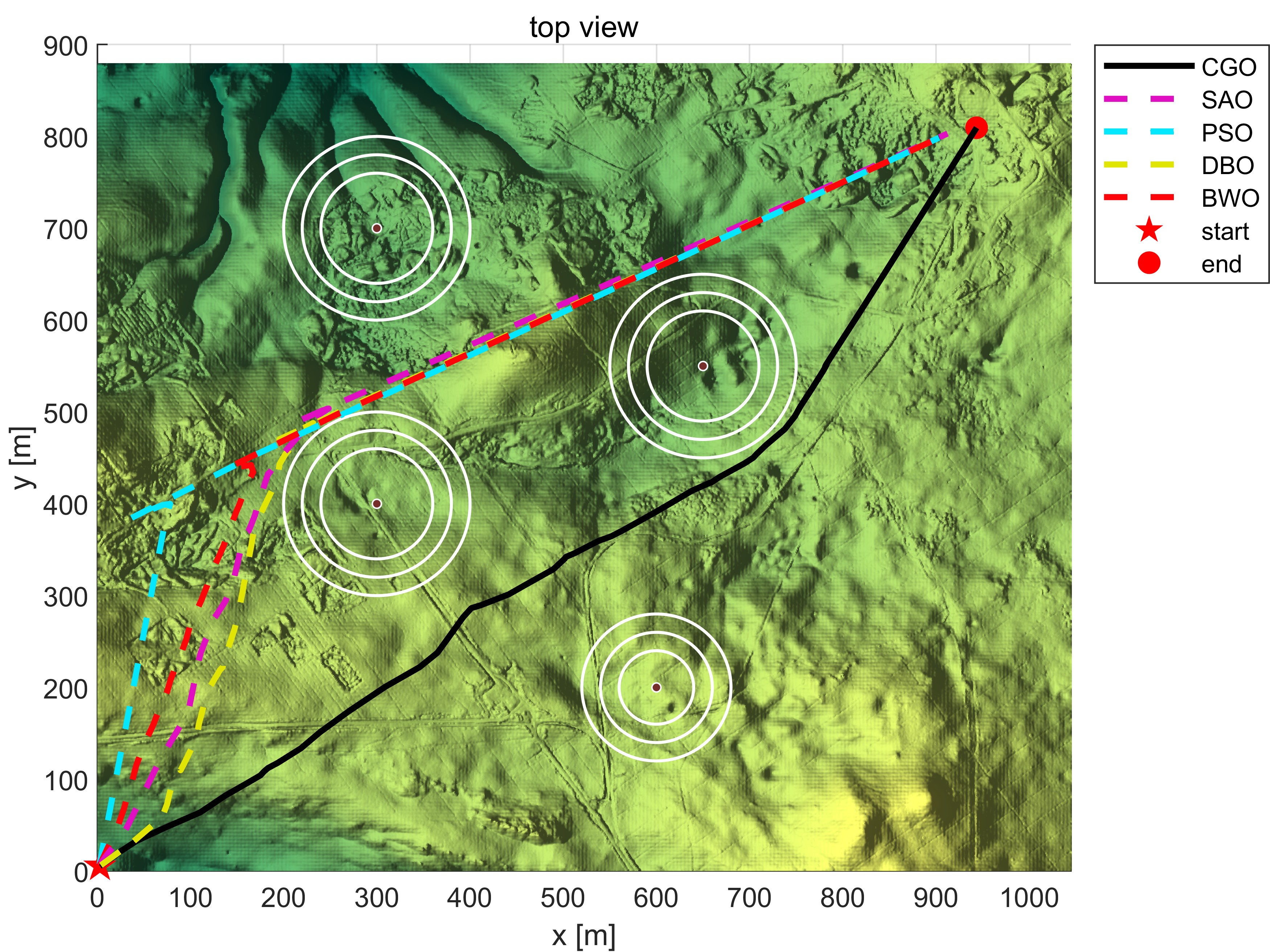}
	  \caption{Top view of UAV path planned by five algorithms under four obstacles}\label{Fig.37}
\end{figure}

In order to further verify the performance of the proposed algorithm, this paper applies the proposed algorithm and four other algorithms to the UAV path planning experiment. Two different obstacle environments are set up to fully verify the performance of the algorithm. Environment 1 sets up seven obstacles to simulate the situation of dense obstacles; Environment 2 sets up four obstacles to simulate a sparse obstacle environment. Figure 32 and Figure 33 respectively show the 3D view of the five algorithms in environment 1 and environment 2; Figure 34 and Figure 35 respectively show the side view of the experimental results in environment 1 and environment 2; Figure 36 and Figure 37 show the top view of environment 1 and environment 2. From the top view and 3D view, it can be seen that the proposed algorithm can effectively avoid obstacles and search the shortest path compared with the other four algorithms. From the side view, the CGO algorithm has higher flight efficiency in climb Angle and dive Angle, which is incomparable to the other four algorithms.

\begin{table*}[htbp]
  \centering
  \caption{The results of five algorithms running independently thirty times in two scenarios}
    \begin{tabular*}{\linewidth}{@{}ccccccc@{}}
    \hline
        &  & SAO & PSO & DBO & BWO & CGO  \bigstrut\\
    \hline
    Scenario one & best & 1056.57 & 1158.48 & 1072.49 & 1121.98 & \textbf{1017.69} \bigstrut[t]\\
      & std & 15.19 & 27.42 & 21.29 & \textbf{11.15} & 19.22  \\
      & mean & 1078.90 & 1231.63 & 1115.37 & 1158.06 & \textbf{1049.02}  \\
    Scenario two & best & 1005.76 & 1080.56 & 1021.21 & 1071.81 & \textbf{965.29}  \\
      & std & \textbf{9.73} & 27.92 & 16.23 & 18.52 & 17.50  \\
      & mean & 1024.22 & 1118.80 & 1.48.91 & 1113.50 & \textbf{992.40} 
    \bigstrut[b]\\
    \hline
    \end{tabular*}%
  \label{tab:addlabel}%
\end{table*}%

Table 8 lists the optimal values, standard deviations, and average values of the five algorithms that are independently run thirty times in two scenarios. It can be seen from the data that the CGO is far ahead of other comparison algorithms in terms of optimal value and average value of the objective function.
\begin{figure}
	\centering
 		\includegraphics[width=0.8\textwidth]{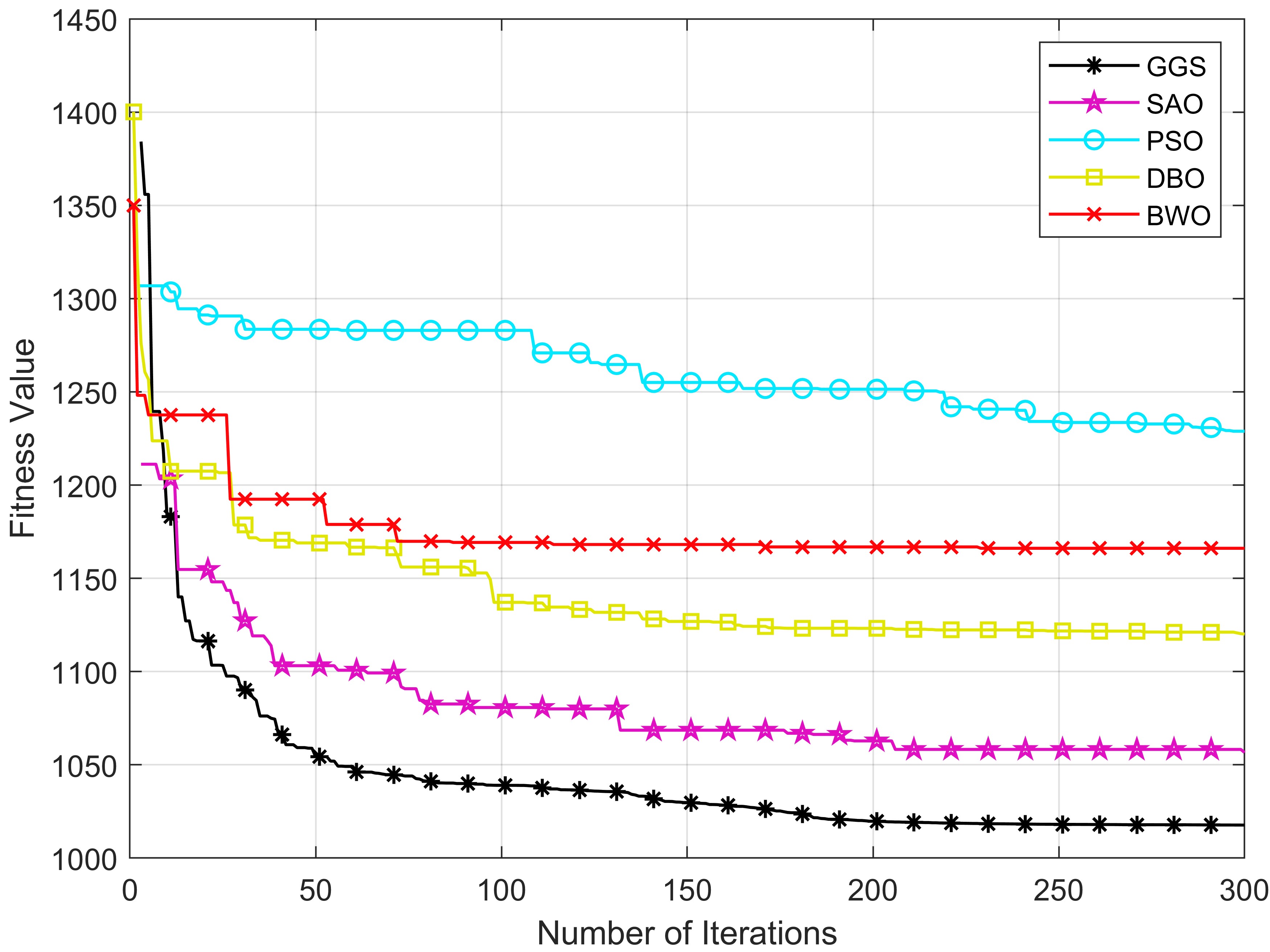}
	  \caption{Objective function convergence curves of five algorithms under seven obstacles}\label{Fig.38}
\end{figure}

\begin{figure}
	\centering
 		\includegraphics[width=0.8\textwidth]{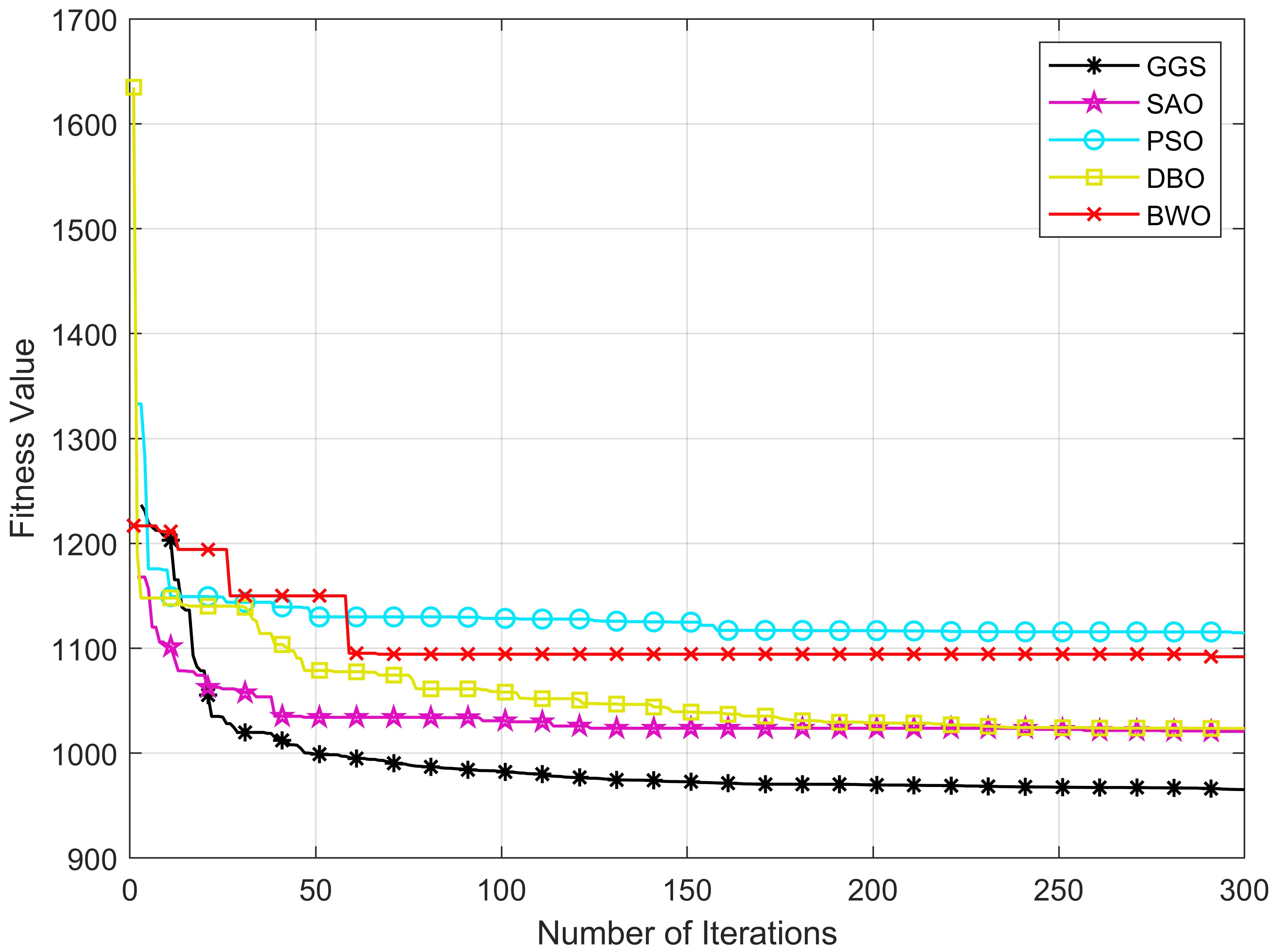}
	  \caption{Objective function convergence curves of five algorithms under four obstacles}\label{Fig.39}
\end{figure}

Figure 39 and Figure 40 reflect the convergence of objective functions of the five algorithms in two scenarios. As you can see from the graph, all algorithms converge in a similar way. However, the CGO algorithm can converge to the optimal value in a shorter time, which proves that the algorithm has excellent optimization ability from the side.

\section{Conclusion}
In this paper, inspired by the game mechanism and player behavior of competitive games, a new meta-heuristic optimization algorithm named competitive game optimization (CGO) algorithm is proposed. The exploration and exploitation phase of the algorithm is achieved by simulating the player's behavior of searching for supplies and fighting, as well as heading for a safe zone. By introducing encounter probability to achieve a balance between exploration and exploitation, the algorithm can jump out of local optimal and prevent premature convergence. 

The performance of the CGO was verified by 29 CEC2017 and 12 CEC2022 unconstrained benchmark functions. The CGO achieved first place on 17 out of 29 CEC2017 test functions and first place on 6 out of 12 CEC2022 test functions. The standard deviation and mean value of other functions also have different degrees of transcendence, which has strong competitiveness among similar algorithms. Finally, the algorithm is successfully applied to eight practical engineering design optimization problems in different fields, and compared with seven similar algorithms. The comparison results show that CGO algorithm has obtained the first place in solving eight display engineering design problems. It shows that CGO algorithm has great advantages in solving practical engineering design problems. 

Finally, the CGO algorithm and four similar algorithms are successfully applied to UAV flight path planning and compared. The comparison results show that the CGO algorithm has strong performance in UAV path planning application. Because of its simple structure and good performance in practical engineering design problems, the CGO algorithm can be extended to more challenging problems in other scientific fields. Therefore, the application of the CGO image segmentation will be the future research direction.

\section*{Declaration of Competing Interest}
All authors have no known conflicts of interest
\section*{Acknowledgments}
This work was supported by the Natural Science Foundation of Henan under grant No.242300421716, the Key Science and Technology Program of Henan Province under grant No.242102220044 and 242102210034, the National Natural Science Foundation of China under Grant 62103379, and the Maker Space Incubation Project under Grant No.2023ZCKJ102.










\bibliographystyle{elsarticle-num}

\bibliography{CGO}



\end{document}